%% file: main.tex
\documentclass{aa}
\usepackage{graphicx}
\usepackage{txfonts}
\usepackage{color}
\usepackage{xspace}
\usepackage{siunitx}
\usepackage{amsmath, amsfonts}
\usepackage{amssymb}
\usepackage{xcolor}
\usepackage{dashbox}
\usepackage{framed}
\usepackage{lipsum}
\usepackage{placeins}
\usepackage{stfloats}
\usepackage{picture}
\usepackage{multirow}
\usepackage{tikz}
\usepackage{hyperref}

\usetikzlibrary{shapes.geometric, arrows.meta, positioning, fit, backgrounds, calc}
\definecolor{procfill}{HTML}{EAF2FB}   
\definecolor{procdraw}{HTML}{2C5F8A}   
\definecolor{iofill}{HTML}{E8E8E8}    
\definecolor{parfill}{HTML}{FFE8CC}    
\definecolor{pardraw}{HTML}{D9730D}   

\hypersetup{
        colorlinks=true, 
        breaklinks=true,
        linkcolor=blue, % color of internal links
        citecolor=blue, % color of links to bibliography
        filecolor=blue, % color of file links
        urlcolor=blue,
        unicode=false, % non-Latin characters in Acrobat bookmarks
        pdftoolbar=true, % show Acrobat toolbar
        pdfmenubar=true, % show Acrobat menu
        pdffitwindow=false, % window fit to page when opened
        pdfstartview={Fit}, % fits the width of the page to the window
        pdftitle={}, % title
        pdfauthor={Alena Rottensteiner}, % author
        pdfsubject={},
        pdfcreator={Alena Rottensteiner}, % creator of the document
        pdfkeywords={},
        pdfnewwindow=true, % links in new window
        pdfdisplaydoctitle=true % display document title instead of file name
}

\begin{document}

% ============================================================================ %
\title{A comprehensive cluster census of Orion}
\subtitle{An application of the Significance Mode Analysis (SigMA) algorithm}

% ============================================================================ %
% authors
\author{Alena Rottensteiner\inst{1}             \and
        Sebastian Ratzenböck\inst{1,2,3}        \and  
        Jo\~ao Alves\inst{1,2}                  \and 
        Stefan Meingast\inst{1}                 \and 
        Sebastian Hutschenreuter\inst{1}
}

% affiliations
\institute{Department of Astrophysics, University of Vienna, Türkenschanzstrasse 17, 1180 Wien, Austria 
\\ \email{alena.kristina.rottensteiner@univie.ac.at}
\and
University of Vienna, Research Network Data Science at Uni Vienna, Kolingasse 14-16, 1090 Wien, Austria
\and
Center for Astrophysics | Harvard \& Smithsonian, 60 Garden St., Cambridge, MA 02138, USA
}

% ============================================================================ %
\date{Received 15 August 2026 / Accepted xxx}

% ============================================================================ %
\abstract
{%Context
Precise astrometric surveys and modern clustering algorithms are working in step to transform our view of star-forming regions. By revealing a much richer substructure than previously accessible, they pave the way for reconstructing star formation histories by accurately resolving and age-dating individual sub-populations. The Orion star-forming complex is the best-studied stellar nursery in the solar neighborhood and the nearest one currently forming massive stars. Even so, a comprehensive characterization of its substructure, including a homogeneous age mapping and extinction analysis, is still incomplete. 
% Aims + methods
Here, we present the most complete census of stellar populations across the entirety of the Orion complex from the newest version of the Significance Mode Analysis (\texttt{SigMA}) clustering algorithm. 
We also outline our additions and improvements to the algorithm that have extended its usage to distant ($>300$ pc) regions. 
% Results
We separate the Orion complex into 47 co-spatial and co-moving stellar groups comprising 11,996 reliable members, with ages ranging from 1.5 to 25~Myr. To evaluate the statistical robustness of each group, we derive cluster persistence values and individual membership probabilities for each source from 10 independent clustering repetitions. Our group memberships agree well with the literature, but \texttt{SigMA} consistently finds a factor of $\sim$2--3 more members. In particular, it resolves more very young populations, such as NGC~2024, RV~Orionis, B~30, NGC~1977, NGC~2068, and NGC~2071, than previous algorithms. In addition to recovering 28 known clusters and three groups previously classified as substructures, we present 16 new co-eval substructure candidates of the Orion star-forming complex.
% Conclusion
This work builds up a new high-resolution time-resolved picture of Orion. This spatio-temporal map allows us to relate its stellar content to the surrounding ISM and paves the way for a detailed analysis of its star formation history in the future.
}

% ============================================================================ %
\keywords{Methods: data analysis -- Stars: kinematics and dynamics -- Stars: pre-main sequence -- (Galaxy:) open clusters and associations: individual: Orion}
% ============================================================================ %
\maketitle
\nolinenumbers  

% ============================================================================ %
\section{Introduction}
\label{sec:Introduction}

The study of star formation regions provides access to crucial astrophysical parameters, from large scales, like molecular cloud formation and dispersal, to small-scale events, such as stellar disk and planet formation timescales. Star formation in a molecular cloud complex can last for millions of years and produce a wealth of individual populations with distinct ages and kinematics. Massive stars, with their ability to strongly influence their gaseous surroundings through feedback, are particularly important for investigating the physical drivers of star formation.

The nearest active stellar nursery producing massive stars is the Orion molecular cloud complex. Situated in the region of its namesake constellation, it spans approximately $10^{\circ}\times20^{\circ}$ on the sky and extends between the molecular cloud edges around 320 and 500 pc \citep{Brown_1994, Kubiak_2017, Grossschedl_2018, Grossschedl_2021}. Housing two giant molecular clouds, Orion A and B, it contains several very young, still-embedded populations such as the Orion Nebula cluster (ONC), NGC 2024, NGC 2068, and NGC 2071, and optically visible populations such as $\lambda$~Ori, $\sigma$~Ori, and NGC~1980. Over the last $12$~Myr alone, more than $10^4$ stars are thought to have formed throughout the region in several sub-populations, while each molecular cloud still contains an estimated $10^5$~M$_{\odot}$ \citep{Bally_2008}.

The structure, number, and formation scenario of the subgroups that comprise the Orion complex have been debated for nearly 80 years. Initially defined by \cite{Blaauw_1964} as the Orion OB~1 association, the region was divided into four groups (a-d) based on its on-sky (2D) stellar density distribution: northwest of the Belt stars (a), the Belt region with Belt stars (b), the sword region (c), and the ONC region including the Trapezium stars (d). They were first thought to have formed sequentially \citep{Blaauw_1964, Elmegreen_1977}, from the dust-free OB~1a ($\sim$ 12 Myr) to the still embedded OB~1d ($<2$~Myr) region. However, it was later discovered that some groups are partially superimposed along the line of sight, and the original formation scenario was contested \citep[e.g.,][]{Brown_1994}.

Since early astrometric surveys like \textsc{Hipparcos} \citep{HIPPARCOS_1997}, and \emph{Gaia} DR1 \citep{Gaia_DR1_2016} lacked the accuracy and stellar quantity to resolve the three-dimensional structure of the region \citep[e.g.,][]{Zari_2017}, most population studies were conducted using only radial velocity data \citep[e.g.,][]{Jeffries_2006, Sacco_2008, Furesz_2008, Tobin_2009, Hernandez_2014, daRio_2016, daRio_2017, Kubiak_2017}, or photometry e.g., \cite{Alves_2012, Bouy_2014}. Owing to the popularity and importance of the region, as well as the diverse study methods and the lack of true 5D astrometry, the populations of Orion have become a much-contested topic. Among the points of contention are, for example, the presence of an extended foreground population of the ONC at $\sim300$~pc (\citealp{Alves_2012, Pillitteri_2013, Bouy_2014}; cf. \citealp{daRio_2016, Fang_2017, Kounkel_2017a}), as well as the presence of multiple populations in the cluster (\citealp{Beccari_2017, Jerabkova_2019}; cf. \citealp{Alzate_2023}). A few studies also measured proper motions in Orion, but their findings were either associated with significant uncertainties \citep{Zacharias_2013} or had a very limited scope \citep{Dzib_2017}. Studies of the molecular clouds were conducted as well \citep[e.g.,][]{Megeath_2012, Hacar_2016, Hacar_2018, Hacar_2024, Socci_2024}.

Only with the revolution of \emph{Gaia}'s second and third data releases \citep{Gaia_DR2_2018, Gaia_DR3_2023} came a better understanding of the structure of Orion and a revival of the discussion around its sub-populations. Large-scale spectroscopic surveys such as APOGEE-2 \citep{APOGEE-2_2020} enabled revolutionary 3D kinematic investigations of the current state and the past of Orion \citep{Kounkel_2017b, Kounkel_2018, Grossschedl_2018, Kounkel_2020, Swiggum_2021, Grossschedl_2021}. Regarding its stellar content and subpopulations, many recent, density-based clustering efforts have been made, utilizing machine learning algorithms and \emph{Gaia}: Using DR2 data, \cite{Kounkel_2018} identified five spatially and kinematically distinct groups in the OB 1 association, whereas \cite{Zari_2019} found 15 groupings. \cite{Chen_2020} defined a total of 22 separate sub-populations, also using DR2. Analyses of \cite{Kerr_2023} explored potential connections between Orion and neighboring star formation sites using \emph{Gaia} DR3. However, their study did not investigate its interior populations. Most recently, \cite{Sanchez_2024} identified 27 stellar groups across the complex. They divide them into 13 large structures, 5 of which are substructured, and 14 small structures distinct from the large structures. Regardless of the study, a lot more populations than initially assumed were found, but no clear pattern between position and age emerged. 

In fact, improved astrometry enabled astronomers to resolve the structure of many nearby star-forming regions and clusters in greater detail than ever before. Even so, no large-scale patterns of star formation, such as age gradients along the positions of different sub-populations, were found \citep{Wright_2023}, with the exception of one region: the Scorpius-Centaurus (Sco-Cen) OB association at $\sim$118-145~pc \citep{deBruijne_1999}. \citet{Kerr_2021} discovered many groups and suspected a sequential formation history. \cite{Ratzenboeck_2023a} developed \texttt{SigMA} (Significance Mode Analysis), a non-parametric, density-based clustering algorithm tailored to the 5D positional and kinematic phase space accessible via \emph{Gaia}. It identified 34 distinct groups in Sco-Cen \citep{Ratzenboeck_2023a}, compared to only three originally defined by \cite{Blaauw_1964}. The method demonstrated sensitivity to stellar volume densities of $\sim$0.01~stars~pc$^{-3}$ and population velocity differences of $\sim$0.5~km~s$^{-1}$. An age analysis of these groups revealed two $\sim$100~pc-long chains of clusters with visible age gradients, providing direct evidence of large-scale spatio-temporal patterns in Sco-Cen's star formation history \citep{Ratzenboeck_2023b, Posch_2023, Posch_2025}.

Here, we apply \texttt{SigMA} to the Orion molecular cloud complex — the nearest active massive star-forming region and one of the most popular testbeds for star and planet formation studies. Doing so requires confronting a challenge absent in the original application: At $d\sim400$~pc and with a line-of-sight (LOS) extent of $\sim200$~pc, Orion covers a substantially larger volume than Sco-Cen, and \emph{Gaia}'s parallax signal-to-noise ratio is both poorer and more variable across the volume. We account for these effects with \texttt{DistantSigMA}, a distance-adaptive extension of \texttt{SigMA} developed for this work, and present the first complete kinematic census of stellar populations across the full Orion complex.
% ============================================================================ %
\section{Data}
\label{sec:Data}

We cluster on an approximately rectangular box with a volume of around $25.5~\times~10^6~\si{pc}^3$, using\emph{Gaia} DR3 catalog data. The bounds are chosen generously to encompass the entire Orion A and B molecular clouds and include other larger-scale structures as well \citep{Beccari_2020, Tian_2020}. The box is defined in a heliocentric Galactic Cartesian coordinate frame as
\begin{equation}
    \label{eq:box-coords}
    \begin{aligned}
        -500~\text{pc} \leq~&\text{X} < -140~\text{pc}\\    %360
        -310~\text{pc} \leq~&\text{Y} < -15~\text{pc}\\     %295
        -250~\text{pc} \leq~&\text{Z} < -10~\text{pc}.\\    %240
    \end{aligned}
\end{equation}
We apply the astrometric quality constraints defined for the \texttt{SigMA} application to the Sco-Cen star-forming region \citep{Ratzenboeck_2023a} to the stars in this volume:
\begin{equation}
\label{eq:box-cuts}
    \begin{aligned}
         &\varpi / \sigma_{\varpi} \equiv \text{parallax signal-to-noise ($S/N_{\varpi}$)} > 4.5\\
         &\texttt{fidelity\_v2} > 0.5
    \end{aligned}
\end{equation}
The term $\varpi / \sigma_{\varpi}$ equals the \texttt{parallax\_over\_error} column in \emph{Gaia} DR3, and \texttt{fidelity\_v2} is a qualifier for the reliability of a star's astrometry. The quality cut values are chosen in accordance with \cite{Rybizki_2022}. The result is a high-fidelity astrometric sample of 910,581 sources with reduced parallax uncertainties. The data retrieval process and more details on the filtering process are provided in Appendix~\ref{Appendix:Data}.

For the clustering (Sect.~\ref{sec:Workflow}), both the observed coordinates provided by \emph{Gaia} and their transformed Galactic Cartesian coordinates in the local standard of rest (LSR) are used. To calculate the LSR, we adopt the solar motion values from \cite{2010Schoenrich}. Cartesian coordinates and stellar distances are computed from the simple parallax-based distance approximation $d~(\text{pc}) = 1000/\varpi~(\text{mas})$. We do not use Bayesian photogeometric distances calculated for \emph{Gaia} data by \cite{Bailer-Jones_2021}. We elaborate on this choice in Appendix~\ref{Appendix:Distances}.

% ============================================================================ %
\section{The clustering process}
\label{sec:Methods}

We perform our clustering under the assumption that stellar clusters manifest as phase-space overdensities, each associated with a unique local density mode \citep{Wishart_1969}. To find them, we employ the mode-finding clustering algorithm \texttt{SigMA}. Throughout this work, we use ``cluster'' in the statistical sense, referring to a density enhancement over a background. We do not seek to make claims about the boundedness or lifetime of the described structures in using this term. An in-depth description of the original algorithm can be found in \citep{Ratzenboeck_2023a}, but we briefly recap its workflow here before describing the additions made in this work to adapt it for use at larger distances.

\subsection{A recap of the SigMA algorithm}
\label{sec:Methods-recap}

\begin{figure*}
  \centering
  \input{workflow1}
  \vspace{-0.5cm}
  \caption{Schematic view of the \texttt{SigMA} workflow, with the tunable parameters highlighted for each stage. Gray-shaded parameters are set to fixed values and are shown only for completeness.}
  \vspace{-0.3cm}
  \label{fig:sigma-workflow}
\end{figure*}
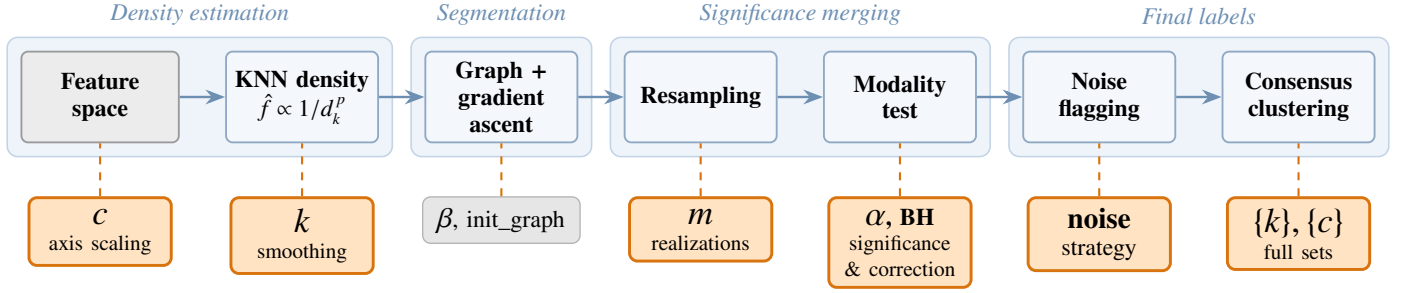

% General intro paragraph
\texttt{SigMA} identifies co-spatial and co-moving stellar populations as overdensities in a multi-dimensional position-velocity phase space, hereafter referred to as the \emph{feature space}. As the method was designed specifically for \emph{Gaia} data, the feature space is typically spanned by 3D stellar positions and their corresponding 2D tangential velocities. These values are available via the five-parameter astrometric solution ($\alpha$, $\delta$, $\varpi$, $\mu_{\alpha^*}$, $\mu_{\delta}$) for 1.46 billion sources in \emph{Gaia} DR3 \citep{Gaia_DR3_2023}. \texttt{SigMA} can operate either directly on the observed coordinates or on transformed quantities. However, different feature spaces require different scaling strategies (Sect.~\ref{sec:Methods-DistantSigMA}). For Sco-Cen, the clustering was performed on heliocentric Galactic Cartesian coordinates ($X$, $Y$, $Z$) and the tangential velocities $v_{\alpha, \text{LSR}}$ and $v_{\delta, \text{LSR}}$.
The algorithm itself is not tailored to specific input spaces, and can be applied to higher or lower dimensions, depending on the available data. 

% Setup of the non-parametric method + density estimation
\texttt{SigMA} works on the assumption that a set of observed stars $X = \{\mathbf{x}_1, \dots, \mathbf{x}_N \}, \mathbf{x}_i \in \mathbb{R}^p$ is generated by an underlying $p$-dimensional density $f$, with $\mathbf{x}_i \sim f$ \citep{Wishart_1969}. In principle, one could therefore trace each data point back to its nearest mode along the density gradient. However, the data generation process is generally complex for real-world data sets, meaning the density distribution must be estimated from the samples $X$. In \texttt{SigMA}, this estimate $\hat{f}$ is obtained using a $k$-nearest neighbor (KNN) estimator on the feature space
\begin{align} 
    \label{eq:knn-density}
    \hat{f}(\mathbf{x}) \propto \frac{1}{d^p_k(\mathbf{x})},  
\end{align}
where $d_k (\mathbf{x})$ denotes the (weighted) Euclidean distance from a point $\mathbf{x} \in \mathbb{R}^p$, to the $k$-th nearest data points in $X$.

% Gradient ascent
After estimating $\hat{f}$, the input data are converted into a $\beta$-skeleton graph with $\beta = 0.99$ \citep{Kirkpatrick_1985}. Every data point is represented by a node carrying its estimated density, and connected to neighboring nodes by graph edges. A graph-based hill-climbing procedure \citep{Koontz_1976} then propagates every node toward the neighbor of highest density, along the steepest ascent, until it reaches a local mode of $\hat{f}$.

% Significance modality test
However, there is a caveat to performing hill-climbing on the estimated density $\hat{f}$: as it is derived from observations, meaning noisy data, $\hat{f}$ is not smooth, and the noise produces additional, spurious density peaks. The gradient ascent, therefore, initially falsely identifies each of these as a true mode, yielding an over-segmented clustering solution. To address this, each pair of neighboring modes is compared using a modality test. If all points belong to a single mode ($H_0$), the clusters are merged, and if the modes are genuinely different ($H_1$), both clusters are retained. Formally, the hypotheses for a normalized path $r(t):= r_t, t \in (0,1)$ between two modes $r_0$ and $r_1$ can be written as
\begin{align*}
    & H_0: f(r_t) \geq \min\{f(r_0),\, f(r_1) \} ~\forall~t \\
    & H_1: \exists~t\,|\,f(r_t) < \min\{f(r_0),\, f(r_1) \} ,
\end{align*}
meaning that $H_0$ is broken if the density of any point $r_t$ is lower than that of the less dense mode. 
Since a higher density corresponds to a smaller $d_k$, the dip condition $\hat{T}$ can be written as
\begin{align*}
    \hat{T}(t) := -p\,\max\{\log d_k(r_0), \log d_k(r_1)\} + p\,\log d_k(r_t),
\end{align*}
which is standard normal under $H_0$ \citep{Burman_2008}. Rather than evaluating the statistic along the entire path, \texttt{SigMA} only performs a single point-wise evaluation at the saddle point between two modes, as it is the shallowest density dip between all paths connecting two nodes. If this minimal dip is significant, the modes are truly separated. The null hypothesis of unimodality is rejected when $\hat{T} > \Phi^{-1}(1-\alpha)$, where $\Phi$ is the standard normal CDF and $\alpha$ determines the significance threshold of the dip. 

Up until this point, each merge decision is based on a single density estimate, which ignores the astrometric measurement uncertainties present in the \emph{Gaia} observables. To propagate these, \texttt{SigMA} resamples the data set $m$ times, by drawing each source from a Gaussian centered on its observed value, with the corresponding error covariance matrix. The modality test is reevaluated at every saddle point for all $m$, and the outcomes are combined into one merge decision using the Cauchy combination test \citep{Liu_2020}. The significance threshold $\alpha$ can also be adaptively adjusted depending on the number of saddle-point tests via the Benjamini-Hochberg (BH) correction \citep{Benjamini_1995}. The merge test is applied successively to all neighboring pairs and repeated until no remaining pair shows a significant density dip.

Following the successful merging of neighboring modes, a noise removal step is applied to each surviving cluster. This is necessary because \texttt{SigMA} traces each star in the input dataset toward its nearest node. This means it performs a complete segmentation of the dataset, in which each star appears in exactly one cluster. From an astronomical point of view, we know that, depending on the observed region and quality cuts, the majority of stars inside a box cut are field stars instead of true cluster members. These stars need to be removed explicitly. The algorithm accepts various noise removal strategies, selected with the parameter \texttt{noise}. Its current methods both remove field stars by fitting a bimodal Gaussian mixture model to the values obtained from the $p$-dimensional density estimate, thereby separating the low-density field from the higher-density cluster population. The ``medium'' method solely removes field stars within each cluster, whereas the ``strict'' method additionally rejects entire sparse clusters.

Two more parameters can greatly influence the clustering: the scale factors $c$ and the neighborhood size $k$. The former are used to reweigh the positional and kinematic subspaces so that neither dominates the $k$-distance metric (see Sect.~\ref{sec:Methods-DistantSigMA}), while the latter sets the degree of smoothing of the $p$-dimensional density. Instead of fixing these parameters to a single value, \texttt{SigMA} runs the full pipeline -- from density estimation through hill-climbing to merging -- over the sets $\{k\}$ and $\{c\}$, producing an ensemble of $n_k \times n_c$ clustering solutions. The runtime of the algorithm scales with $\mathcal{O}(N\log N)$ multiplied by the number of iterations $\{c\} \cdot \{k\}$. A consensus result over all iterations is obtained by linking clusters that strongly overlap across the ensemble, retaining only those that persist across many parameter choices and discarding solutions that randomly fragment or merge. The consensus is applied in two stages: first across the scaling-factor variations at each fixed $k$, and then across the resulting solutions over all $k$. The pipeline workflow and the tunable parameters influencing each stage are shown in Fig.~\ref{fig:sigma-workflow}.

\subsection{The scale factor and distance application problem}
\label{sec:Methods-DistantSigMA}

As stated in Eq.~\ref{eq:knn-density}, \texttt{SigMA} calculates the $p$-dimensional distance $d_k(\mathbf{x})$ between data points $\mathbf{x} \in \mathbb{R}^p$. Since the phase space combines positional and kinematic coordinates, its subspaces generally have different units and numerical ranges, which would cause the Euclidean distance to be dominated by whichever axis has the largest spread. \texttt{SigMA} therefore applies multiplicative scale factors $c$ to the phase space axes, scaling them to similar value ranges and, more importantly, to similar variances among the members of a cluster along each axis. As \citet{Ratzenboeck_2023a} clustered in Cartesian coordinates, they scaled the velocity subspace relative to the position subspace. The scaling relationship between the subspaces was empirically determined for the targeted region using past cluster extractions \citep{Gagne2018, Cantat-Gaudin2020}. Ten scale factors were drawn from a Bayesian posterior predictive model calibrated on these clusters, and the same factor $c_v$ was applied to both velocity axes per iteration.

\begin{figure}[t]
    \centering
    \includegraphics{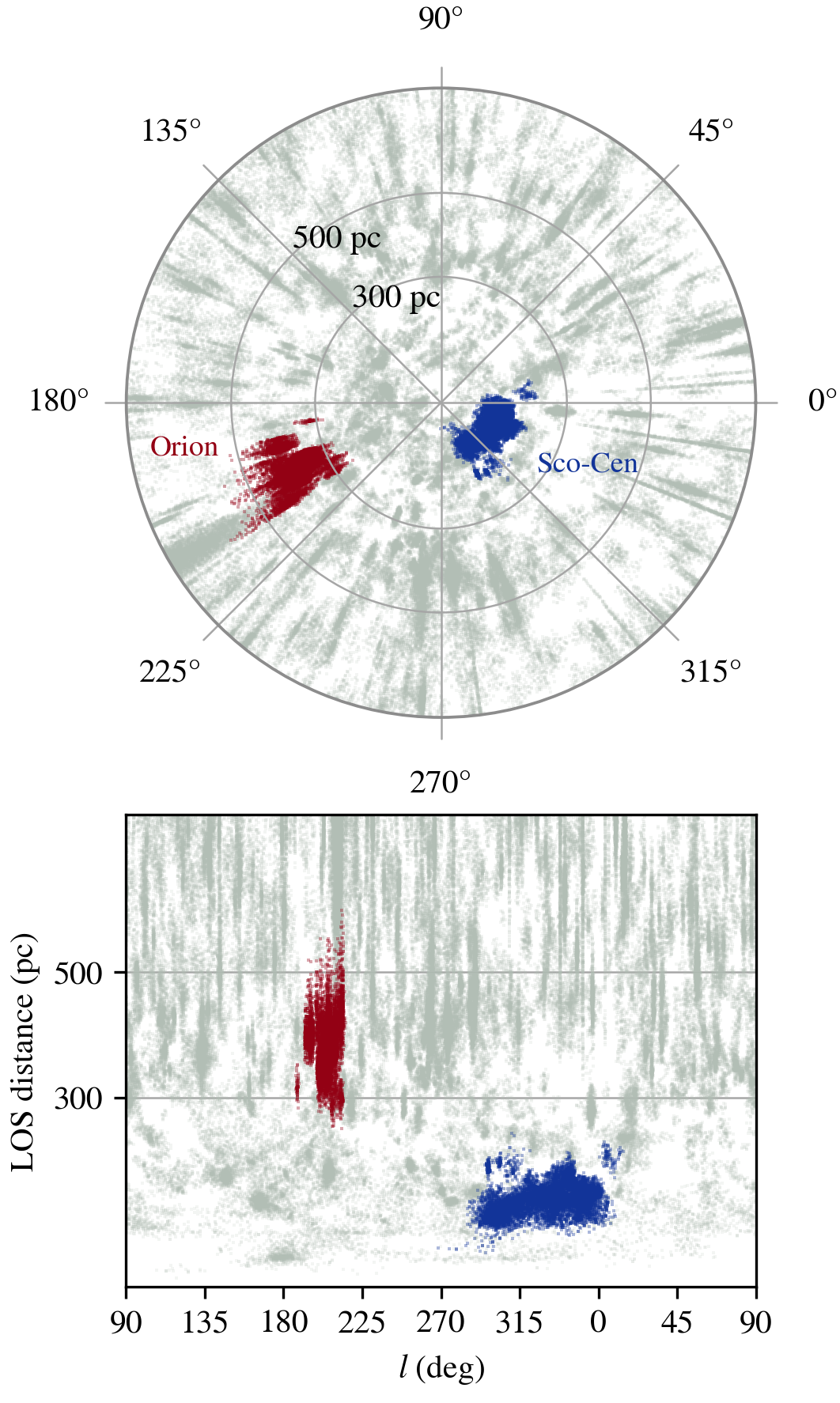}
    \caption{Stretching effect of clusters along the LOS as a function of distance, shown for the open clusters in the solar neighborhood \citep{Hunt_2024}, the Sco-Cen association \citep{Ratzenboeck_2023a}, and the Orion complex (this work). \emph{Top:} Top-down, heliocentric view of the clusters and associations in polar coordinates. \emph{Bottom:} Unraveled version of the upper panel along galactic longitude.}
    \label{fig:LOS-cluster-stretching}
\end{figure}

For more distant stars, however, the absolute value of the measured parallax becomes comparable to its own measurement uncertainty ($\varpi \sim \sigma_{\varpi}$), thereby degrading or even erasing the distance information of individual stars. This causes the appearance of the so-called fingers-of-God in \emph{Gaia} data, which correspond to an apparent stretching along the LOS direction from the Sun. Figure~\ref{fig:LOS-cluster-stretching} illustrates this effect, showing open clusters in the solar neighborhood \citep{Hunt_2024} in gray and members of the Orion (this work) and Sco-Cen star-forming regions \citep{Ratzenboeck_2023a} in red and blue, respectively. Sco-Cen, at $d\sim150$~pc, shows no visible LOS stretching, whereas Orion ($\sim$300--500~pc) displays significant distortion, with the near edge being less affected than the far edge. 

As the distortion effects vary across the region, no single set of scale factors can describe the entire clustering volume. Thus, the Sco-Cen scaling solution is not applicable to Orion. Instead, we generalize the strategy so that, in addition to equalizing the value ranges along the phase-space axes, it also mitigates the parallax-error-induced stretching effect. It does so by accounting for the expected geometry of clusters at a given position in the survey volume. We note that this requirement is not specific to \texttt{SigMA}, but to any density- or centroid-based clustering algorithm relying on a global metric. Such methods implicitly assume that clusters have similar geometries throughout the surveyed volume. With \emph{Gaia} DR3 astrometry, this assumption breaks down beyond $\sim300$~pc, and these methods can lose sensitivity to the very subpopulation limits they target.

% ============================================================================ %
\section{The \texttt{DistantSigMA} pipeline}
\label{sec:Workflow}

We have developed an extension to the existing clustering algorithm tailored to applications beyond $\gtrsim 300$~pc -- \texttt{DistantSigMA}. First, the elongation primarily affects one axis of the five-parameter solution provided by \emph{Gaia}. Converting the values to galactic Cartesian coordinates would put all positional coordinates into the same units and ranges, but would smear these effects across the three coordinate axes simultaneously. Therefore, we deploy \texttt{DistantSigMA} directly on the observed coordinates. This has the benefit of isolating the error influence to one axis, which moreover has an SNR orders of magnitude lower than that of the remaining coordinates. However, the 5D space is now a mixture of angular units (deg, deg, mas, mas/yr, mas/yr), requiring all axes to be scaled independently.
Thus, we designed an entirely new strategy for calculating the scale factors. While the core procedure of \texttt{SigMA} in Fig.~\ref{fig:sigma-workflow} remains unchanged, we have added modules to the pipeline to compute the scale factors directly from the input data. The algorithm automatically computes scale factors using simulated cluster observations. To generate these simulations, we take ideal spherical Gaussian clusters and convolve them with realistic Gaia measurement uncertainties at a specific Galactic position. We then optimize the scale factors along each observable. This maps the stretched, error-distorted clusters back into a spherical shape under a weighted Euclidean distance metric.

\subsection{Coarse clustering}
\label{sec:Workflow-coarse}

 Since the clustering volume of Orion extends from $\approx$300--500~pc along the LOS, the elongation effects also vary significantly across the box and require different scale factors. Therefore, we first partitioned the 910,581 \emph{Gaia} sources in the box into a small number of segments. As \texttt{SigMA} naturally produces mutually exclusive partitions of an input dataset along low-density boundaries, it can be directly used for this step. Since the goal is to break the box into large subsets spanning hundreds of parsecs, the exact scaling of the subspaces is less important for the segmentation step than for recovering exact clusters. Therefore, the coarse clustering was done in Heliocentric Galactic Cartesian coordinates and LSR velocities. The LSR velocity columns ($v_{\alpha,\text{LSR}}$, $v_{\delta,\text{LSR}}$) were scaled using the Bayesian posterior predictive velocity scaling of \cite{Ratzenboeck_2023a}. 

We tested a range of large $k$-values to determine the most stable partition of segments with comparable chunk sizes that remained consistent across repetitions. The parameter space exploration is outlined in Appendix~\ref{Appendix:Segments}. The best partition was achieved with $k=950$, yielding five segments.

\subsection{Fine-grained clustering}
\label{sec:Workflow-fine}

The fine-grained clustering is always performed in the \emph{Gaia} data space ($\alpha$, $\delta$, $\varpi$, $\mu_{\alpha,*}$, $\mu_\delta$), without applying any prior coordinate transformation. It is set up in three stages: 
\begin{enumerate}
    \item a preliminary run -- where only the parallax axis was scaled to mitigate predominant distortion effects -- to identify robust cluster cores (Sect.~\ref{sec:Workflow-fine-preliminary})
    \item the derivation of data-driven scale factors from those clusters (Sect.~\ref{sec:Workflow-scalefactors})
    \item and a final re-clustering using those scale factors (Sect.~\ref{sec:Workflow-fine-final})
\end{enumerate}

The pipeline parameters used for the fine-grained clustering are four different density smoothing values $k\in \{15, 20, 25, 30\}$, a resampling number $m = 10$, a significance threshold of $\alpha = 0.01$, and \texttt{noise}= strict. Further information on their choice and the stability of the results under parameter variations is provided in Appendix~\ref{Appendix:Parameter-exploration}.

\subsubsection{Preliminary clustering}
\label{sec:Workflow-fine-preliminary}

In the preliminary clustering step, we sought to identify a set of reliable clusters to serve as the basis for deriving the data-driven scale factors. We did not aim for completeness or even for the highest possible merge/split sensitivity at this point, since the exact number of clusters does not affect the scale factor derivation. First, all axes were robustly normalized using their medians and median absolute deviations. Then, the parallax axis was scaled to mitigate elongation effects. Using simple SNR estimates, we used 10 plausible scale factors $c \in [0.1,0.55]$ linearly spaced to down-weight its 5D-distance influence. Then we performed the clustering following the pipeline workflow shown in Fig.~\ref{fig:sigma-workflow}, with the BH-correction enabled.

\subsubsection{Data-driven scale factors}
\label{sec:Workflow-scalefactors}

The preliminary clusters identified in the previous step were used to derive data-driven, per-region scale factor bounds for the final clustering via a basic cluster simulation: For each detected cluster, we computed the position and velocity covariances separately using the Minimum Covariance Determinant (MCD) estimator \citep{Rousseeuw_1984, Butler_1993, Rousseeuw_1999} on the Galactic Cartesian data ($X$, $Y$, $Z$, $v_{\alpha,\text{LSR}}$, $v_{\delta,\text{LSR}}$), and determined the cluster center from the MCD location estimate. The position covariance was then rewritten in diagonal form, with the median eigenvalue $\lambda_{\mathrm{med}}$ as the diagonal entries. In doing so, each cluster is assumed to be spherical in position space. This isotropic assumption was then combined with the unchanged-velocity block to form an idealized five-dimensional covariance matrix.

Next, a synthetic clone of each cluster was generated by drawing member stars from a multivariate Gaussian defined by this covariance and the MCD-based center. We drew $1.5$ times the number of observed members to account for the likely underestimation of the member count due to the settings of the preliminary clustering. To replicate the astrometric uncertainties at a given cluster distance, the synthetic stellar positions were then error-convolved with typical \emph{Gaia} measurement uncertainties:
First, the simulated Cartesian positions were converted to observed coordinates ($\alpha$, $\delta$, $\varpi$). Then, positional uncertainties ($\sigma_\alpha$, $\sigma_\delta$, $\sigma_\varpi$) were conditionally sampled from the Gaia catalog conditioned on the cluster mean position (see Appendix~\ref{Appendix:Data} for more information on the sampling source).  New $\alpha$, $\delta$, and $\varpi$ values were each drawn from a Gaussian centered on their noise-free value, with their respective sampled uncertainty as standard deviation. Proper motions were intentionally left unmodified to avoid compounding the astrometric errors. Simulated sources with negative resampled parallaxes were discarded.

The strict parameter settings of the preliminary clustering meant that the cluster sample was likely biased against small populations. To mitigate this, we added a small artificial cluster with $N = \min(\texttt{KNN\_list})$ members at the global centroid of all detected clusters, using the position covariance of the smallest detected cluster scaled by a factor of 0.5 in the position subspace. It received the same error treatment as the real clusters. 

The scale factors for all five coordinates ($\alpha$, $\delta$, $\varpi$, $\mu_{\alpha,*}$, $\mu_\delta$) were computed from the standard deviations of individual cluster samples along each axis $x \in \{\alpha, \delta, \varpi, \mu_{\alpha,*}, \mu_\delta\}$. Thus, for each cluster in our simulated sample, we get one $\sigma_x$. Scale factors are defined as $c_x:= 1/\sigma_x$, as the standard deviation $\sigma_x$ sets the natural scale of a cluster along an axis $x$. This means that in an unweighted Euclidean metric, the density estimate would be dominated by the axes with the largest spread. Rescaling by $1/\sigma_x$ equalizes the axes, so that the clustering is carried out in a space in which the clusters are close to spherical in absolute terms. By comparing the standard deviations for all simulated clusters per region, we computed scale factor bounds as $[1/\sigma_{x,\rm max},\, 1/\sigma_{x,\rm min}]$ for each dimension.

\subsubsection{Final clustering}
\label{sec:Workflow-fine-final}

For the final clustering, \texttt{SigMA} sampled the calculated scale factor bounds using a Sobol low-discrepancy sequence \citep{Sobol_1967} implemented in \texttt{scipy}. It is well-suited for evenly sampling high-dimensional spaces. Since the sequence can only draw multiples of 2, we opted for $n_c = 16$ combinations of scale factors. We assigned the same scale factor to $\alpha$ and $\delta$ in each iteration, derived from the intersection of their respective scale factor bounds. The remaining three axes each received individual scale factors, meaning we sampled the Sobol sequence in $d_c=4$ dimensions. We justify this choice as clusters at the investigated distance can be assumed to be spatially constrained, but keeping $\mu_{\alpha,*}$ and $\mu_\delta$ independent allows greater flexibility in non-symmetric velocity behavior. We discuss the influence of the choice of $n_c$ and $d_c$ in Appendix~\ref{Appendix:Parameter-exploration}. Clustering was again performed following the workflow shown in Fig.~\ref{fig:sigma-workflow}. In contrast to the preliminary clustering, we disabled the BH correction in the final step to discourage too aggressive merging and obtain more fine-grained cluster boundaries.

\subsection{Repetition stability}
\label{sec:Workflow:repetition-stability}
We repeated the final clustering of each segment $n_{\text{rep}}=10$ times using different seeds to sample the scale factors. This allowed us to cover the scale factor parameter space more thoroughly than just with a higher $n_c$ (see Appendix~\ref{Appendix:Parameter-exploration}). At the same time, repeating the clustering enabled us to quantify the \emph{persistence}, or lifetime, of each recovered cluster across all runs, and assign per-star memberships. We emphasize that all quantities reported for the clusters from this procedure were derived from a combined consensus-component analysis across all repetitions -- meaning no single repetition was privileged as a reference.

\subsubsection{Cluster persistence and membership stability}
\label{sec:Workflow-persistence}
\begin{figure}[t]
    \centering
    \includegraphics[width=1\linewidth]{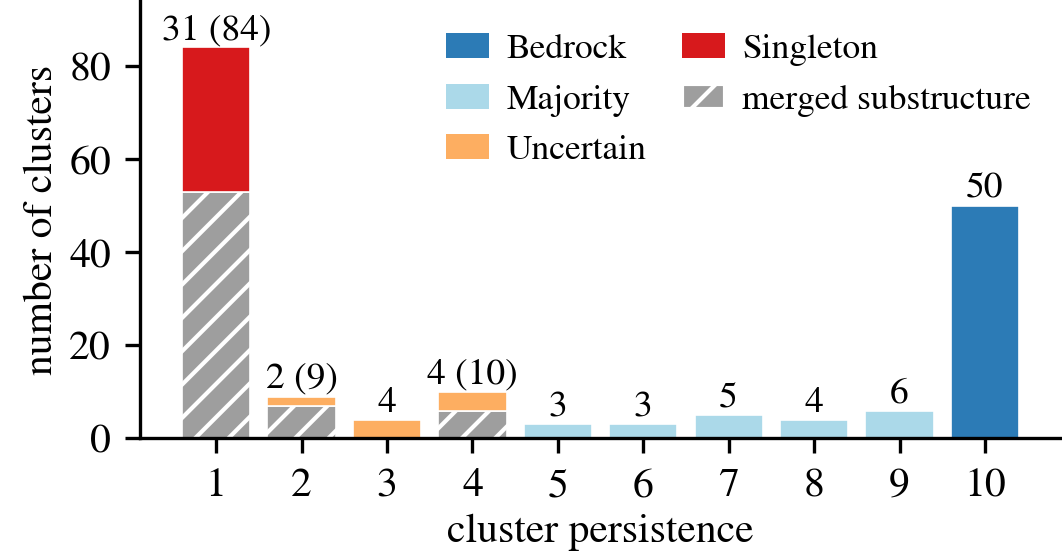}
    \caption{Number of consensus clusters as a function of cluster persistence. The hatched gray area shows the components that were merged back into a more persistent parent via the substructure merge. The bar colors correspond to the respective persistence tier. Counts above the bars are surviving components, with the total bar height shown in parentheses. There are 81 non-singleton clusters at the consensus stage, but the final catalog contains 76, due to the merging of six split clusters ($-3$ at 10/10, $-1$ at 8/10, $-2$ at 7/10) and the split of the ISF ($+1$ at 9/10).}
    \label{fig:rep-stability}
\end{figure}

We first discarded entries that were classified as noise across all runs. From the remaining dataset, we built a graph in which each node represents the set of \emph{Gaia} source identifiers associated with a given (\texttt{cluster\_label}, \texttt{rep\_id}) pair. Next, we built a sparse membership matrix $n_{\text{nodes}} \times n_{\text{stars}}$ and computed the pairwise Jaccard similarity between each pair of nodes $A$ and $B$:
\begin{align}
    \label{eq:Jaccard}
    J \equiv \frac{|A \cap B|}{|A \cup B|}
\end{align}
Typically, a value $\geq 0.5$ indicates a high similarity between two sets \citep{Ratzenboeck_2023a}.
We added edges between all nodes where $J \geq 0.5$, provided they had different rep\_ids, and, in doing so, we connected the same cluster across different repetitions. We identified all connected components, meaning groups of nodes that are linked to one another by edges but not to any node outside the group, in the graph. Then, we verified that each node was assigned to exactly one component. We then counted the number of distinct repetitions in each component. This cluster persistence score was binned into four tiers: Bedrock (10/10), Majority (5--9/10), Uncertain (2--4/10), and Singletons (1). The distribution of connected components and their persistence score are shown in Fig.~\ref{fig:rep-stability}. We investigated this distribution at different $J$ thresholds and found that it remains largely similar across the threshold range 0.3--0.8.

For each consensus cluster (connected component), we then assigned stellar memberships. The membership of a star is defined as the number of repetitions in which it was placed in a given cluster, divided by the number of repetitions in which the cluster was recovered at all. Normalizing by a cluster's lifespan rather than by the total number of repetitions means the membership is bounded at 1 and is decoupled from the stability tier: a star that is recovered in a cluster every time that cluster appears reaches membership 1, even for a low-persistence cluster. Because a star could be recovered in different clusters across repetitions, it may carry a non-zero membership in more than one component (see Appendix~\ref{Appendix-tables}).

\subsubsection{Substructure}
\label{sec:Workflow-substructure}
The Jaccard similarity evaluation of Eq.~\ref{eq:Jaccard} works well for identifying shared sets in approximately equally sized clusters, but it is not suited for identifying subsets of clusters: If a small part of a cluster is split off a bigger parent in a few ($< 5$) repetition runs, Eq.~\ref{eq:Jaccard} would not be fulfilled and it would not be merged into its more stable parent. To do exactly this, we tested each cluster with a detection rate $< 5$ as a potential child of a more stable cluster. To qualify as a parent cluster, it needed to 1) be persistent through 5 or more repetitions, 2) be at least twice as large as the potential splinter, and 3) share at least 90\% of the splinter members. If all three requirements were met, the splinter was merged into its parent, and the star memberships of the newly added sources were updated. The cluster's stability tier was held fixed at the parent's pre-merge persistence.

\begin{figure*}[t]
    \centering
    \includegraphics[width=\linewidth]{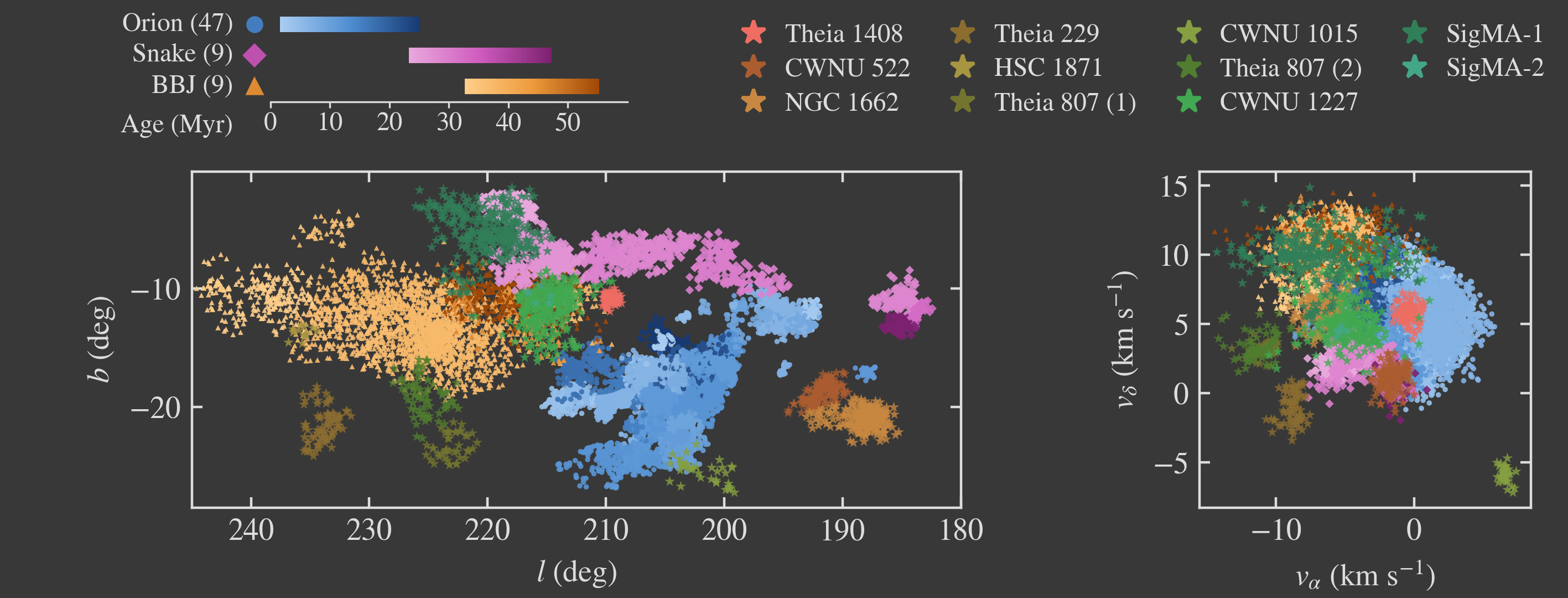}
    \caption{Galactic on-sky positions of the 76 clusters detected by \texttt{SigMA} across the investigated volume. Three larger structures comprising multiple clusters of similar age and velocity can be identified: The Orion complex, consisting of 47 clusters (\emph{blue circles}), the Snake structure \citep{Tian_2020} (\emph{pink diamonds}), and the BBJ group \citep{Beccari_2020} (\emph{orange triangles}), comprising 9 clusters each. The color coding of these three groups indicates the estimated cluster age. The remaining 11 groups are not connected to these structures.}
    \label{fig:overview}
\end{figure*}

\subsection{Postprocessing and the special case of the ISF}
\label{sec:Workflow-postprocessing}
% Soft probabilities
The consensus components and per-star memberships were combined into a single membership catalog by assigning each star to the component with the highest membership percentage. Singleton clusters (Fig.~\ref{fig:rep-stability}) were discarded. If a star had the same maximum membership probability for more than one component, it was assigned to the cluster for which the 5D distance to the center was minimized. These 472 sources are flagged in the final catalog. We report the full set of per-component memberships as soft probabilities for all stars that appeared in more than one component with $p>0$ in a supplementary table (Appendix~\ref{Appendix-tables}). 
% outlier flag + spurious clusters
Due to the complexity of the consensus function and the repetition stability workflow, we observed that a few consensus clusters contained obvious 5D outliers. We flagged such stars using a robust Mahalanobis criterion: For each cluster, we fit an MCD covariance in the 5D feature space and flagged a star if its robust Mahalanobis distance exceeded the empirically determined threshold of $8\sigma$. 
We define stars with a membership probability $p\geq 0.5$ and not flagged as outliers as ``reliable members''.

% front and back merge
Upon analyzing the clustering solution, we found that 5 clusters were split into a front and back part, and one cluster was divided laterally, across two coarse segments (Sect.~\ref{sec:Workflow-coarse}). We identified them by analyzing their connectivity, 3D positions, 2D velocity overlaps, and color-magnitude diagrams (CMDs), and manually merged them in a post-processing step. 

% the ISF
There was one unusual cluster in our selection -- the Integral-shaped filament (ISF) in Orion~A, which includes the ONC. This cluster was one of those initially split into a front and back part. Given its enormous size ($N \sim 5,000$) and high member density, and the large contamination fraction visible in on-sky positions and in the CMD, we determined that \texttt{SigMA} could not break it with standard settings. Therefore, we repeated the final clustering (Sect.~\ref{sec:Workflow-fine-final}) for this cluster, after merging the front and back parts. Since the cluster spanned two segments, we used the union of scale factor ranges calculated for the two segments and drew $n_c = 32$ samples to account for the broader range. We ran the standard ten repetitions with strict noise removal and combined them with the repetition stability procedure described in Sect.~\ref{sec:Workflow:repetition-stability}. This separates the input into two components: the ISF (ONC) cluster (2264 stars) on the Bedrock tier and the NGC~1977 cluster (113 stars), recovered in 9/10 repetitions. The remaining 2,488 sources (51\%) are rejected as noise. This decomposition remains stable for different scale factor samplings (16 vs. 32) and ranges (intersection vs. union), with the boundary against the field population stable to 3\% (2253--2321 stars), whereas the member number of NGC~1977 varies between 113 and 192 stars. The adopted parameter setting yields the most conservative estimate. We further discuss the special case of the ISF cluster in the context of the algorithm limitations in Sect.~\ref{sec:Discussion-limitations}.

\subsection{Cluster ages}
\label{sec:Workflow-ages-distances}

For all parameter derivations, we considered only reliable cluster members. We estimated the cluster ages with the age fitting algorithm described in \cite{Ratzenboeck_2023b} (\textsc{Chronos} package), using PARSEC isochrones \citep{Bressan_2012, Chen_2014, Marigo_2017}. We note the known discrepancy between various isochronal models, which has been addressed in previous works \citep{Ratzenboeck_2023b}. The same limitations apply here. Details about the age-fitting procedure, along with the best-fit isochrones, are provided in Appendix~\ref{Appendix-CMDs}.

% ============================================================================ %
\section{Results}
\label{sec:Results}

In this section, we present all  (statistical) clusters detected with \texttt{SigMA} in the full clustering volume (Eq.\ref{eq:box-coords}), before conducting an in-depth study of the clusters associated with the Orion star-forming complex. All results are based on the tables described in Appendix~\ref{Appendix-tables}.

\subsection{Overview of all clusters in the box volume}
\label{sec:Results-all}

\begin{figure*}
    \centering
    \includegraphics[width=\linewidth]{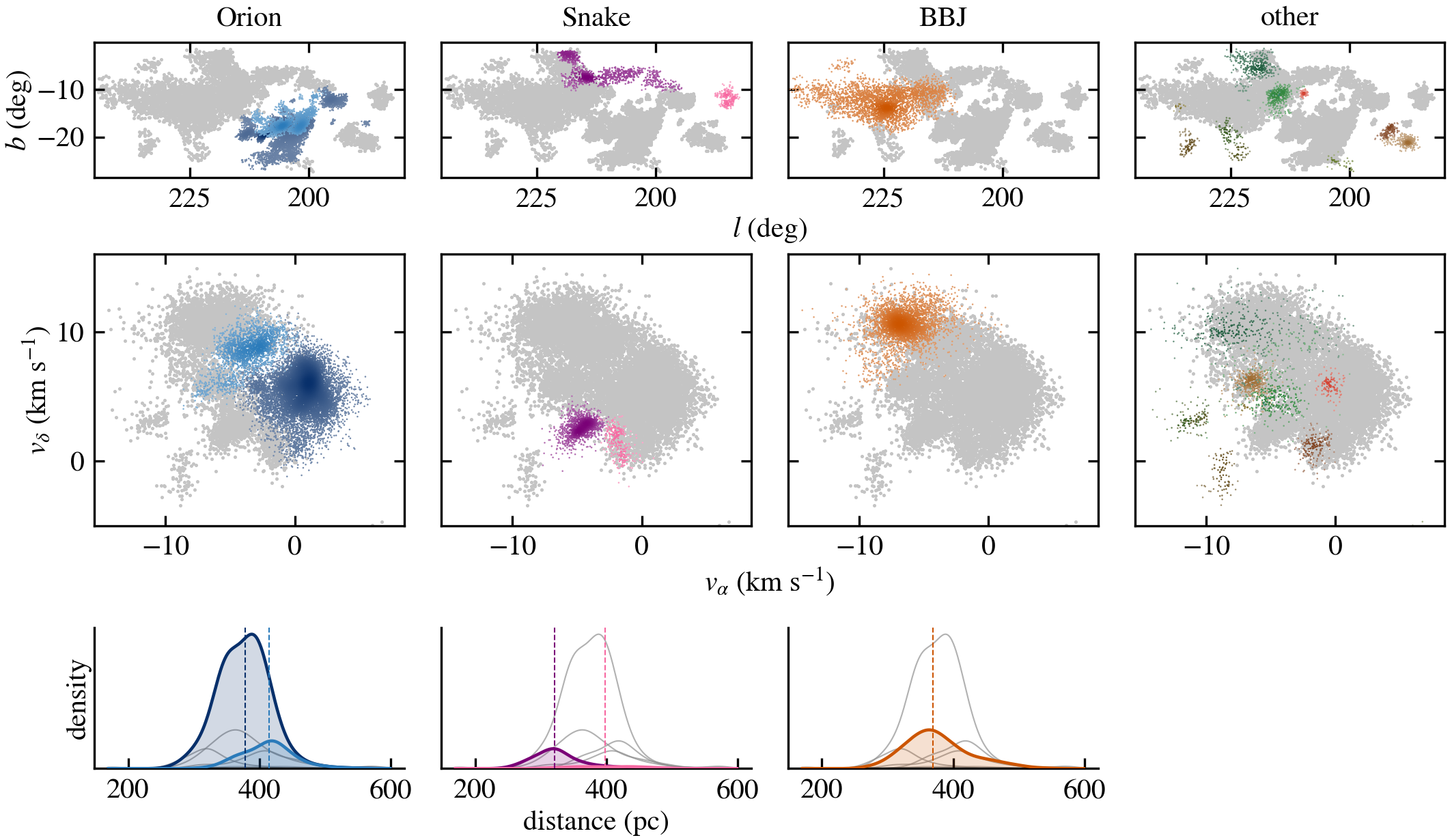}
    \caption{Separation of the different regions using transversal velocities. The \emph{top row} shows the titular structure highlighted in Galactic on-sky positions, the \emph{middle row} shows its corresponding transversal velocity components, and the \emph{bottom row} shows the KDE of the member distances for the three structures, with the median indicated by a dashed line. The full sample is displayed in gray in the background of the first two rows.}
    \label{fig:transversal-velocities}
\end{figure*}

% Stats
\texttt{SigMA} identified 19,862 stars (2.18\%) of the initial box of 910,581 sources as cluster members, grouped into 76 clusters. Only 1,861 sources have membership probabilities $p < 0.5$, and 312 (128 with $p\geq 0.5$) are flagged as outliers. We define a reliable sample of 17,873 stars ($90$\%), which we use to derive mean cluster parameters (Tab.~\ref{tab:overview-orion}), ages, and extinctions. Of the clusters, 47/76 ($\approx 62$\%) are on the Bedrock, 19 ($\approx 25$\%) on the Majority, and 10 ($\approx 13$\%) on the Uncertain stability tier, respectively.

% Description overview figs
In Fig.~\ref{fig:overview} we show the Galactic on-sky positions and transversal velocities of all clusters. While the clusters are distributed across a large section of galactic longitudes, they are quite tightly grouped in velocity space and sometimes overlap along the line of sight. Using their velocity information, we can identify three larger groupings, as shown in Fig.~\ref{fig:transversal-velocities}. We find two new individual clusters, which we name SigMA-1 and SigMA-2, respectively.

% First group: Orion 
The richest grouping is the Orion star-forming region. It comprises 47 clusters and forms two distinct velocity overdensities, shown in different shades of blue in Fig.~\ref{fig:transversal-velocities}. Both share the same spatial volume, although the smaller component is farther away along the LOS, as shown in the third row of the figure. The Orion solution will be discussed in detail in the next sections.

The second grouping we identify contains 9 clusters and corresponds to the stellar Snake structure \citep{Tian_2020}, forming a mostly straight spine-like feature between $l \in [185,220]$~deg at $b\approx -10$~deg. We note, however, that one cluster along its extent is not recovered by \texttt{SigMA}. The stellar content of the structure in our clustering solution comprises 1,546 stars, distributed across 9 groups. Eight groups have very similar ages ranging from 23.3 to 32.6 Myr, and one group is aged 47.3~Myr. This makes most of them slightly younger but still close to previous estimates of 30--40 Myr \citep{Tian_2020, Wang_2022}. We observe a tentative age gradient from the head cluster Snake-1 to the tail clusters Snake-7 and 8. It is not entirely clean, as Snake-5 is older than Snake-3 and 4 but lies between them, but the age estimates for these three clusters lie within their mutual uncertainty bounds. We find 897 of the originally published members of the structure \citep{Tian_2020}, but miss another 1,092, of which 810 are in our clustering volume. The missed stars are preferentially close to the near edge of the structure along the LOS, with most constituting the bridge between our Snake-6 and Snake-7. The proximity to the box edges may have degraded the algorithm performance and explain part of the discrepancy. Our membership selection should therefore not be viewed as complete. 

We find 654 sources not contained in their catalog, mostly towards the back of the structure. We also decided to include the clusters Snake-8 and Snake-9 into our definition of the structure, although they are further away along the LOS and are not included in \citet{Tian_2020}. Snake-9 is also $\sim15$~Myr older than the other groups. However, their velocities match that of Snake-7, which is included in the previous catalog, very well. We highlight this tail of the structure (Snake-7--9), which forms its own velocity overdensity, in pink in Fig.~\ref{fig:transversal-velocities}.

We identify a third, more diffuse structure comprising 9 groups with similar transverse velocities, which we associate with the BBJ groups \citep{Beccari_2020}. We estimate their ages to be 32.6--37.1~Myr for the 6 younger groups and 44.4--55.2~Myr for the three older ones. The ages of the groups generally increase as $l$ decreases across the entire structure, but BBJ-5 and BBJ-7 break this tentative gradient. We again emphasize that these memberships should not be viewed as complete, since this grouping is quite close to the borders of our box, and the algorithm was not tuned to find these clusters, but picked them up as a byproduct of the Orion clustering.

The remaining clusters (Fig.~\ref{fig:transversal-velocities}, fourth column) do not share spatio-kinematic coherence with the three identified groupings, nor with one another.

\subsection{The \texttt{SigMA} view of the Orion complex}
\label{sec:Results-Orion}

We identify clusters that are a part of the Orion complex almost exclusively from the transverse velocity and spatial separation of its members from other clusters and structures in Fig.~\ref{fig:transversal-velocities}, with one exception: The cluster Theia~1408, which we estimated to be 164.4~Myr old, shares the velocity overdensity with the majority of the Orion clusters (shown in red in panel 4, middle row of Fig.~\ref{fig:transversal-velocities})  and needs to be discarded due to its age. 

\begin{figure*}[t]
    \centering
    \includegraphics[width=\textwidth]{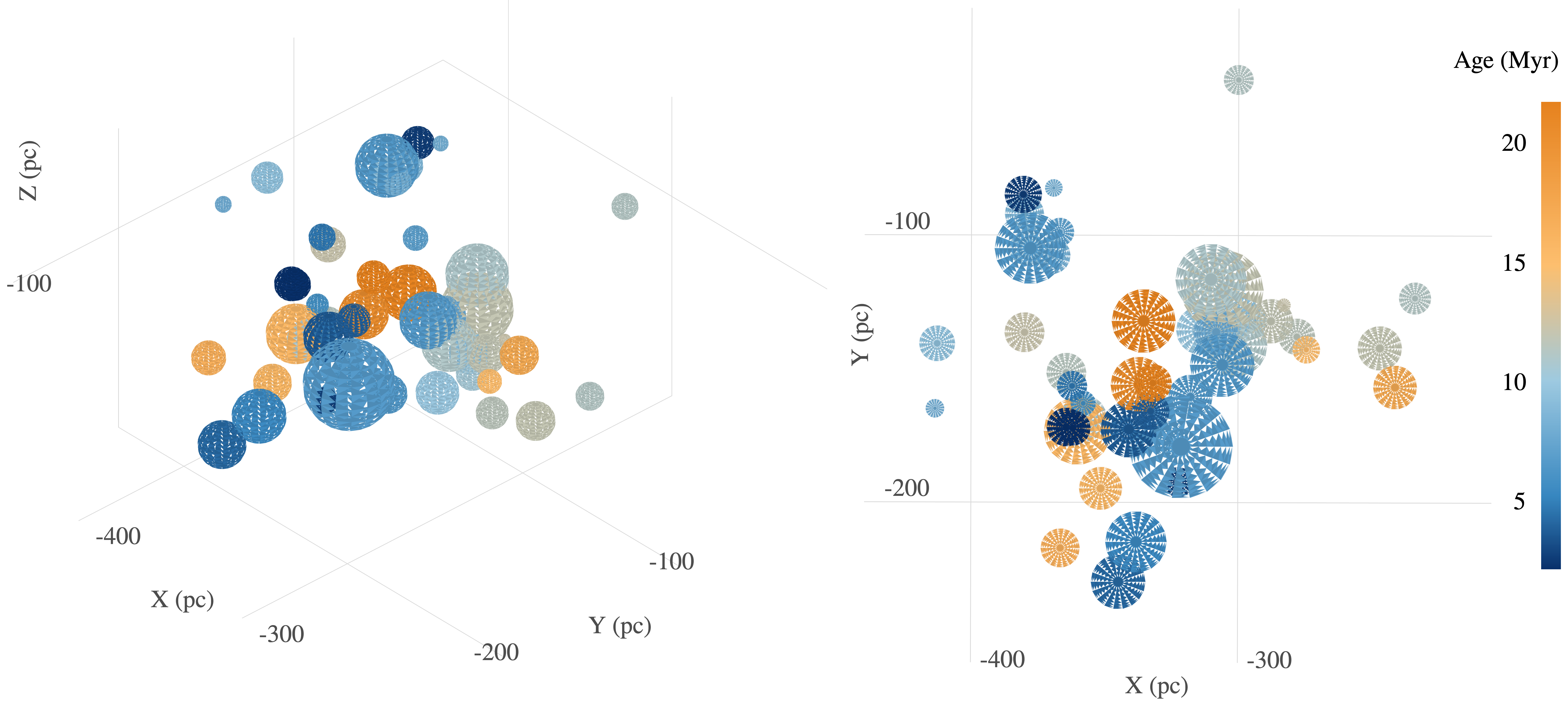}
    \caption{Overview of the 47 \texttt{SigMA} clusters comprising the Orion complex, shown in Galactic Cartesian coordinates in the heliocentric reference frame. \emph{Left:} Frontal view of the clusters looking toward Orion~A. \emph{Right:} Top-down view of the clusters. The clusters are shown as spheres centered at the median coordinates of each cluster, scaled with their member count as $\propto N^{1/3}$. An interactive version of this figure is available online*.}
    \label{fig:Orion-master}
\end{figure*}

\begin{figure*}[t]
    \centering
    \includegraphics[width=\textwidth]{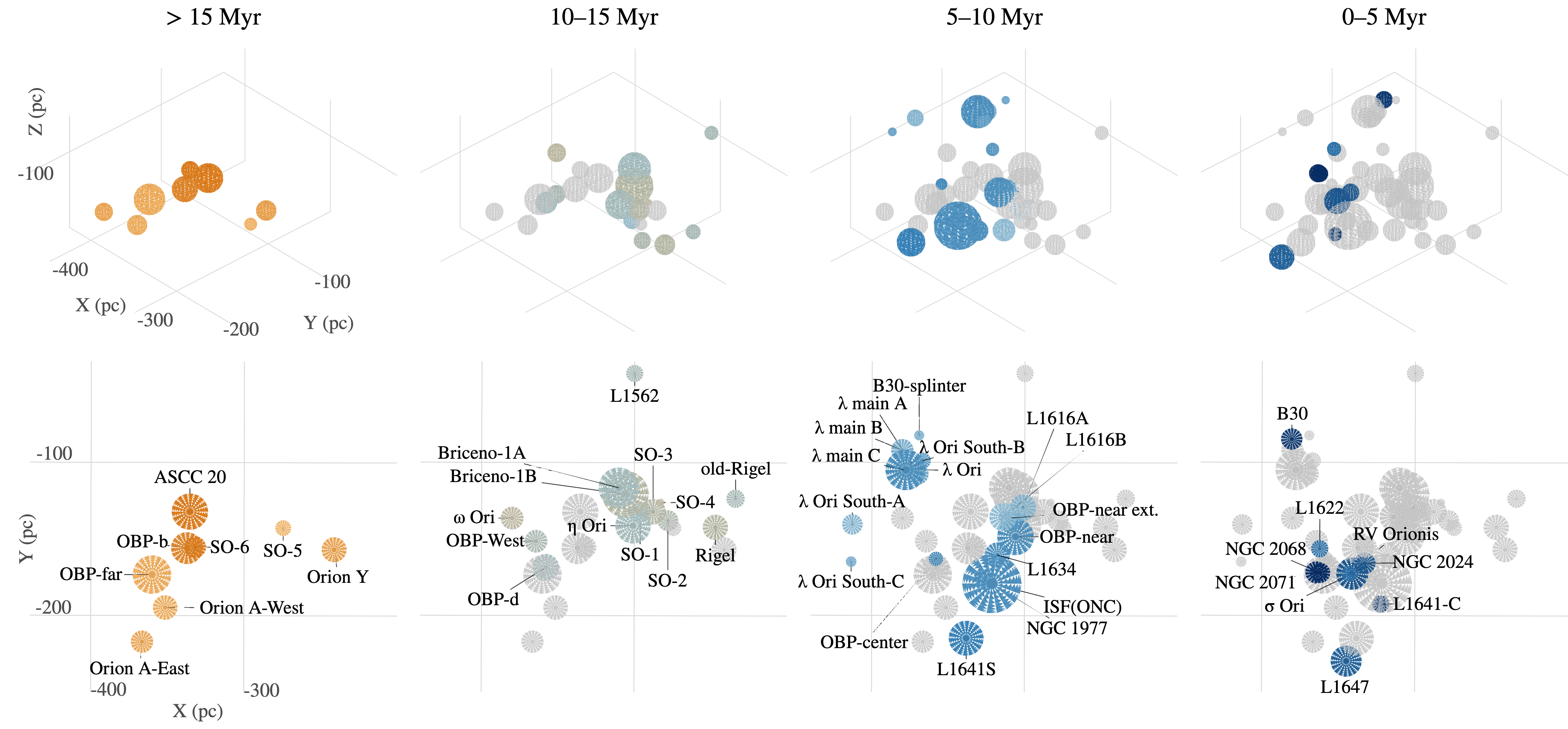}
    \caption{Sequential view of the star-forming history of Orion. The \emph{top row} shows the frontal view of the complex, and the \emph{bottom row} shows the top-down view projected onto the $(X, Y)$-plane. The color-coding is the same as in Fig.~\ref{fig:Orion-master}.}
    \label{fig:Orion-age-panels}
\end{figure*}

We define the \texttt{SigMA} Orion solution as 47 clusters, comprising 13,224 stars in total, and 11,996 reliable stars ($>90\%$). An overview of the mean cluster parameters is given in Tab.~\ref{tab:overview-orion}, and in Appendix~\ref{Appendix-CMDs} we show the cluster CMDs and best-fitting isochrones, and conservatively estimate the contamination of the full sample ($10.4$\%) and the subset used for age fitting ($5.9$\%). We relate 31 groups to known clusters or substructures from previous clustering solutions, while 16 have not been found previously. Where possible, we adapt the cluster names from the literature\footnote{We do not adopt the ``Rigel'' cluster name \citep{Chen_2020, Sanchez_2024}, as we attribute the namesake star to a different cluster than the literature. We call the new \texttt{SigMA} cluster ``Rigel'' and the literature cluster ``old-Rigel''.}, name new groups for known stars they contain, and label the others \texttt{SigMAOrion--\#} (SO--\#). New clusters are printed in boldface in the table. 34/47 clusters are on the Bedrock tier, 10 are on the Majority tier, and only three new ones are on the Uncertain tier. We classify all these groups as independent structures, and none are substructures of one another. 

The 3D cluster distribution in Cartesian coordinates is shown in Fig.~\ref{fig:Orion-master}. Due to strong elongation effects, we visualize the Orion clusters as spheres centered at their respective median cluster coordinates. The sphere sizes scale with their member numbers as $\propto N^{1/3}$. The color-coding of the spheres corresponds to the ages determined with isochrone fitting, and the cluster CMDs are shown in Fig.~\ref{fig:CMDs}. In the interactive version of the figure, cluster tiers and extinction can also be selected as a color scale.

The cluster ages range from 1.5 Myr for NGC~2071 to 21.8~Myr for the known ASCC~20 cluster or 25~Myr for the cluster SO-6 (Uncertain tier). Almost all older populations ($>15$~Myr) are located toward the back of the complex along the LOS, the $\lambda$~Ori region is an agglomeration of subgroups at a higher $Z$ than the rest, and the youngest clusters in the main complex trace the Orion~A and B molecular clouds, respectively.

In Fig.~\ref{fig:Orion-age-panels}, we again show the clusters of Orion, but this time split into four age bins that are added sequentially to the figure, starting from clusters $>15$~Myr and then equally spaced in 5~Myr steps. Similar to Fig.~\ref{fig:Orion-master}, we can see that the oldest clusters almost all formed along a diagonal line in the top-down view, at the back of the complex. Next, the 10--15~Myr old groups formed at an obtuse angle to the initial sub-populations. In the 5--10~Myr window, the $\lambda$~Ori region at the back and above the main cloud complex starts forming stars, while the Orion~A molecular cloud produces more massive clusters in a parallel line in front of the $>15$~Myr old clusters. Lastly, at $<5$~Myr, the Orion~B clusters appear, as do the youngest clusters of Orion~A and $\lambda$~Ori.

\begin{figure}[t]
    \centering
    \includegraphics[width=1\linewidth]{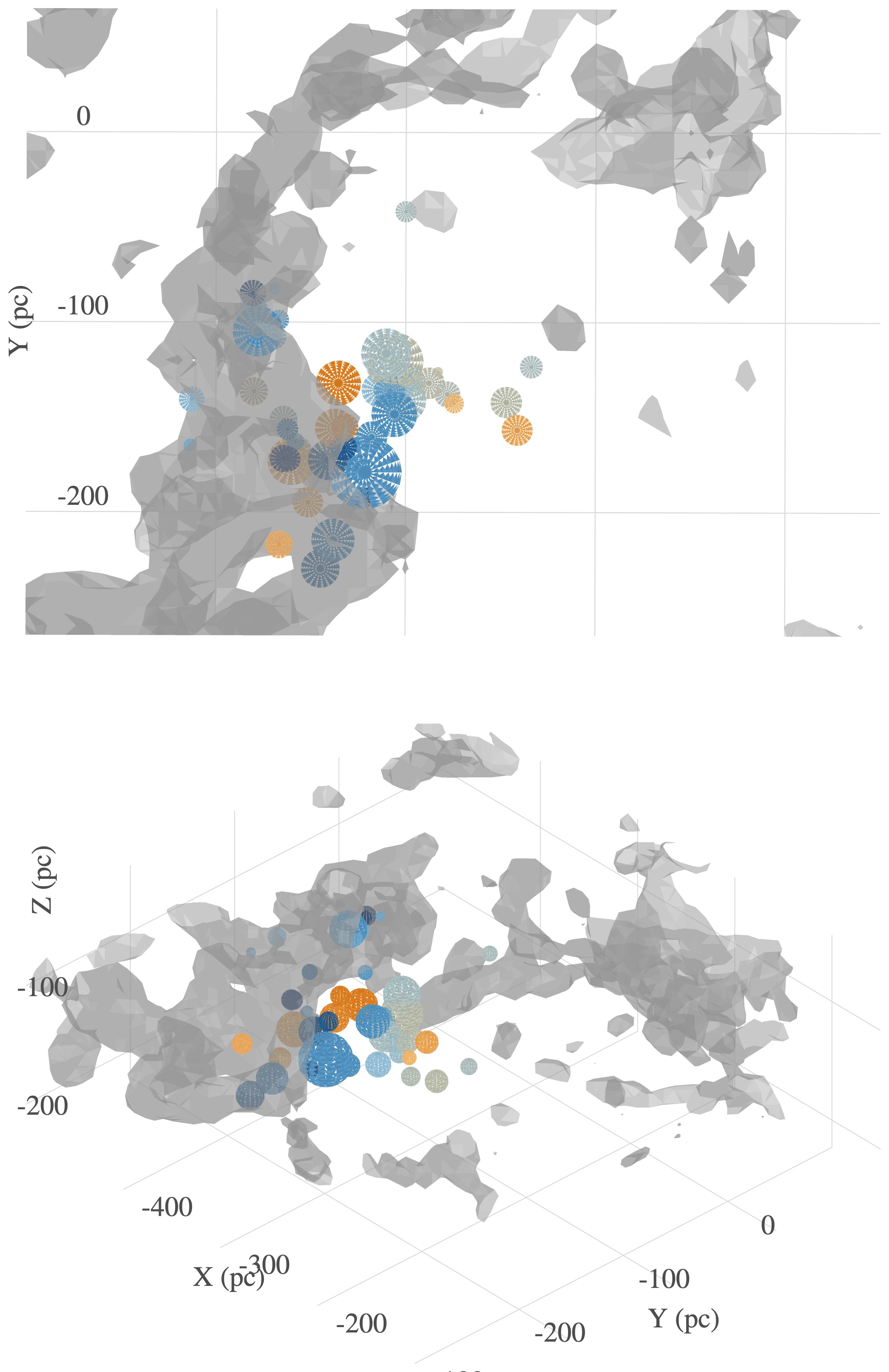}
    \caption{Similar to Fig.~\ref{fig:Orion-master}, but now also showing the 95th percentile dust contour from the dustmap by \cite{2024Edenhofer}. An interactive version of this figure is available online*. }
    \label{fig:orion-master-dust}
\end{figure}

Fig.~\ref{fig:orion-master-dust} again shows the Orion clusters, but this time we also plot the dust in the volume as traced by the 3D dustmap of \cite{2024Edenhofer}. The 3D contours are drawn between the 95th percentile contour of dust density. This allows us to visualize the borders of the Orion-Eridanus shell by tracing the denser dust ridges rather than the very diffuse components. 

Most Orion clusters reside in the cavity inside the dust shell. No clusters older than 16~Myr appear embedded in the dust envelope of this contour, with Orion~A-East (16.7~Myr) notably sitting in a smaller cavity of the dust. Orion-A West (15.8~Myr) is the only cluster older than 10~Myr whose center appears embedded, and also has the highest derived extinction for this age bin ($A_V = 0.6$~mag). OBP-far also appears partially embedded in the dust contour, but has an estimated extinction of $A_V = 0.1$~mag. 

The $\lambda$~Ori group is located at the edge of the dust contour and is still partially embedded, judging from its extinction estimates ($A_V = 0.3-1.75$~mag). The clusters sit inside a secondary, smaller dust cavity that partially extends over them at higher $Z$. $\lambda$~Ori South-A and $\lambda$~Ori South-C have relatively high fitted extinction values, but their centers sit at the far edge of the dense dust ridge.

From the fitted extinctions, the young clusters NGC~2071 ($A_V=1.6$~mag), NGC~2068 ($A_V=1.3$~mag), L1641-C ($A_V=2.3$~mag), L1647 ($A_V=1.2$~mag), and RV~Orionis ($A_V=1.6$~mag) are clearly still (partially) embedded, and their extinction and age values should be viewed as conservative limits as the isochrone fits do not account for differential extinction and missing embedded members. 

For the two young clusters L1641S ($A_V=0$~mag) and L1622 ($A_V=0.3$~mag), the fit produces low extinctions, contrary to known literature values \citep{Kun_2008, Meingast_2018} and the fact that their centers are within the dust contour. Their fit posteriors are regular, so the low extinctions may be driven by a bias towards optically visible members or by the known age-extinction degeneracy in isochrone fitting. NGC~2024 also has a comparatively high age estimate and a low extinction fraction, likely due to the absence of an estimated $>50\%$ still-embedded members \citep{Rottensteiner_2026}. However, contrary to the other two clusters, its uncertainty intervals are very large.
\begin{figure*}[ht!]
        \centering
    \includegraphics[width=\linewidth]{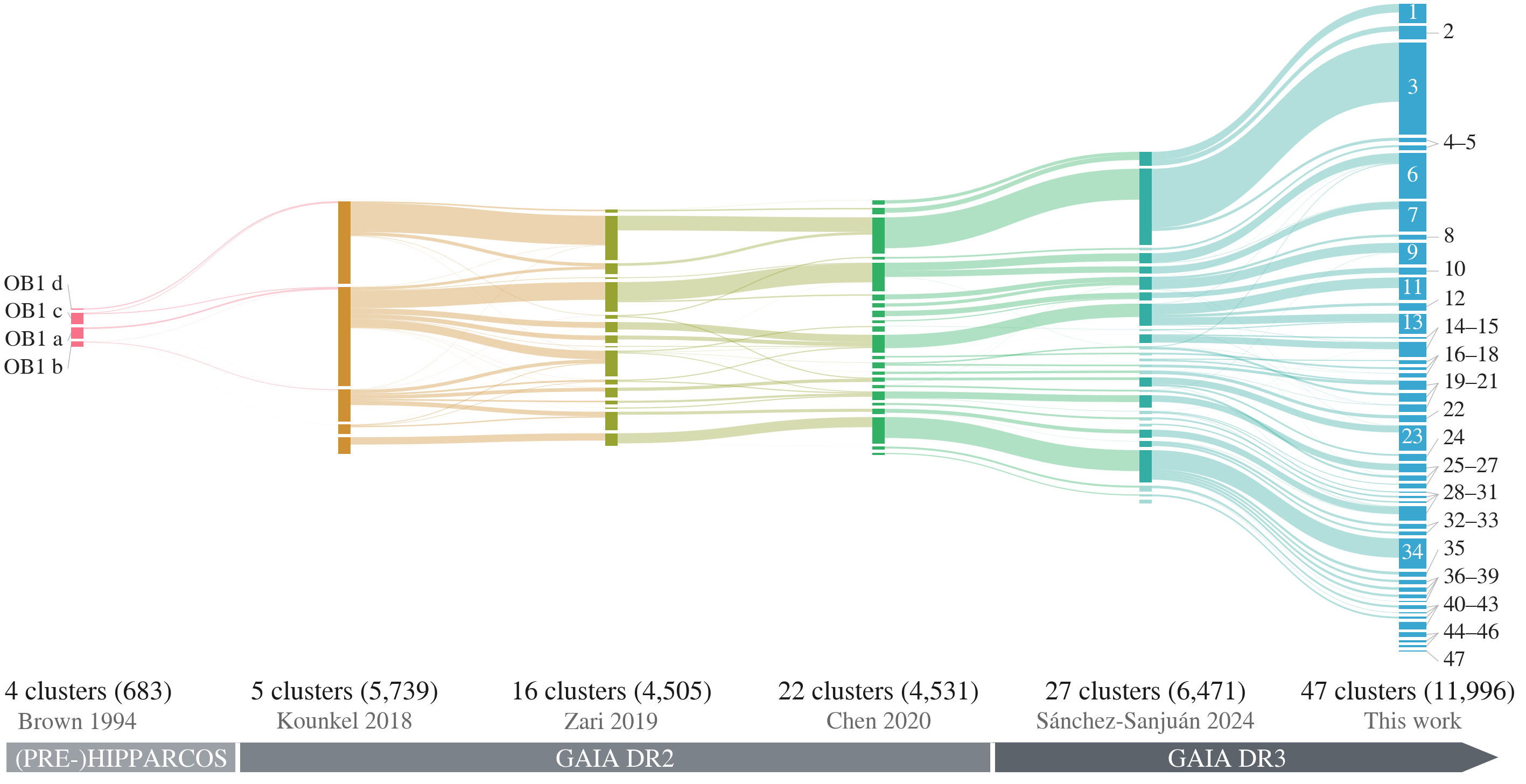}
    \vspace{-0.5cm}
    \caption{Overview of the clusters found in Orion throughout the years. An interactive, labeled version of this figure is available online*.\\
      \textbf{1} L1641S; \textbf{2} L1647; \textbf{3} ISF(ONC);
      \textbf{4} NGC~1977; \textbf{5} Orion~A-East; \textbf{6} Briceno-1B;
      \textbf{7} Briceno-1A; \textbf{8} $\omega$~Ori; \textbf{9} ASCC~20;
      \textbf{10} Orion~Y; \textbf{11} OBP-near; \textbf{12} OBP-near~ext.;
      \textbf{13} $\eta$~Ori; \textbf{14} RV~Orionis;
      \textbf{15} $\sigma$~Ori; \textbf{16} SO-2; \textbf{17} old-Rigel;
      \textbf{18} SO-1; \textbf{19} L1616B; \textbf{20} L1616A;
      \textbf{21} SO-3; \textbf{22} L1634; \textbf{23} OBP-far;
      \textbf{24} Orion~A-West; \textbf{25} OBP-d; \textbf{26} OBP-West;
      \textbf{27} NGC~2024; \textbf{28} OBP-center; \textbf{29} L1562;
      \textbf{30} $\lambda$~Ori~South-B; \textbf{31} OBP-b;
      \textbf{32} NGC~2071; \textbf{33} NGC~2068; \textbf{34} $\lambda$~Ori;
      \textbf{35} $\lambda$~main~B; \textbf{36} B30;
      \textbf{37} $\lambda$~main~A; \textbf{38} $\lambda$~main~C;
      \textbf{39} B30-splinter; \textbf{40} $\lambda$~Ori~South-A;
      \textbf{41} $\lambda$~Ori~South-C; \textbf{42} L1622;
      \textbf{43} Rigel; \textbf{44} SO-6; \textbf{45} L1641-C;
      \textbf{46} SO-5; \textbf{47} SO-4.
    }
    \label{fig:Literature-clusterings}
\end{figure*}

% ============================================================================ %
\section{Discussion}
\label{sec:Discussion}

To this day, Orion remains one of the best-studied regions of the sky. Its properties as the nearest active star-forming region producing massive stars, its youth, and richness in stars, proplyds, planetary disks, gas, and dust clouds have attracted investigations from a broad tapestry of research areas \citep[e.g.,][]{Hacar_2018, Tobin_2020, Aru_2024}. Naturally, among those studies, various clustering efforts can be found, going back to the earliest dates when astrometric measurements became available. Below, we connect our solution to this historical framework, discuss its robustness and limitations, and investigate the extinction properties of the Orion clusters. 

\subsection{Literature comparison}
\label{sec:Discussion-literature}

In Fig.~\ref{fig:Literature-clusterings}, we display a qualitative overview of the most pivotal and recent clustering efforts in Orion, along with the datasets they used. The bar heights correspond to the number of unique members published in each group. However, only sources with a \emph{Gaia} DR3 entry were considered for crossmatches between results. Thus, especially the overlap between \textsc{Hipparcos} and \emph{Gaia}, which targeted different brightness ranges, is low.

% Brown
The earliest definition of groups in the Ori OB~1 association was provided by \cite{Blaauw_1964}, who separated the complex into four subgroups (a-d). Using literature catalogs and observational data, \cite{Brown_1994} curated a photometric census of OB~1 member candidates and calculated distances and ages for these four groups. This catalog served as input for the \textsc{Hipparcos} mission \citep{HIPPARCOS_1997}, which for the first time provided stellar astrometry on a large scale. The \textsc{Hipparcos} data quality was found to be too poor for conclusive clustering efforts in the Orion region due to its position near the galactic anti-center and its small relative motions, but provisional membership selections were shown to overlap the photometric member list by 96\% \citep{deZeeuw_1999}. We show the (probable) members of the groups selected by \cite{Brown_1994} and observed by \textsc{Hipparcos} in the left column of Fig.~\ref{fig:Literature-clusterings}.

% Kounkel, Zari, Chen
Two decades later and utilizing the much increased data quantity and quality provided by \emph{Gaia} DR2 \citep{Gaia_DR2_2018}, \cite{Kounkel_2018} found 5 groups with almost ten times more stars (5,739) than previously reported. The first four were named (A-D) after the literature, while the fifth corresponded to the $\lambda$~Ori region. 
With the same dataset, \cite{Zari_2019} and \cite{Chen_2020} found a comparable number of member stars ($\sim 4,500$), but three and four times more clusters than \cite{Kounkel_2018}, respectively. 

%Sanchez-Sanjuan
The most recent clustering effort in Orion was done by \cite{Sanchez_2024}, who found 13 ``big'' and 34 ``small'' structures. 14 of the latter are classified as individual clusters, 8 as duplicates of big structures, and 12 as substructures of the big regime. In Fig.~\ref{fig:Literature-clusterings}, we show the 27 unique clusters. 

% SigMA comparison
The \texttt{SigMA} Orion clusters, shown in the rightmost column of Fig.\ref{fig:Literature-clusterings}, exhibit a very good agreement with previous clustering efforts. In the reliable subset, we recover 85.3\% of the members published by \citet{Chen_2020} (3865/4531 stars) and 84.3\% of those of \citet{Sanchez_2024} (5458/6471 stars), and 81.9\% of the union of the two catalogs (6460/7885 stars). The two literature solutions agree less well with each other than either of them does with our solution: they share only 3117 stars, which is 68.8\% of the \citet{Chen_2020} members and 48.2\% of those of \citet{Sanchez_2024}. For the intersection between the two catalogs, \texttt{SigMA} recovers 91.9\% (93.5\% of the full sample). This means the stars ``missing'' in our solution are mostly groups that earlier studies also disagree about.

The 25 clusters shared between \citet{Chen_2020, Sanchez_2024} and \texttt{SigMA} are mostly the same objects, only more richly populated in this work. Only the 91-star OBP-East group of \citet{Sanchez_2024} is absent from our solution at any membership probability. Additionally, Orion~B~East of \citet{Chen_2020} is recovered only marginally, with 3 of its 53 stars. Due to the decision of \cite{Sanchez_2024} to treat substructure differently, \texttt{SigMA} at times agrees more with \cite{Chen_2020} clusters in cases like L1641S and L1647, or ASCC~20 and $\omega$~Ori. The biggest resolution of previously unified clusters and groupings that we find with \texttt{SigMA} is in $\lambda$~Ori and $\lambda$~Ori East. \cite{Sanchez_2024} find three substructures, whereas we find 6, and only B30 is in exact agreement. Additionally, we find more fine structure in the Orion~B region than either of the other two solutions. \cite{Chen_2020} could not stably recover NGC~2024 despite it being the second-largest active stellar nursery in Orion after the ONC, and \cite{Sanchez_2024} could only recover it merged with NGC~2068 and NGC~2071. In this work, we can split all three clusters and additionally identify the newly reported RV~Orionis group \citep{Zerjal_2024}. We report a big overlap between their solution and ours for NGC~2024 and RV~Orionis, but we do not find their Flame group.

We also compare our results to the all-sky cluster catalog of \citet{Hunt_2024}, restricting the crossmatch region to our clustering volume (Sect.~\ref{sec:Data}). Of their clusters, 29 match the Orion complex in position, velocity, and age. We find very good agreement with their solution, recovering 28 of those clusters and 3,273/3,587 stars (91.2\%) for our reliable sample (3,786/4,125 for the full sample) with \texttt{SigMA}. For 24 of the recovered clusters, more than 90\% of the crossmatched stars fall into a single \texttt{SigMA} cluster, whereas \texttt{SigMA} splits the others into subgroups. The only cluster we miss with the reliable sample is HSC~1523, which is very small (N=17) and falls below the CST~$>5\sigma$ significance threshold for their high-quality sample.

\subsection{Robustness of the resolved Orion substructure}
\label{sec:Discussion-robustness}

The increase from the $\sim27$ Orion populations identified in recent \emph{Gaia}-based studies to the 47 groups recovered here naturally raises the question whether the additional substructure is reproducible or reflects numerical fragmentation introduced by the clustering procedure. We evaluate the robustness of our solution using the internal stability of the \texttt{SigMA} solution (Sect.~\ref{sec:Workflow:repetition-stability}), external catalog comparisons, and the color--magnitude diagrams of the individual groups. A validation using the star formation history will be the subject of an upcoming work.

In terms of internal stability, 44 of the 47 Orion groups ($\approx 94\%$) are recovered in at least half of the repetitions, each produced from different scaling factors. Of the 16 groups without a clear counterpart in the catalogs considered here, only three are on the Uncertain tier. Most of the newly proposed structure is therefore reproducible under the resampling and consensus procedure adopted in this work. This does not by itself establish that the groups are physically distinct, but it excludes fragmentation introduced by the stochastic elements of the procedure.

% source
At the source level, the substantial agreement with the independent analyses of \citet{Chen_2020, Hunt_2024} and \citet{Sanchez_2024} demonstrates strong continuity among the input populations recovered by different methods, although it does not, by itself, show that the stars are partitioned into equivalent subgroups. However, where direct counterparts exist, the partitions do agree with, or correspond to finer subdivisions of previously identified structures, as shown in Fig.~\ref{fig:Literature-clusterings} and discussed for the comparison to \citet{Hunt_2024}. For these structures, \texttt{SigMA} generally finds more members than previous algorithms. For instance, it recovers three times as many cluster members as the all-sky HDBSCAN \citep{HDBSCAN_2017} method employed by \cite{Hunt_2024}. These are nearly the same statistics as those for the Sco-Cen comparison with this catalog \citep{Ratzenboeck_2023b}. Even compared to the dedicated Orion clustering of \citet{Sanchez_2024}, which also uses HDBSCAN, it finds 1.85 times as many stars.

A comparison with the cluster CMDs shows that the larger member numbers in the \texttt{SigMA} clusters cannot be simply attributed to field contamination. The contamination estimate of the full Orion sample is 10.4\% (Appendix~\ref{Appendix-CMDs}), which is too small to account for a factor of 1.85--3 more members. The individual CMDs of the groups (Fig.~\ref{fig:CMDs}) provide an additional, photometrically independent qualitative validation of the segmentation because photometry was not used in the clustering process of \texttt{SigMA}. They generally display stellar sequences consistent with approximately coeval populations, as expected if the astrometric subdivisions trace physically distinct groups rather than arbitrary numerical fragments. Although we do not quantify the significance of this CMD coherence here, its presence across the recovered groups supports the adopted segmentation. The CMDs therefore suggest that \texttt{SigMA} offers a richer, more complete view of the same clusters seen by the unsupervised HDBSCAN methods, rather than a more contaminated one.

\subsection{Limitations}
\label{sec:Discussion-limitations}

As with all current clustering algorithms, the main limitation of \texttt{SigMA} is the quality of the available data, specifically, the stellar parallax errors at larger LOS distances. While we mitigated their influence and stretching effects as much as possible in the \texttt{DistantSigMA} extension, we would achieve better performance with smaller errors and could cluster on more sources to begin with, since the parallax-over-error quality cut is very restrictive.

The ISF cluster illustrates a different limitation. We could not separate this cluster into its known subpopulations (NGC~1980 and ONC), apart from NGC~1977, even with a second clustering run (Sect.~\ref{sec:Workflow-postprocessing}). While there is photometric evidence for multiple populations in the ONC region \citep[e.g.,][]{Beccari_2017, Jerabkova_2019}, \texttt{SigMA} cannot reliably extract further substructure, which is likely why the derived group age of 6.5~Myr is significantly higher than the 1--3~Myr estimates for the ONC \citep{Hillenbrand_2001, Schoettler_2020}. The reason for the inseparability lies in the special phase space structure of the group: the filament produced thousands of young stars in a very constrained volume \citep[e.g.,][]{Bally_2008}, and the group's velocity distribution in the \texttt{SigMA} solution is essentially a large two-dimensional Gaussian with no resolvable substructure. The astrometric complexity of the region is well documented \citep{Sanchez_2024}. Moreover, the proximity of its members to the ionized gas of the ONC nebula and other very bright sources may degrade their astrometric solution. 

A third limitation, specific to the youngest clusters, is the lack of observations of very young stars. Their most deeply embedded members are likely found near the cluster centers, where the density peak would lie. This was shown for NGC~2024, for example, by \cite{Rottensteiner_2026}. As they are missing from the optical \emph{Gaia} data, the density distribution of these clusters shows a hole in the middle, and the algorithm cannot perform as intended. However, we highlight that, despite their likely missing-member problem, \texttt{SigMA} recovered the very young clusters NGC~1977, NGC~2024, NGC~2068, NGC~2071, RV~Orionis, and B~30, outperforming all previous clustering efforts.

Better data quality, such as expected from the upcoming \emph{Gaia} data release 4, will likely alleviate the first limitation. Whether it also resolves the ISF depends on whether its apparent smoothness is intrinsic or owed to the current data quality. For the very young, dust-shrouded clusters, one may need to go a step further and incorporate information on embedded member stars from infrared surveys \citep[e.g., VISIONS;][]{Meingast_2023a} to circumvent the missing density core problem.

\subsection{Extinction evolution in Orion}
\label{sec:Discussion-Extinction}

\begin{figure}
    \centering
    \includegraphics[width=1\linewidth]{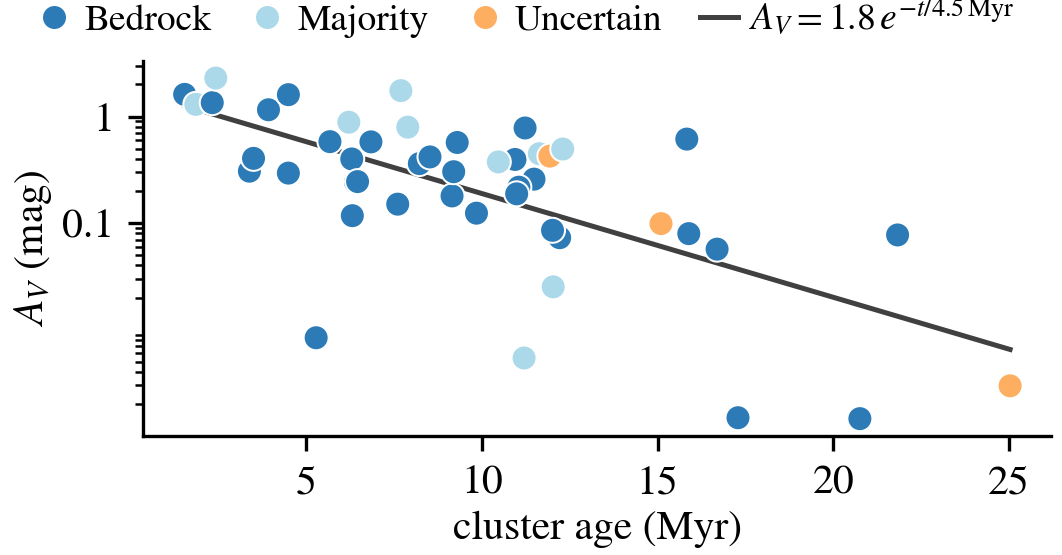}
    \caption{Extinction $A_V$ against cluster age for the 47 Orion clusters, colored by membership tier. The stability tiers are shown to demonstrate that the fit is not steered by low-confidence clusters.} 
    \label{fig:cluster-av}
\end{figure}

To investigate the extinction properties of the clusters in more depth than by comparing to the 3D dustmap, we plot cluster extinction values as a function of age in Fig.~\ref{fig:cluster-av}. An exponential fit in log-space using a least-squares technique yields $A = 1.8 \pm 0.7$~mag and $\tau = 4.5 \pm 0.7$~Myr. The anticorrelation is significant and independent of the assumed functional form (Spearman $\rho = -0.61$, $p < 10^{-5}$). We did not use a weighting strategy, as the fit is purely qualitative,  and we cannot account for possible correlations between ages and extinctions produced by the isochrone fitting.  The quoted uncertainties are the formal fit errors, meaning the decay timescale should be read as an indicative trend rather than a precise measure.

The decay from the fit is consistent with rapid gas expulsion within the first few million years of a cluster's lifetime. However, we stress that our sample is biased towards optically revealed clusters and therefore likely underestimates the extinction of very young clusters. The relation is also consistent with the qualitative picture of Fig.~\ref{fig:orion-master-dust}: The clusters identified as clearly embedded (NGC~2071, NGC~2068, L1641-C, L1647, and RV~Orionis) all lie above the fitted curve, by a median factor of 1.6, and the partially embedded groups follow the same trend. L1641S and L1622, whose fitted extinctions disagree with the dustmap and the literature, are below the curve, with L1641S the most discrepant point in the diagram. This again implies that the fits are affected by an age-extinction degeneracy or missing members.

The trend shown in Fig.~\ref{fig:cluster-av} is best interpreted as an empirical description of the optically accessible, post-emergence populations rather than of the complete clearing process. Within that limitation, however,  it supplies a low-extinction endpoint that can be placed in the broader context of studies targeting younger and more deeply embedded clusters. Our data provide the most detailed homogeneous view to date of how low levels of extinction evolve among the resolved stellar populations of Orion.

Comparison with previous studies may indicate that extinction evolution is not described by a single exponential over the full $A_V$ range. For compact embedded clusters with foreground-corrected extinctions reaching $A_V \sim 30~\mathrm{mag}$, \citet{Rawat2024} obtained $A_V = 26.3\,\exp(-1.26t)$. The corresponding e-folding time of $\approx 0.8~\mathrm{Myr}$ is substantially shorter than the $4.5 \pm 0.7~\mathrm{Myr}$ derived here in the predominantly low-extinction regime. Interestingly, the two relations intersect near $t \sim 2.6~\mathrm{Myr}$ and $A_V \sim 1~\mathrm{mag}$. This may suggest two regimes: rapid dispersal of dense natal material during the embedded phase, followed by a slower post-emergence decline in residual or diffuse extinction. This is qualitatively consistent with the resolved MYStIX study by \citet{Getman2014}, where the youngest subclusters occupy highly obscured molecular cloud regions, intermediate-age populations are revealed in clusters, and the oldest populations are spatially distributed. Spatially resolved H\,{\sc ii}-region studies similarly place the onset of gas clearing at approximately $1$--$2~\mathrm{Myr}$ and the partially exposed phase at $2$--$3~\mathrm{Myr}$ \citep{Hannon2022}, while near-infrared-selected clusters indicate typical emergence times of $3$--$5~\mathrm{Myr}$ \citep{Messa2021}. At extinctions comparable to those studied here, \citet{Grossi2010} also reported decreasing $A_V$ with age for clusters and associations in M33 spanning $A_V=0.3$--$1.0~\mathrm{mag}$ and ages of $2$--$15~\mathrm{Myr}$.

These comparisons motivate, but do not establish, a transition near $A_V \sim 1~\mathrm{mag}$ and an age of a few Myr. Differences in foreground-extinction corrections, selection wavelength, cluster mass, and covariance between fitted age and extinction prevent the published decay timescales from being compared directly.

% ============================================================================%
\section{Conclusion}
\label{sec:Conclusion}

We present the most complete census of stellar populations across the Orion star-forming complex to date, based on \emph{Gaia}~DR3 astrometry and on \texttt{DistantSigMA}, the distance-adaptive extension of the \texttt{SigMA} clustering algorithm. \texttt{DistantSigMA} addresses the cluster elongation effects introduced by observational errors at larger distances ($\gtrsim 300$~pc) by implementing a data-driven scaling scheme of the axes of the clustering parameter space. Thanks to multiple repetitions of the clustering process, we also provide cluster persistence scores and membership stabilities for all presented groups.

Our main result is a stellar membership catalog that resolves the Orion complex into 47 groups, with ages ranging from 1.5 to 25~Myr. They comprise 13,224 stars, of which 11,996 are reliable members. This is a factor of $\sim$2--3 more members and a substantially finer subdivision than the 27--29 populations of recent \emph{Gaia}-based studies. We relate 31 to known groups or substructures and report 16 as new structure candidates for the Orion complex. We recover 85\% of the members published by \citet{Chen_2020}, 84\% of those of \citet{Sanchez_2024}, and 91\% of the \citet{Hunt_2024} members. About $94\%$ of the groups are stable across at least half of the ten repetitions, and only three of the 16 new groups are classified as Uncertain. The cluster CMDs show coherent sequences for each group. Notably, we are more successful in recovering the youngest known populations than previous clustering efforts.

Using the homogeneously derived ages and extinctions of this cluster sample, we show that extinction decays with cluster age in Orion. The fitted extinction follows $A_V = 1.8\,e^{-t/4.5\,\mathrm{Myr}}$, consistent with rapid gas expulsion in the first few million years. 

We also find clusters beyond Orion in our box, reporting a total of 76 clusters (17,873 reliable members) within the survey volume. Of the other clusters found, nine each trace the stellar Snake \citep{Tian_2020} (1,546 stars, 23.3--47.3~Myr) and the BBJ groups \citep{Beccari_2020} (32.6--55.2~Myr). Neither was targeted, and their memberships should not be read as complete, but both are recovered as coherent kinematic structures with resolved internal age sequences.

We are not able to fully and reliably separate all components of the integral-shaped filament, due to its apparent smoothness in the 5D phase space. It resists decomposition beyond the split of NGC~1977, which is likely why its fitted age exceeds the accepted age of the ONC (Sect.~\ref{sec:Discussion-limitations}). Another limitation of the algorithm is the missing-member problem for the youngest, still embedded stars at the centers of young clusters, which may affect mode-seeking clustering algorithms such as \texttt{SigMA}.
Improved astrometry from \emph{Gaia}~DR4 will likely yield even more precise clustering results, and incorporating information about the embedded population itself might improve membership completeness in the youngest groups.
  
This catalog provides the most complete, homogeneous map of both cluster membership and age across an entire star-forming complex, including the most complete census yet of its youngest clusters, obtained using a clustering algorithm on \emph{Gaia} data. The spatio-temporal pattern visible in Fig.~\ref{fig:Orion-age-panels} is the kind of signal that such a map makes accessible, and reconstructing the star formation history of Orion from it will be the subject of a dedicated forthcoming work.

% ============================================================================%
\section*{Data availability}
The full version of Tab.~\ref{tab:overview-orion}, the machine-readable version of Tab.~\ref{tab:member-catalog}, and the table of soft probabilities for ambiguous cluster members will be made available in electronic form at the CDS.

% ============================================================================%
\begin{acknowledgements}
Co-Funded by the European Union (ERC, ISM-FLOW, 101055318). Views and opinions expressed are, however, those of the author(s) only and do not necessarily reflect those of the European Union or the European Research Council Executive Agency. Neither the European Union nor the granting authority can be held responsible for them.
This work has made use of data from the European Space Agency (ESA) mission
\emph{Gaia} (\url{https://www.cosmos.esa.int/gaia}), processed by the \emph{Gaia}
Data Processing and Analysis Consortium (DPAC,
\url{https://www.cosmos.esa.int/web/gaia/dpac/consortium}). Funding for the DPAC
has been provided by national institutions, in particular the institutions
participating in the \emph{Gaia} Multilateral Agreement.
The following Python libraries were used in this work without specifically being referenced in the main text: \texttt{numpy} \citep{numpy}, \texttt{pandas} \citep{pandas1, Pandas2}, 
\texttt{scipy} \citep{scipy}, \texttt{scikit-learn} \citep{scikit-learn}, \texttt{scikit-optimize} \citep{skopt},
\texttt{astropy} \citep{astropy:2013, astropy:2018, astropy:2022}, \texttt{numba} \citep{numba}, \texttt{networkx} \citep{networkx}, \texttt{plotly} \citep{plotly}, \texttt{matplotlib} \citep{matplotlib}, \texttt{arviz} \citep{arviz}, \texttt{nglpy} (github/maljovec), \texttt{pynverse} (github/alvarosg) and \texttt{chronos} (github/sebastianratzenboeck).
\end{acknowledgements}

% ============================================================================%
\bibliography{ref}
\clearpage
% ============================================================================%
\appendix
\nolinenumbers  

% ============================================================================%
\section{\emph{Gaia} DR3 data retrieval and filtering}
\label{Appendix:Data}

All data for the clustering were downloaded from the \emph{Gaia} DR3 archive via the \texttt{Gaia} TAP+ service, using the \texttt{astroquery} \citep{astroquery:2019} and \texttt{pyvo} \citep{pyvo_TAP_2019} packages. To stay within the archive row-count limit, the right ascension range was split into three equal sub-queries as follows:
\begin{align*}
    &\texttt{SELECT * FROM gaiadr3.gaia\_source}\\
    &\texttt{WHERE ra >= ra\_lo AND ra < ra\_hi}\\
    &\texttt{AND dec >= -35.0 AND dec <= 35}\\
    &\texttt{AND parallax >= 1.5 AND parallax <= 6}\\
    &\texttt{AND parallax\_over\_error > 4.5}
\end{align*}
Here, \texttt{ra\_lo} and \texttt{ra\_hi} denote the respective sub-range boundaries of each query. The three queries together covered $\alpha \in [40^\circ, 120^\circ)$, $\delta \in [-35^\circ, 35^\circ]$, and $\varpi \in [1.5, 6]$~mas. We applied the \texttt{parallax\_over\_error} cut, which approximates a signal-to-noise cut in the parallax ($S/N_\varpi$), directly at the query time to reduce the retrieved data volume. The parameter choice of $S/N_\varpi > 4.5$ ensured consistency with the \texttt{fidelity\_v2} classifier, which uses this threshold to switch between high- and low-$S/N_\varpi$ internal models \citep{Rybizki_2022}. The total number of sources in the query was 4,172,950. We then queried the \texttt{fidelity\_v2} source reliability parameter from the \texttt{gedr3spur} catalog of \cite{Rybizki_2022} and cross-matched it with the initial dataset using the \emph{Gaia} unique source identifier: 
\begin{align*}
     &\texttt{SELECT *}\\
     &\texttt{FROM gedr3spur.main AS gaia}\\
     &\texttt{JOIN TAP\_UPLOAD.t1 AS mine}\\
     &\texttt{USING (source\_id)}
\end{align*}    
We applied \texttt{fidelity\_v2}$~> 0.5$ \citep{Ratzenboeck_2023a, Rybizki_2022, Zari_2021} to filter reliable sources in our dataset, which reduced the initial selection to 3,831,476 sources. Similar to \cite{Ratzenboeck_2023a}, we tested a more stringent threshold of \texttt{fidelity\_v2}$~> 0.9$, but found that only 5\% of our dataset fell between the values 0.5 and 0.9, and opted for the less conservative value. We computed Galactic Cartesian coordinates using the inverted parallax as a distance proxy (see Appendix~\ref{Appendix:Distances}) and applied the positional cuts listed in Eq.~\ref{eq:box-coords} to isolate the Orion region. Our final dataset is a roughly rectangular selection of 910,581 stars with reliable 5D astrometry.

The sampling data for the cluster simulations (Sect.~\ref{sec:Workflow-scalefactors}) was retrieved from the \emph{Gaia} DR3 archive with the following ADQL query:
\begin{align*}
    &\texttt{SELECT}\\\
    &\hspace{1.7em}\texttt{source\_id, ra, ra\_error, dec, dec\_error,}\\
    &\hspace{1.7em}\texttt{parallax, parallax\_error, parallax\_over\_error,}\\
    &\hspace{1.7em}\texttt{pmra, pmra\_error, pmdec, pmdec\_error, l, b,}\\
    &\hspace{1.7em}\texttt{radial\_velocity, radial\_velocity\_error,}\\
    &\hspace{1.7em}\texttt{phot\_g\_mean\_mag, bp\_rp, random\_index}\\
 &\texttt{FROM gaiadr3.gaia\_source}\\
&\texttt{WHERE parallax > 0}\\
   &\texttt{AND 1000.0 / parallax <= 750}\\
   &\texttt{AND parallax\_over\_error > 4.5}\\
   &\texttt{AND MOD(random\_index, 20) = 0}
\end{align*}
The condition \texttt{MOD(random\_index, 20) = 0} returns a reproducible 5\% random subsample. We additionally applied the \texttt{fidelity\_v2} $~> 0.5$ cut, ensuring that the resulting catalog within 750~pc is subjected to the same quality criteria as the clustering sample.

% ============================================================================%
\section{Notes on the distance approximation}
\label{Appendix:Distances}
\begin{figure}[t]
    \centering
    \includegraphics{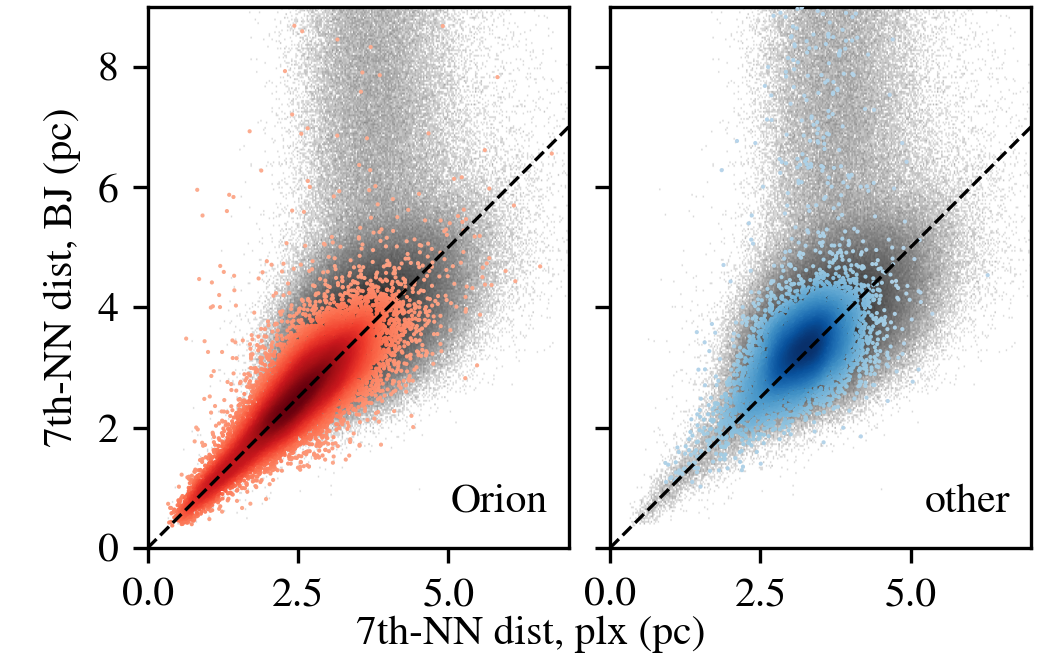}
    \caption{Comparison of the 3D Galactic Cartesian 7th nearest neighbor distances calculated using either the inverse of the \emph{Gaia} DR3 parallax (plx), or photogeometric distance estimates \citep[][abbreviated as BJ]{Bailer-Jones_2021}. The full dataset is shown in gray, the \texttt{SigMA} Orion cluster sample is highlighted in red (\emph{left panel}), and other clusters found in this work are shown in blue (\emph{right panel}).
    }
    \label{fig:distances}
\end{figure}

\noindent Using inverted stellar parallaxes as proxies for their distances is subject to limitations, depending on both the absolute source distance and the relative measurement error. A popular alternative approach is to use photogeometric distances for stars, estimated for \emph{Gaia} DR3 sources using a Bayesian model by \cite{Bailer-Jones_2021}. 

We investigated the relationship between parallax-based distances and photogeometric distances for the stars in our box to test if there are any discernible differences between the two distance parameters. We show the relation of the 3D Galactic Cartesian 7th nearest neighbor distance for the two variables in Fig.~\ref{fig:distances}. We chose this parameter as it should yield a good density estimate without over-smoothing. The Cartesian space was chosen to avoid mixing different units. The full box is shown in gray, and the colored subsets show the Orion member stars as well as the non-Orion cluster members identified with \texttt{SigMA} in this work. The Orion clusters follow the 1:1 correlation line closely and symmetrically, while the remaining clusters show a slightly larger scatter toward higher distances, but still have high symmetry around the 1:1 line. This means that, especially for Orion, but also for many non-Orion clusters, the choice of distance estimate should not significantly affect the clustering. Neither population significantly populates the vertical tail in the full dataset distribution, where the two distance metrics begin to deviate strongly (7th-NN (BJ) $\gtrsim 5.2$~pc).
When considering only the reliable cluster members (Majority tier and $p \geq 0.5$), the scatter around the 1:1 relation and toward larger 7th-NN distances decreases in both panels (not shown).

% ============================================================================%
\section{Splitting the initial dataset}
\label{Appendix:Segments}

\begin{figure*}[ht!]
    \centering
    \includegraphics{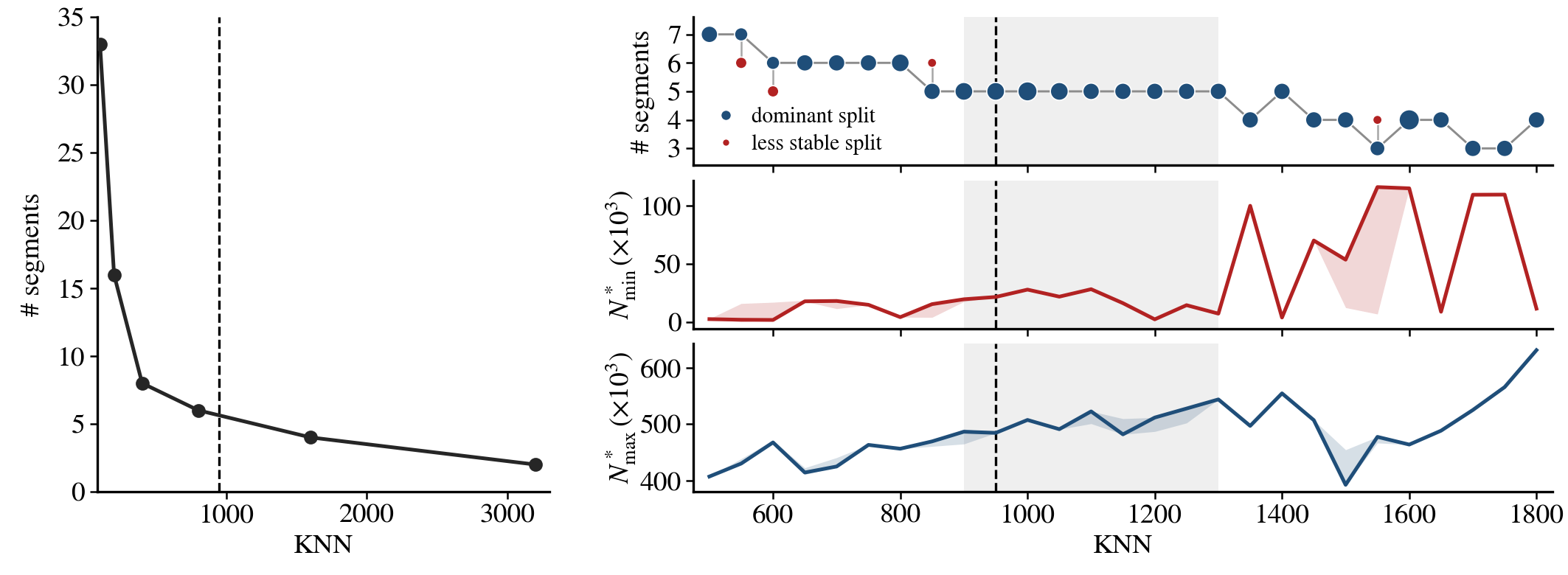}
    \caption{Coarse segmentation of the initial dataset. 
    \emph{Left:} Segmentation evaluation on a coarse logarithmic scan of $k$ values. 
    \emph{Right, first row:} Zoom-in of the segments in the region of interest. The marker size scales with the number of resamplings per $k$. Blue marks the dominant (most frequent) splits and red the less frequent ones. 
    \emph{Right, second and third row:} Median sizes of the smallest ($N^*_{\mathrm{min}}$) and largest ($N^*_{\mathrm{max}}$) segment across resamples, with the shaded band showing the full resample-to-resample range. The dashed line indicates the optimal value $k=950$, and the five-segment plateau is shaded in gray.}
    \label{fig:KNN-coarse}
\end{figure*}

To calculate appropriate scale factors for the large volume considered in the Orion extraction and to keep computational costs manageable, we performed a coarse preliminary segmentation of the initial dataset into a few stable subsets. \texttt{SigMA} is well suited for this partitioning step, as it naturally segments a dataset along low-density boundaries and produces mutually exclusive groups. Running it with a large $k$-nearest neighbors (KNN) value on a dataset dominated by field stars should, in theory, separate the volume into a small number of extended regions without artificially disrupting true clusters. 

We evaluated the number of partitions for different $k$ values using heliocentric Galactic Cartesian coordinates and LSR tangential velocities. The velocity columns were scaled using the Bayesian posterior predictive strategy of \cite{Ratzenboeck_2023a} with $n_c = 10$, and the positional columns were left unscaled. We applied internal resampling ($m = 10$) and the Benjamini--Hochberg correction for multiple hypothesis testing \citep{Benjamini_1995}, and used an initial significance level $\alpha = 0.01$. The labels for each scale-factor solution were first aligned with those of a common reference solution via one-to-one Jaccard-similarity matching. We then trained one Random Forest classifier per aligned solution (\texttt{sklearn}, 100 trees), taking the five phase-space coordinates as features and the aligned segment labels as the target class, so that each forest learns the segment boundaries of a given scale factor. Each forest was fitted to a random 80\% of the sources, and its predictions were extended to the full dataset, smoothing over random edge effects. The final per-source segment assignment was determined by majority vote across the $n_c$ Random Forest predictions. Segments smaller than $k$ were merged into the spatially nearest segment, so that no segment is smaller than the chosen neighborhood size.

We first sampled a logarithmic parameter space $k \in [100, 3200]$ and then focused on the subset $k \in [500, 1800]$, which produced 3 to 7 segments (Fig.~\ref{fig:KNN-coarse}, panels 1 and 2). We found that the dataset partition can still vary across individual clustering runs, even with fixed parameter settings and internal resampling. For this reason, we performed several segmentations at each $k$ value for the second subset. The most stable plateau was the five-segment plateau for $k \in [900,1300]$. Comparing the sizes of the smallest and largest segments (Fig.~\ref{fig:KNN-coarse}, panels 3 and 4), we selected the segmentation achieved by $k=950$ as the most balanced partition, but note that the difference in the partitions between $k=950$ and $k=1050$ is negligible.

% ============================================================================%
\section{Parameter stability grid}
\label{Appendix:Parameter-exploration}
\begin{figure*}[ht!]
        \centering
    \includegraphics[width=\linewidth]{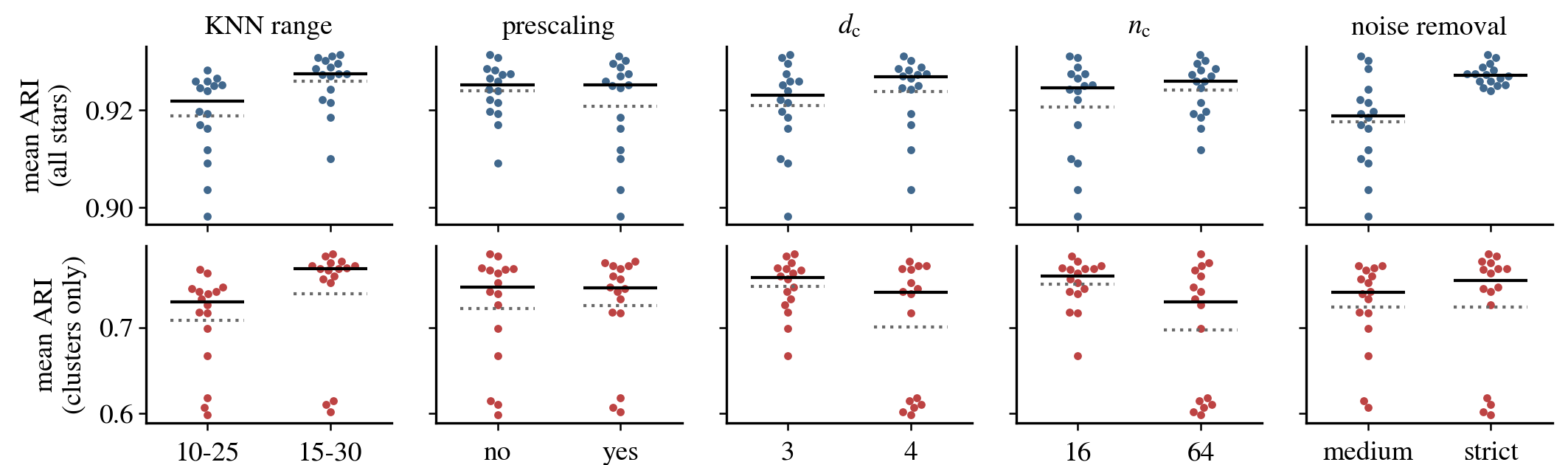}
    \caption{Marginalized plots of the isolated effect of each grid parameter on the overall clustering results. The top row shows the results for the full dataset, whereas the bottom row shows the results with global noise removed. The solid line is the median, and the dashed line shows the mean.}
    \label{fig:Grid-ARIs}
\end{figure*}

To test the solution's dependence on our chosen pipeline parameters, we evaluated a grid of different parameter settings on a partition of our dataset. We chose one of the three medium-sized segments ($\sim 100,000$ stars), which contained most of the Orion population. The parameter grid consisted of two sets of KNN values (15--30, 10--25), which affect both the preliminary clustering (Sec.~\ref{sec:Workflow-fine-preliminary}) and the final run (Sec.~\ref{sec:Workflow-fine-final}), the prescaling of the axes (yes/no), affecting only the preliminary run, and the number of scale factors (16, 64), noise removal method (medium, strict) and imposing the same scale factor for the proper motion axes or not (yes/no) (dimension), which all affect the final runs. The grid spanned by these parameter combinations has 32 nodes. 

We calculated the clustering results for each grid node setting and compared the results between different nodes using the adjusted rand score (ARI), which is the Rand index adjusted for chance. The Rand Index quantifies the similarity between two clustering solutions by comparing pairwise label assignments \citep{Hubert_1985}. This value indicates the nominal agreement between two labeled datasets and is normalized to unity. We stress that the ARI score does not provide any information about the physical reliability or goodness of a given solution. It is, however, well-suited to showcase the isolated influence of each parameter setting on the clustering solution. To do so, we consider marginalized plots, in which we compare ARI scores between the two settings of the same parameter, averaged over all other parameters. The marginalized plots are displayed in Fig.~\ref{fig:Grid-ARIs}, with the solid lines indicating the median, and the dashed lines indicating the mean ARI for each parameter setting. The colored dots correspond to the ARI score at each grid point and are shown to represent the spread in the data. The top row shows the results for the full dataset, whereas the bottom row shows the results after removing global noise, i.e., stars labeled as noise at every grid point.

As can be seen, the presence of large noise clusters artificially inflates the ARI scores. Therefore, to evaluate the influence of different parameters on the actual clusters, we refer to the second row. The KNN range slightly affects the ARI scores, with the 15--30 range yielding a higher score. We adopt this range in the final run. Prescaling does not affect agreement across clustering runs, with median values nearly identical in both settings. Imposing the same scale factors on the proper motions yields slightly higher agreement among the different solutions, whereas allowing the proper motions to deviate from isotropic conditions yields high ARI scores for half the nodes and lower ones for the other half. However, the difference in median is $\sim 0.02$. We chose $d_{\text{c}} = 4$ in our final parameter settings, as we prefer not to force the proper motions of a cluster to be isotropic. Increasing the number of scale factors also decreases the mutual agreement between the solutions by $0.03$, at 4$\times$ the computation time. Moreover, Sobol sampling is designed so that $n_{\text{c}} = 16$ already spans the full sampling range that $n_{\text{c}} = 64$ can cover. We chose a compromise for this parameter by using only 16 samples per run, but performing 10 repetitions with different sampling seeds to more comprehensively cover the scale factor parameter space (Sect.~\ref{sec:Workflow:repetition-stability}). Lastly, the noise removal method has only a slight effect on the results after global noise is removed.

% ============================================================================%
\section{Cluster CMDs and contamination estimates for the Orion clusters}
\label{Appendix-CMDs}

\begin{figure}[ht!]
    \centering
    \includegraphics[width=\linewidth]{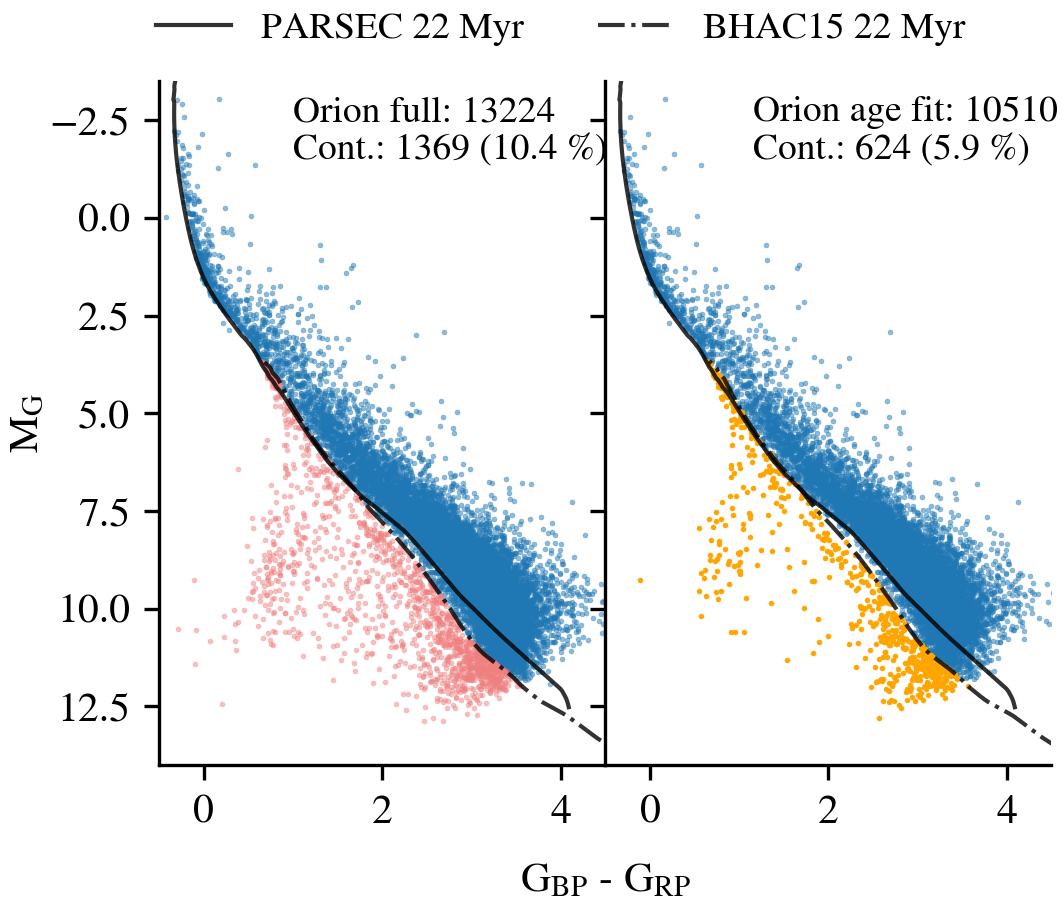}
    \caption{\emph{Gaia} $\text{M}_{\text{G}}$ vs. $\text{G}_{\text{BP}}-\text{G}_{\text{RP}}$ CMD showing the \texttt{SigMA}-selected Orion members alongside 22~Myr model isochrones. Contaminants are identified using the BHAC15 isochrone as a cutoff threshold and are displayed in coral and orange, respectively.
    \emph{Left}: Contamination fraction when considering all cluster members. \emph{Right}: Contamination fraction when considering the reliable members.}
    \label{fig:Orion-contamination}
\end{figure}

\begin{figure*}[ht]
        \centering
    \includegraphics[width=\linewidth]{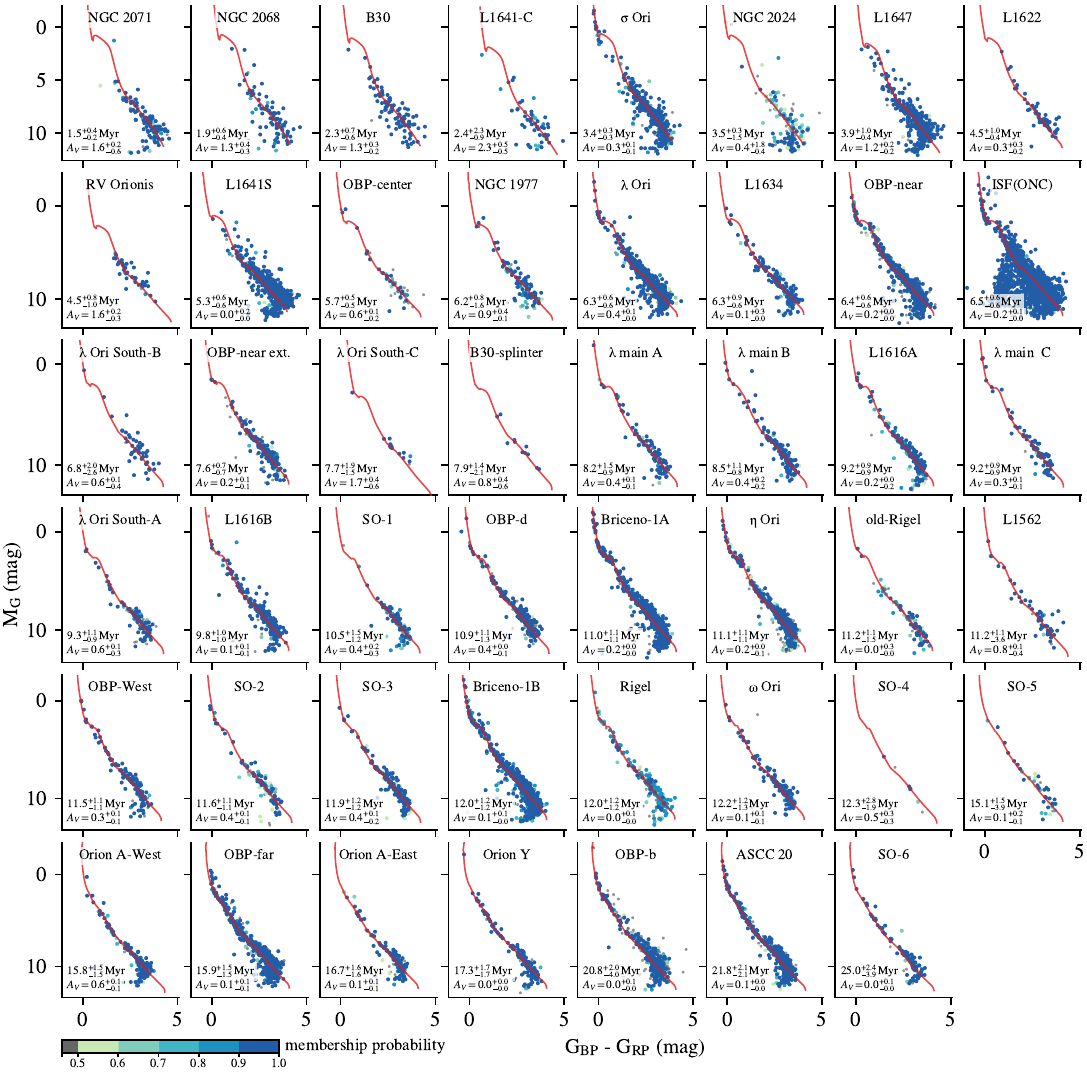}
    \caption{Cluster CMDs and best-fit PARSEC isochrones for all clusters associated with the Orion star-forming region, ordered by age. The cluster members are colored according to their assigned membership probability.}
    \label{fig:CMDs}
\end{figure*}

We present the CMDs and best-fit PARSEC isochrones for the 47 clusters associated with the Orion star-forming complex in Fig.~\ref{fig:CMDs}. Only stars with a membership probability $\geq 0.5$ that were not flagged as outliers were considered in the fit. We implemented photometric quality cuts for each Gaia passband $X$, following \cite{Ratzenboeck_2023b}:
\begin{equation}
\label{eq:photometric-cuts}
\begin{aligned}
&\text{G}_{\text{err},~X} = 1.0857~/~\text{phot\_}X\texttt{\_mean\_flux\_over\_error} \\
      &\text{G}_{\text{err, BP}}  < 0.15 \\
      &\text{G}_{\text{err, G}}  < 0.007 \\
      &\text{G}_{\text{err, RP}}  < 0.03 
\end{aligned}
\end{equation} 
Ages were fitted with the \textsc{CHRONOS} algorithm \citep{Ratzenboeck_2023b}, which employs a Bayesian framework with a skewed Cauchy fit function. The shape of the fit function is controlled by hyperparameters \texttt{skewness} and \texttt{scale} and the fit parameters were as follows: age $~\in [1,200]$~Myr, \texttt{skewness}~$\in [0,0.99]$, \texttt{scale}~$\in [0.001,0.5]$, and $A_V \in [0,3]$~mag. All priors were flat, and the metallicity was fixed to solar values. The dynamical range for the fit was truncated at $\text{M}_{\text{G}} = 10$~mag, as model isochrones diverge from the observed data for the lower main sequence \citep[e.g.,][]{2024Rottensteiner}. The fits were computed on the $\text{G}_{\text{BP}}-\text{G}_{\text{RP}}$ CMDs using a Bayesian Monte Carlo Markov Chain fit with 80 walkers and 800 steps (150 burn-in steps). The values in the bottom right corner of each panel are the mode and $1\sigma$ high-density interval (HDI) of a 100-bin histogram of each sampled posterior. If the HDI-based age uncertainty is smaller than the logarithmic model isochrone grid resolution (0.04~dex), the grid resolution is quoted as the uncertainty on each side. The posteriors were inspected visually for each cluster.

We estimate the contamination fraction of our cluster solution, following \cite{Ratzenboeck_2023a}, using the CMD and model isochrones. In the left panel of Fig.~\ref{fig:Orion-contamination}, we display the full sample of 13,224 cluster members, along with the 22~Myr isochrones from PARSEC \citep{Bressan_2012} and BHAC15 \citep{Baraffe_2015}.\footnote{The oldest cluster in our solution is SO-6, which is 25~Myr old, but on the Uncertain tier. Hence, we chose the ASCC~20 cluster (21.8~Myr) as a conservative age boundary.} Like \cite{Ratzenboeck_2023a}, we use the latter to estimate contamination, as the former have been shown to diverge more strongly from observations in the lower main sequence \citep[see e.g.,][]{2024Rottensteiner}. The right panel of the figure shows the same set of isochrones, but this time only the sources consulted for the age fit are shown. In addition to excluding outliers and stars with membership probability $< 0.5$, these also had to meet photometric quality cuts (see Eq.~\ref{eq:photometric-cuts} in Appendix~\ref{Appendix-CMDs}). As shown, the contamination fraction is generally low, and applying the membership and photometric quality cuts described in this work halves it.

\clearpage

% ============================================================================%
\section{Auxiliary tables}
\label{Appendix-tables}

Our auxiliary datasets include the full version of Tab.~\ref{tab:overview-orion}, the machine-readable version of Tab.~\ref{tab:member-catalog}, and a table of soft probabilities for sources initially assigned to more than one consensus cluster. The full version of Tab.~\ref{tab:overview-orion} includes all \texttt{SigMA} clusters in the volume, not only Orion, as well as the full astrometric parameter set in observed and Galactic Cartesian coordinates and LSR velocities. All tables will be made available via the CDS.

\input{tables/main_table}
\input{tables/machine-readable-table}

% ============================================================================%
\end{document}

%% file: workflow1.tex
\definecolor{procfill}{HTML}{F4F7FB}   
\definecolor{procdraw}{HTML}{8FA9C4}   
\definecolor{iofill}{HTML}{ECECEC}
\definecolor{iodraw}{HTML}{9A9A9A}
\definecolor{parfill}{HTML}{FFE3C2}  
\definecolor{pardraw}{HTML}{D9730D}
\definecolor{grpfill}{HTML}{EEF4FA}  
\definecolor{grpdraw}{HTML}{C3D4E6}

\begin{tikzpicture}[
  node distance = 7mm and 6mm,
  font = \small,
  proc/.style = {rectangle, rounded corners=2.5pt, draw=procdraw, line width=0.7pt,
                 fill=procfill, text width=18mm, align=center, inner sep=3pt,
                 minimum height=12mm},
  io/.style   = {rectangle, rounded corners=2.5pt, draw=iodraw, line width=0.7pt,
                 fill=iofill, text width=18mm, align=center, inner sep=4pt,
                 minimum height=12mm},
  flow/.style = {-{Stealth[length=2.6mm]}, line width=0.9pt, draw=procdraw!130},
  param/.style = {rectangle, rounded corners=3pt, draw=pardraw, line width=1.1pt,
                  fill=parfill, align=center, inner sep=4pt, text width=16mm},
  param2/.style = {rectangle, rounded corners=3pt, draw=gray!60, line width=.5pt,
                  fill=gray!20, align=center, inner sep=4pt, text width=18mm},
  tap/.style  = {dashed, line width=0.8pt, draw=pardraw},
  band/.style = {rounded corners=4pt, fill=grpfill, draw=grpdraw, line width=0.6pt},
  bandlbl/.style = {font=\footnotesize\itshape, text=procdraw!130}
]

% ---------- spine ----------
\node[io]   (input)  {\textbf{Feature\\space}};
\node[proc] (dens)   [right=of input] {\textbf{KNN density}\\[1pt]$\hat{f}\!\propto\!1/d_k^{\,p}$};
\node[proc] (hill)   [right=of dens]  {\textbf{Graph +\\gradient ascent}};
\node[proc] (resamp) [right=of hill]  {\textbf{Resampling}};
\node[proc] (merge)  [right=of resamp]{\textbf{Modality\\test}};
\node[proc] (noise)  [right=of merge]{\textbf{Noise\\flagging}};
\node[proc] (cons)   [right=of noise] {\textbf{Consensus\\clustering}};

\draw[flow] (input)--(dens);  \draw[flow] (dens)--(hill);
\draw[flow] (hill)--(resamp); \draw[flow] (resamp)--(merge);
\draw[flow] (merge)--(noise); \draw[flow] (noise)--(cons); 

% ---------- parameter boxes ----------
\node[param] (c)    [below=of input]  {{\fontsize{13}{13}\selectfont\bfseries $c$}\\{\fontsize{8}{9.5}\selectfont axis scaling}};
\node[param] (k)    [below=of dens]   {{\fontsize{13}{23}\selectfont\bfseries $k$}\\{\fontsize{8}{9.5}\selectfont smoothing}};
\node[param2] (beta) [below=of hill]   {{\fontsize{11}{11}\selectfont\bfseries $\beta$}{\fontsize{8}{9.5}, init\_graph}};
\node[param] (m)    [below=of resamp] {{\fontsize{13}{13}\selectfont\bfseries $m$}\\ {\fontsize{8}{9.5}\selectfont realizations}};
\node[param] (ab)   [below=of merge]  {{\fontsize{13}{13}\selectfont\bfseries $\alpha$}{\bfseries, BH}\\{\fontsize{8}{9.5}\selectfont significance\\\& correction}};
 \node[param] (nr)   [below=of noise]  {{\fontsize{11}{12}\selectfont\bfseries
  noise}\\{\fontsize{8}{9.5} strategy}};
  plus one tap line:
  \draw[tap] (nr)--(noise);
\node[param] (grid) [below=of cons]   {{\fontsize{13}{13}\selectfont\bfseries $\{k\},\{c\}$}\\{\fontsize{8}{9.5}\selectfont full sets}};

\draw[tap] (c)--(input);   \draw[tap] (k)--(dens);    \draw[tap] (beta)--(hill);
\draw[tap] (m)--(resamp);  \draw[tap] (ab)--(merge);  \draw[tap] (grid)--(cons);

% ---------- background shading ----------
\begin{scope}[on background layer]
  \node[band, fit=(input)(dens),   inner sep=5pt] (b1) {};
  \node[band, fit=(hill),          inner sep=5pt] (b2) {};
  \node[band, fit=(resamp)(merge), inner sep=5pt] (b3) {};
  \node[band, fit=(noise)(cons),          inner sep=5pt] (b4) {};
\end{scope}
\node[bandlbl, above=1pt of b1] {Density estimation};
\node[bandlbl, above=1pt of b2] {Segmentation};
\node[bandlbl, above=1pt of b3] {Significance merging};
\node[bandlbl, above=1pt of b4] {Final labels};

\end{tikzpicture}

%% file: tables/main_table.tex
\begin{table*}[!ht] 
\begin{center}
\begin{small}
\caption{Overview of the cluster parameters of the 47 \texttt{SigMA} clusters in Orion. New clusters are indicated in boldface.}
\renewcommand{\arraystretch}{1.3}
\resizebox{0.99\textwidth}{!}{%
\begin{tabular}{lcrrcrrrcrrrrrrrrrrr}
\hline \hline
 Name & N\tablefootmark{a} & \multicolumn{1}{c}{age} & \multicolumn{1}{c}{$A_V$} & \multicolumn{1}{c}{Tier}\tablefootmark{b} & \multicolumn{1}{c}{$l$} & \multicolumn{1}{c}{$b$} & \multicolumn{1}{c}{$d$} & \multicolumn{1}{c}{$n_{\text{RV}}$} & \multicolumn{1}{c}{$X$} & \multicolumn{1}{c}{$Y$} & \multicolumn{1}{c}{$Z$} & \multicolumn{1}{c}{$U$} & \multicolumn{1}{c}{$V$} & \multicolumn{1}{c}{$W$} \\
      &   & \multicolumn{1}{c}{(Myr)} & \multicolumn{1}{c}{(mag)} & & \multicolumn{1}{c}{(deg)} & \multicolumn{1}{c}{(deg)} & \multicolumn{1}{c}{(pc)} & & \multicolumn{1}{c}{(pc)} & \multicolumn{1}{c}{(pc)} & \multicolumn{1}{c}{(pc)} & \multicolumn{1}{c}{(km, s$^{-1})$} &
\multicolumn{1}{c}{(km, s$^{-1})$} &
\multicolumn{1}{c}{(km, s$^{-1})$} \\
\hline

\textbf{NGC 2071} & 115 & 1.5$^{+0.4}_{-0.2}$ & 1.61$^{+0.19}_{-0.56}$ & B & 205.18$^{+0.15}_{-0.15}$ & -14.31$^{+0.26}_{-0.26}$ & 415$^{+17}_{-17}$ & 3 & -364$^{+15}_{-15}$ & -172$^{+7}_{-7}$ & -104$^{+5}_{-5}$ & 4.7$^{+39.2}_{-39.2}$ & 8.5$^{+18.1}_{-18.1}$ & 2.3$^{+12.5}_{-12.5}$ \\
NGC 2068 & 89 & 1.9$^{+0.6}_{-0.4}$ & 1.30$^{+0.43}_{-0.30}$ & M & 205.32$^{+0.14}_{-0.14}$ & -14.36$^{+0.17}_{-0.17}$ & 414$^{+18}_{-18}$ & 2 & -362$^{+15}_{-15}$ & -172$^{+8}_{-8}$ & -103$^{+5}_{-5}$ & -- & -- & -- \\
B30 & 104 & 2.3$^{+0.7}_{-0.6}$ & 1.35$^{+0.33}_{-0.24}$ & B & 192.57$^{+0.16}_{-0.16}$ & -11.71$^{+0.40}_{-0.40}$ & 398$^{+14}_{-14}$ & 6 & -381$^{+14}_{-14}$ & -85$^{+4}_{-4}$ & -81$^{+5}_{-5}$ & -13.4$^{+7.1}_{-7.1}$ & 4.1$^{+1.3}_{-1.3}$ & 1.9$^{+2.8}_{-2.8}$ \\
\textbf{L1641-C} & 54 & 2.4$^{+2.3}_{-0.9}$ & 2.30$^{+0.52}_{-0.53}$ & M & 210.90$^{+0.20}_{-0.20}$ & -19.37$^{+0.10}_{-0.10}$ & 398$^{+17}_{-17}$ & 4 & -322$^{+14}_{-14}$ & -192$^{+9}_{-9}$ & -133$^{+5}_{-5}$ & -2.6$^{+7.4}_{-7.4}$ & 2.6$^{+4.7}_{-4.7}$ & 1.2$^{+2.4}_{-2.4}$ \\
$\sigma$ Ori & 359 & 3.4$^{+0.3}_{-0.3}$ & 0.31$^{+0.08}_{-0.06}$ & B & 206.83$^{+0.27}_{-0.27}$ & -17.30$^{+0.30}_{-0.30}$ & 400$^{+15}_{-15}$ & 25 & -341$^{+13}_{-13}$ & -173$^{+7}_{-7}$ & -119$^{+5}_{-5}$ & -13.2$^{+9.9}_{-9.9}$ & -2.1$^{+4.0}_{-4.0}$ & 0.1$^{+3.0}_{-3.0}$ \\
NGC 2024 & 107 & 3.5$^{+0.3}_{-1.5}$ & 0.40$^{+1.81}_{-0.40}$ & B & 206.51$^{+0.17}_{-0.17}$ & -16.36$^{+0.07}_{-0.07}$ & 387$^{+31}_{-31}$ & 3 & -333$^{+25}_{-25}$ & -166$^{+14}_{-14}$ & -109$^{+9}_{-9}$ & -4.2$^{+1.8}_{-1.8}$ & 1.7$^{+0.1}_{-0.1}$ & 1.6$^{+0.1}_{-0.1}$ \\
L1647 & 320 & 3.9$^{+1.0}_{-0.4}$ & 1.16$^{+0.19}_{-0.17}$ & B & 213.64$^{+0.52}_{-0.52}$ & -19.77$^{+0.21}_{-0.21}$ & 439$^{+31}_{-31}$ & 18 & -345$^{+24}_{-24}$ & -230$^{+16}_{-16}$ & -149$^{+10}_{-10}$ & -2.8$^{+5.1}_{-5.1}$ & 0.4$^{+4.4}_{-4.4}$ & 1.8$^{+3.0}_{-3.0}$ \\
L1622 & 55 & 4.5$^{+1.0}_{-0.4}$ & 0.29$^{+0.27}_{-0.21}$ & B & 203.25$^{+0.21}_{-0.21}$ & -12.30$^{+0.13}_{-0.13}$ & 405$^{+13}_{-13}$ & 5 & -362$^{+11}_{-11}$ & -156$^{+5}_{-5}$ & -86$^{+3}_{-3}$ & -8.7$^{+3.3}_{-3.3}$ & 1.2$^{+1.6}_{-1.6}$ & -0.8$^{+0.6}_{-0.6}$ \\
RV Orionis & 41 & 4.5$^{+0.8}_{-1.0}$ & 1.60$^{+0.16}_{-0.27}$ & B & 206.82$^{+0.17}_{-0.17}$ & -16.99$^{+0.07}_{-0.07}$ & 401$^{+11}_{-11}$ & 4 & -343$^{+9}_{-9}$ & -173$^{+5}_{-5}$ & -117$^{+3}_{-3}$ & -13.2$^{+7.2}_{-7.2}$ & -2.9$^{+4.4}_{-4.4}$ & 2.2$^{+1.5}_{-1.5}$ \\
L1641S & 454 & 5.3$^{+0.6}_{-0.6}$ & 0.01$^{+0.18}_{-0.01}$ & B & 212.46$^{+0.46}_{-0.46}$ & -18.99$^{+0.23}_{-0.23}$ & 424$^{+25}_{-25}$ & 16 & -338$^{+21}_{-21}$ & -215$^{+13}_{-13}$ & -138$^{+9}_{-9}$ & -2.6$^{+5.1}_{-5.1}$ & 3.3$^{+3.5}_{-3.5}$ & 2.5$^{+2.3}_{-2.3}$ \\
OBP-center & 32 & 5.7$^{+0.5}_{-0.5}$ & 0.58$^{+0.10}_{-0.18}$ & B & 204.48$^{+0.21}_{-0.21}$ & -16.33$^{+0.14}_{-0.14}$ & 410$^{+6}_{-6}$ & 5 & -358$^{+6}_{-6}$ & -163$^{+3}_{-3}$ & -115$^{+3}_{-3}$ & -10.0$^{+20.3}_{-20.3}$ & 2.0$^{+9.7}_{-9.7}$ & -0.3$^{+6.2}_{-6.2}$ \\
\textbf{NGC 1977} & 104 & 6.2$^{+0.8}_{-1.6}$ & 0.88$^{+0.41}_{-0.14}$ & M & 208.48$^{+0.10}_{-0.10}$ & -19.12$^{+0.11}_{-0.11}$ & 392$^{+12}_{-12}$ & 6 & -326$^{+9}_{-9}$ & -176$^{+6}_{-6}$ & -129$^{+4}_{-4}$ & -5.5$^{+6.2}_{-6.2}$ & -0.1$^{+3.1}_{-3.1}$ & 1.4$^{+2.5}_{-2.5}$ \\
$\lambda$ Ori & 719 & 6.3$^{+0.6}_{-0.6}$ & 0.40$^{+0.05}_{-0.02}$ & B & 195.45$^{+0.87}_{-0.87}$ & -11.81$^{+0.79}_{-0.79}$ & 401$^{+15}_{-15}$ & 44 & -378$^{+15}_{-15}$ & -105$^{+9}_{-9}$ & -82$^{+7}_{-7}$ & -9.8$^{+4.6}_{-4.6}$ & 1.6$^{+2.4}_{-2.4}$ & 2.5$^{+1.7}_{-1.7}$ \\
L1634 & 173 & 6.3$^{+0.9}_{-0.6}$ & 0.12$^{+0.29}_{-0.03}$ & B & 206.77$^{+0.61}_{-0.61}$ & -21.69$^{+0.82}_{-0.82}$ & 383$^{+11}_{-11}$ & 15 & -318$^{+10}_{-10}$ & -160$^{+5}_{-5}$ & -142$^{+7}_{-7}$ & -6.8$^{+5.5}_{-5.5}$ & -0.6$^{+3.1}_{-3.1}$ & 1.4$^{+3.0}_{-3.0}$ \\
OBP-near & 532 & 6.4$^{+0.6}_{-0.6}$ & 0.24$^{+0.03}_{-0.04}$ & B & 205.88$^{+1.07}_{-1.07}$ & -17.07$^{+0.53}_{-0.53}$ & 356$^{+15}_{-15}$ & 56 & -306$^{+12}_{-12}$ & -148$^{+10}_{-10}$ & -104$^{+6}_{-6}$ & -4.9$^{+3.9}_{-3.9}$ & 0.9$^{+2.3}_{-2.3}$ & 2.9$^{+1.8}_{-1.8}$ \\
ISF(ONC) & 2197 & 6.5$^{+0.6}_{-0.6}$ & 0.25$^{+0.05}_{-0.04}$ & B & 209.11$^{+0.53}_{-0.53}$ & -19.43$^{+0.28}_{-0.28}$ & 390$^{+14}_{-14}$ & 137 & -321$^{+13}_{-13}$ & -179$^{+7}_{-7}$ & -130$^{+5}_{-5}$ & -8.9$^{+5.6}_{-5.6}$ & 0.3$^{+3.4}_{-3.4}$ & 1.4$^{+2.8}_{-2.8}$ \\
$\lambda$ Ori South-B & 46 & 6.8$^{+2.0}_{-2.6}$ & 0.58$^{+0.10}_{-0.39}$ & B & 195.16$^{+0.17}_{-0.17}$ & -17.09$^{+0.13}_{-0.13}$ & 397$^{+16}_{-16}$ & 4 & -367$^{+16}_{-16}$ & -99$^{+4}_{-4}$ & -115$^{+5}_{-5}$ & 3.6$^{+5.8}_{-5.8}$ & 1.0$^{+0.5}_{-0.5}$ & 4.1$^{+1.5}_{-1.5}$ \\
\textbf{OBP-near ext.} & 185 & 7.6$^{+0.7}_{-0.7}$ & 0.15$^{+0.09}_{-0.08}$ & B & 203.94$^{+0.40}_{-0.40}$ & -17.80$^{+0.35}_{-0.35}$ & 355$^{+13}_{-13}$ & 18 & -308$^{+11}_{-11}$ & -136$^{+6}_{-6}$ & -109$^{+5}_{-5}$ & -3.2$^{+4.1}_{-4.1}$ & 2.8$^{+2.3}_{-2.3}$ & 3.0$^{+1.3}_{-1.3}$ \\
\textbf{\boldmath $\lambda$ Ori South-C} & 13 & 7.7$^{+1.9}_{-1.5}$ & 1.75$^{+0.43}_{-0.57}$ & M & 201.79$^{+0.25}_{-0.25}$ & -11.69$^{+0.35}_{-0.35}$ & 454$^{+10}_{-10}$ & 1 & -414$^{+10}_{-10}$ & -165$^{+5}_{-5}$ & -91$^{+6}_{-6}$ & -- & -- & -- \\
\textbf{B30-splinter} & 11 & 7.9$^{+1.4}_{-2.1}$ & 0.79$^{+0.40}_{-0.59}$ & M & 192.50$^{+0.24}_{-0.24}$ & -11.39$^{+0.12}_{-0.12}$ & 386$^{+7}_{-7}$ & 1 & -369$^{+7}_{-7}$ & -82$^{+1}_{-1}$ & -77$^{+3}_{-3}$ & -- & -- & -- \\
\textbf{\boldmath $\lambda$ main A} & 102 & 8.2$^{+1.5}_{-0.9}$ & 0.36$^{+0.15}_{-0.08}$ & B & 196.29$^{+0.31}_{-0.31}$ & -12.17$^{+0.34}_{-0.34}$ & 394$^{+14}_{-14}$ & 12 & -370$^{+13}_{-13}$ & -108$^{+5}_{-5}$ & -83$^{+4}_{-4}$ & -9.8$^{+2.9}_{-2.9}$ & 0.5$^{+1.1}_{-1.1}$ & 2.2$^{+1.3}_{-1.3}$ \\
\textbf{\boldmath $\lambda$ main B} & 119 & 8.5$^{+1.1}_{-0.8}$ & 0.42$^{+0.17}_{-0.23}$ & B & 193.77$^{+0.79}_{-0.79}$ & -12.80$^{+0.25}_{-0.25}$ & 401$^{+19}_{-19}$ & 11 & -380$^{+17}_{-17}$ & -92$^{+7}_{-7}$ & -88$^{+5}_{-5}$ & -12.5$^{+1.8}_{-1.8}$ & 1.4$^{+0.9}_{-0.9}$ & 2.0$^{+0.6}_{-0.6}$ \\
L1616A & 198 & 9.2$^{+0.9}_{-0.9}$ & 0.18$^{+0.04}_{-0.17}$ & B & 203.33$^{+0.70}_{-0.70}$ & -21.25$^{+0.47}_{-0.47}$ & 352$^{+21}_{-21}$ & 24 & -301$^{+18}_{-18}$ & -129$^{+8}_{-8}$ & -127$^{+8}_{-8}$ & -3.1$^{+6.7}_{-6.7}$ & 4.0$^{+2.3}_{-2.3}$ & 2.3$^{+2.6}_{-2.6}$ \\
$\lambda$ main  C & 80 & 9.2$^{+0.9}_{-0.9}$ & 0.30$^{+0.10}_{-0.05}$ & B & 195.35$^{+0.20}_{-0.20}$ & -12.29$^{+0.24}_{-0.24}$ & 403$^{+14}_{-14}$ & 12 & -379$^{+13}_{-13}$ & -105$^{+5}_{-5}$ & -86$^{+4}_{-4}$ & -14.4$^{+1.6}_{-1.6}$ & 0.1$^{+0.5}_{-0.5}$ & 1.9$^{+0.8}_{-0.8}$ \\
$\lambda$ Ori South-A & 92 & 9.3$^{+1.1}_{-0.9}$ & 0.57$^{+0.11}_{-0.31}$ & B & 198.86$^{+0.32}_{-0.32}$ & -11.58$^{+0.27}_{-0.27}$ & 445$^{+20}_{-20}$ & 7 & -413$^{+18}_{-18}$ & -141$^{+7}_{-7}$ & -88$^{+5}_{-5}$ & -16.5$^{+4.0}_{-4.0}$ & 2.6$^{+2.2}_{-2.2}$ & -6.6$^{+1.4}_{-1.4}$ \\
L1616B & 230 & 9.8$^{+1.0}_{-1.0}$ & 0.12$^{+0.09}_{-0.07}$ & B & 203.58$^{+0.57}_{-0.57}$ & -24.07$^{+0.70}_{-0.70}$ & 374$^{+30}_{-30}$ & 27 & -314$^{+25}_{-25}$ & -136$^{+11}_{-11}$ & -150$^{+14}_{-14}$ & -2.8$^{+6.3}_{-6.3}$ & 4.3$^{+3.3}_{-3.3}$ & 0.4$^{+2.7}_{-2.7}$ \\
\textbf{SO-1} & 101 & 10.5$^{+1.5}_{-1.2}$ & 0.37$^{+0.20}_{-0.29}$ & M & 204.19$^{+0.69}_{-0.69}$ & -22.55$^{+0.44}_{-0.44}$ & 351$^{+25}_{-25}$ & 2 & -296$^{+21}_{-21}$ & -132$^{+8}_{-8}$ & -134$^{+9}_{-9}$ & -- & -- & -- \\
OBP-d & 204 & 10.9$^{+1.1}_{-1.3}$ & 0.40$^{+0.05}_{-0.12}$ & B & 205.26$^{+0.16}_{-0.16}$ & -18.30$^{+0.28}_{-0.28}$ & 417$^{+12}_{-12}$ & 18 & -358$^{+10}_{-10}$ & -169$^{+6}_{-6}$ & -130$^{+5}_{-5}$ & -11.6$^{+4.5}_{-4.5}$ & 0.8$^{+2.3}_{-2.3}$ & -1.0$^{+1.5}_{-1.5}$ \\
Briceno-1A & 713 & 11.0$^{+1.1}_{-1.1}$ & 0.19$^{+0.04}_{-0.01}$ & B & 200.61$^{+1.10}_{-1.10}$ & -16.59$^{+0.51}_{-0.51}$ & 346$^{+10}_{-10}$ & 79 & -310$^{+10}_{-10}$ & -116$^{+8}_{-8}$ & -98$^{+5}_{-5}$ & -5.5$^{+4.6}_{-4.6}$ & 4.0$^{+1.9}_{-1.9}$ & 3.7$^{+1.5}_{-1.5}$ \\
$\eta$ Ori & 483 & 11.1$^{+1.1}_{-1.1}$ & 0.22$^{+0.02}_{-0.09}$ & B & 205.10$^{+0.67}_{-0.67}$ & -19.37$^{+0.72}_{-0.72}$ & 353$^{+14}_{-14}$ & 54 & -301$^{+13}_{-13}$ & -141$^{+6}_{-6}$ & -117$^{+5}_{-5}$ & -5.7$^{+3.6}_{-3.6}$ & 1.5$^{+2.1}_{-2.1}$ & 1.2$^{+1.8}_{-1.8}$ \\
old-Rigel & 66 & 11.2$^{+1.1}_{-1.5}$ & 0.01$^{+0.27}_{-0.01}$ & M & 207.66$^{+0.49}_{-0.49}$ & -24.12$^{+0.81}_{-0.81}$ & 290$^{+14}_{-14}$ & 10 & -234$^{+12}_{-12}$ & -123$^{+5}_{-5}$ & -118$^{+5}_{-5}$ & -0.4$^{+6.4}_{-6.4}$ & 0.6$^{+3.7}_{-3.7}$ & -0.2$^{+4.0}_{-4.0}$ \\
L1562 & 54 & 11.2$^{+1.1}_{-3.6}$ & 0.78$^{+0.08}_{-0.43}$ & B & 188.01$^{+0.27}_{-0.27}$ & -17.02$^{+0.32}_{-0.32}$ & 317$^{+11}_{-11}$ & 7 & -300$^{+11}_{-11}$ & -42$^{+3}_{-3}$ & -93$^{+4}_{-4}$ & 0.4$^{+15.3}_{-15.3}$ & 4.9$^{+0.7}_{-0.7}$ & 3.1$^{+5.2}_{-5.2}$ \\
OBP-West & 125 & 11.5$^{+1.1}_{-1.1}$ & 0.26$^{+0.06}_{-0.10}$ & B & 202.50$^{+0.58}_{-0.58}$ & -18.68$^{+0.39}_{-0.39}$ & 417$^{+17}_{-17}$ & 19 & -364$^{+16}_{-16}$ & -151$^{+8}_{-8}$ & -133$^{+7}_{-7}$ & -13.4$^{+3.3}_{-3.3}$ & 4.4$^{+1.6}_{-1.6}$ & 0.8$^{+1.5}_{-1.5}$ \\
\textbf{SO-2} & 98 & 11.6$^{+1.1}_{-1.1}$ & 0.45$^{+0.08}_{-0.13}$ & M & 206.51$^{+0.77}_{-0.77}$ & -24.45$^{+0.58}_{-0.58}$ & 341$^{+19}_{-19}$ & 16 & -278$^{+15}_{-15}$ & -138$^{+7}_{-7}$ & -142$^{+9}_{-9}$ & -1.9$^{+3.8}_{-3.8}$ & 3.3$^{+2.3}_{-2.3}$ & 1.1$^{+2.4}_{-2.4}$ \\
\textbf{SO-3} & 177 & 11.9$^{+1.2}_{-1.2}$ & 0.42$^{+0.06}_{-0.23}$ & U & 205.15$^{+1.19}_{-1.19}$ & -20.74$^{+0.82}_{-0.82}$ & 338$^{+19}_{-19}$ & 20 & -288$^{+16}_{-16}$ & -132$^{+8}_{-8}$ & -120$^{+8}_{-8}$ & -4.2$^{+4.5}_{-4.5}$ & 4.0$^{+3.6}_{-3.6}$ & 1.9$^{+2.2}_{-2.2}$ \\
Briceno-1B & 1081 & 12.0$^{+1.2}_{-1.2}$ & 0.09$^{+0.06}_{-0.02}$ & B & 201.46$^{+0.81}_{-0.81}$ & -18.60$^{+0.81}_{-0.81}$ & 347$^{+13}_{-13}$ & 125 & -306$^{+12}_{-12}$ & -121$^{+7}_{-7}$ & -111$^{+6}_{-6}$ & -5.6$^{+4.0}_{-4.0}$ & 4.0$^{+1.6}_{-1.6}$ & 2.7$^{+1.6}_{-1.6}$ \\
\textbf{Rigel} & 173 & 12.0$^{+1.2}_{-1.2}$ & 0.03$^{+0.10}_{-0.03}$ & M & 209.60$^{+1.08}_{-1.08}$ & -24.22$^{+0.82}_{-0.82}$ & 313$^{+22}_{-22}$ & 30 & -247$^{+17}_{-17}$ & -142$^{+12}_{-12}$ & -129$^{+9}_{-9}$ & 0.2$^{+4.1}_{-4.1}$ & 5.7$^{+2.5}_{-2.5}$ & 1.1$^{+2.4}_{-2.4}$ \\
$\omega$ Ori & 127 & 12.2$^{+1.2}_{-1.3}$ & 0.07$^{+0.12}_{-0.07}$ & B & 199.73$^{+0.65}_{-0.65}$ & -14.76$^{+0.61}_{-0.61}$ & 418$^{+21}_{-21}$ & 14 & -380$^{+18}_{-18}$ & -136$^{+10}_{-10}$ & -108$^{+7}_{-7}$ & -15.2$^{+6.2}_{-6.2}$ & 5.2$^{+2.4}_{-2.4}$ & -0.9$^{+1.7}_{-1.7}$ \\
\textbf{SO-4} & 5 & 12.3$^{+2.8}_{-1.9}$ & 0.50$^{+0.28}_{-0.30}$ & M & 204.01$^{+0.07}_{-0.07}$ & -20.13$^{+0.12}_{-0.12}$ & 330$^{+5}_{-5}$ & 1 & -283$^{+4}_{-4}$ & -126$^{+2}_{-2}$ & -113$^{+3}_{-3}$ & -- & -- & -- \\
\textbf{SO-5} & 42 & 15.1$^{+1.5}_{-3.9}$ & 0.10$^{+0.17}_{-0.10}$ & U & 207.50$^{+0.49}_{-0.49}$ & -21.55$^{+0.40}_{-0.40}$ & 333$^{+13}_{-13}$ & 6 & -274$^{+10}_{-10}$ & -143$^{+6}_{-6}$ & -122$^{+5}_{-5}$ & -3.5$^{+1.3}_{-1.3}$ & 3.1$^{+0.7}_{-0.7}$ & 1.8$^{+0.9}_{-0.9}$ \\
\textbf{Orion A-West} & 159 & 15.8$^{+1.5}_{-1.5}$ & 0.61$^{+0.08}_{-0.13}$ & B & 208.86$^{+1.02}_{-1.02}$ & -18.80$^{+0.47}_{-0.47}$ & 425$^{+24}_{-24}$ & 15 & -351$^{+18}_{-18}$ & -195$^{+16}_{-16}$ & -137$^{+8}_{-8}$ & -14.0$^{+4.7}_{-4.7}$ & 2.2$^{+3.5}_{-3.5}$ & -3.8$^{+1.7}_{-1.7}$ \\
OBP-far & 608 & 15.9$^{+1.5}_{-1.5}$ & 0.08$^{+0.10}_{-0.08}$ & B & 205.64$^{+1.12}_{-1.12}$ & -17.53$^{+0.71}_{-0.71}$ & 419$^{+22}_{-22}$ & 78 & -360$^{+18}_{-18}$ & -173$^{+13}_{-13}$ & -126$^{+10}_{-10}$ & -14.0$^{+4.3}_{-4.3}$ & 4.2$^{+2.1}_{-2.1}$ & -2.2$^{+2.2}_{-2.2}$ \\
Orion A-East & 118 & 16.7$^{+1.6}_{-1.6}$ & 0.06$^{+0.10}_{-0.06}$ & B & 210.67$^{+0.79}_{-0.79}$ & -16.12$^{+0.59}_{-0.59}$ & 444$^{+26}_{-26}$ & 13 & -367$^{+21}_{-21}$ & -217$^{+15}_{-15}$ & -121$^{+9}_{-9}$ & -12.9$^{+3.0}_{-3.0}$ & 1.7$^{+3.1}_{-3.1}$ & -5.2$^{+1.1}_{-1.1}$ \\
Orion Y & 166 & 17.3$^{+1.7}_{-1.7}$ & 0.00$^{+0.02}_{-0.00}$ & B & 212.91$^{+0.64}_{-0.64}$ & -16.64$^{+0.81}_{-0.81}$ & 300$^{+9}_{-9}$ & 30 & -241$^{+8}_{-8}$ & -157$^{+5}_{-5}$ & -86$^{+5}_{-5}$ & -1.5$^{+3.3}_{-3.3}$ & 3.1$^{+2.1}_{-2.1}$ & 2.4$^{+0.9}_{-0.9}$ \\
OBP-b & 345 & 20.8$^{+2.0}_{-4.0}$ & 0.00$^{+0.07}_{-0.00}$ & B & 204.84$^{+0.67}_{-0.67}$ & -17.03$^{+0.55}_{-0.55}$ & 389$^{+23}_{-23}$ & 34 & -337$^{+20}_{-20}$ & -156$^{+10}_{-10}$ & -113$^{+8}_{-8}$ & -14.3$^{+4.6}_{-4.6}$ & 0.3$^{+2.8}_{-2.8}$ & -3.5$^{+1.6}_{-1.6}$ \\
ASCC 20 & 512 & 21.8$^{+2.1}_{-2.1}$ & 0.08$^{+0.04}_{-0.03}$ & B & 201.52$^{+0.66}_{-0.66}$ & -17.20$^{+0.98}_{-0.98}$ & 378$^{+34}_{-34}$ & 73 & -335$^{+31}_{-31}$ & -132$^{+12}_{-12}$ & -112$^{+7}_{-7}$ & -15.8$^{+4.1}_{-4.1}$ & 3.6$^{+1.7}_{-1.7}$ & -2.1$^{+1.6}_{-1.6}$ \\
\textbf{SO-6} & 108 & 25.0$^{+2.4}_{-3.9}$ & 0.00$^{+0.09}_{-0.00}$ & U & 205.02$^{+0.87}_{-0.87}$ & -14.27$^{+0.86}_{-0.86}$ & 379$^{+21}_{-21}$ & 15 & -332$^{+19}_{-19}$ & -155$^{+9}_{-9}$ & -92$^{+10}_{-10}$ & -17.6$^{+3.9}_{-3.9}$ & 3.8$^{+0.7}_{-0.7}$ & -3.4$^{+1.2}_{-1.2}$ \\

\hline
\end{tabular}
} % end resize box
\renewcommand{\arraystretch}{1}
\label{tab:overview-orion}
\tablefoot{Cols.~3--4 list the mode and $1\sigma$ high-density interval (HDI) of each sampled posterior. Cols.~6--8 and 10--15 list the median parameters for each cluster from the members $N$, and the uncertainties represent the 1$\sigma$ scatter around the median.
\tablefoottext{a}{Number of cluster members with $p\geq 0.5$ and not classified as outliers.} 
\tablefoottext{b}{Persistence tiers: B = Bedrock (10), M = Majority (5--9), U = Uncertain (2--4).}
\tablefoottext{c}{Formerly known as Rigel cluster \citep{Chen_2020, Sanchez_2024}, but newer data show that Rigel is a member of another cluster instead.}
}
\end{small}
\end{center}
\end{table*}

%% file: tables/machine-readable-table.tex
\begin{table*}[!ht] 
\begin{center}
\begin{small}
\caption{Catalog of the 19,862 stars labeled for cluster membership as identified with \texttt{SigMA} in the Orion volume.}
\renewcommand{\arraystretch}{1.2}
\resizebox{0.95\textwidth}{!}{%
\begin{tabular}{lll}
\hline \hline
\multicolumn{1}{l}{Column name} &
\multicolumn{1}{l}{Unit} &
\multicolumn{1}{l}{Column description} \\
\hline
\multicolumn{2}{c}{\emph{Cluster memberships}} & \\
\hline
\texttt{source\_id} &  & \emph{Gaia} DR3 source ID\\
\texttt{SigMA} &  & \texttt{SigMA} membership label for each source, as defined in Table~\ref{tab:overview-orion}   \\
\texttt{Name} & & Cluster name  \\
\texttt{Association} & & Designated association (Orion, Snake, BBJ, none)\\
\texttt{membership} & &Membership stability normalized to unity \\
\texttt{rep\_stability} & & Cluster persistence across the 10 pipeline runs (0--10) \\
\texttt{tier} & & Cluster persistence grouped into tiers: Bedrock (10), Majority (5--9), Uncertain (2--4) \\
\texttt{robust\_outlier} & [0/1] & Source was flagged as outlier with a 5D Mahalanobis distance $> 8\sigma$ \\
\hline
\multicolumn{2}{c}{\emph{Astrometry}}  & \\
\hline
\texttt{ra} & deg  & \emph{Gaia} DR3 Right ascension  \\
\texttt{dec} & deg  & \emph{Gaia} DR3 Declination \\
\texttt{l} & deg  & Galactic longitude  \\
\texttt{b} & deg  & Galactic latitude \\
\texttt{parallax} & mas  & \emph{Gaia} DR3 parallax  \\
\texttt{$\mu_{\alpha}^*$} & mas\,yr$^{-1}$  & \emph{Gaia} DR3 proper motion in Right ascension \\
\texttt{$\mu_{\delta}$} & mas\,yr$^{-1}$  & \emph{Gaia} DR3 proper motion in Declination \\
\texttt{distance} & pc  & Distance derived from inverse of the parallax   \\
\texttt{radial\_velocity} & km\,s &  \emph{Gaia} DR3 radial velocity \\
\texttt{X} & pc  & Heliocentric Galactic Cartesian $X$ coordinate, increasing toward the Galactic center   \\
\texttt{Y} & pc  & Heliocentric Galactic Cartesian $Y$ coordinate, increasing in direction of Galactic rotation   \\
\texttt{Z} & pc  & Heliocentric Galactic Cartesian $Z$ coordinate, increasing toward the Galactic North-pole   \\
\texttt{v\_alpha\_lsr} & km\,s$^{-1}$  & Tangential velocity in the direction of $\alpha$, relative to the LSR   \\
\texttt{v\_delta\_lsr} & km\,s$^{-1}$  & Tangential velocity in the direction of $\delta$, relative to the LSR   \\
\hline
\multicolumn{2}{c}{\emph{Photometry}} & \\
\hline
\texttt{Gmag} &mag&\emph{Gaia} DR3 G-band magnitude\\
\texttt{BPmag} &mag&\emph{Gaia} DR3 BP-band magnitude\\
\texttt{RPmag} &mag&\emph{Gaia} DR3 RP-band magnitude\\
\texttt{absG} & mag  & absolute G-band magnitude\\
\texttt{G\_${\text{err}}$} &mag& photometric G-band error (Eq.~\ref{eq:photometric-cuts})\\
\texttt{G\_${\text{BPerr}}$} &mag&photometric BP-band error (Eq.~\ref{eq:photometric-cuts})\\
\texttt{G\_${\text{RPerr}}$} &mag& photometric RP-band error (Eq.~\ref{eq:photometric-cuts})\\
\hline
\multicolumn{2}{c}{\emph{Fit parameters and flags}} & \\
\hline
\texttt{age} & Myr &isochronal age, defined as the mode of the posterior distribution\\
\texttt{age\_lo} & Myr &  lower limit of the 1$\sigma$ highest density interval (HDI) of the marginalized posterior age PDF\tablefootmark{a}\\
\texttt{age\_hi} &Myr&  upper limit of the 1$\sigma$ HDI of the marginalized posterior age PDF\tablefootmark{a}\\
\texttt{av} &mag& extinction $A_V$, defined as the mode of the posterior distribution\\
\texttt{av\_lo} &mag&  lower limit of the 1$\sigma$ HDI of the posterior $A_V$ distribution \\
\texttt{av\_hi}&mag&  upper limit of the 1$\sigma$ HDI of the posterior $A_V$ distribution \\
\texttt{soft\_prob\_member} & [0/1] & Source has membership probability $p>0$ for more than one cluster \\
\texttt{ambiguous\_max} & [0/1] & Source has the same membership maximum for more than one cluster \\
\texttt{used\_for\_age} & [0/1] & Source was used for age estimation\\
\texttt{fidelity\_v2} & & Fidelity value assigned by \cite{Rybizki_2022} \\
\hline
\end{tabular}
} % end resize box
\renewcommand{\arraystretch}{1}
\label{tab:member-catalog}
\tablefoot{The full machine-readable version of the catalog is given online, while a column overview is given here. We list all relevant derived parameters. The full \emph{Gaia} DR3 parameters can be queried from the Gaia Archive by using the Gaia \texttt{source\_id}.
 \tablefoottext{a}{Where the HDI for the age fit is narrower than the resolution of the isochrone
  grid, the grid resolution ($0.04$~dex in $\log t$) is
  adopted as the minimum uncertainty.}
}
\end{small}
\end{center}
\end{table*}

%% file: ref.bib
@incollection{Bally_2008,
       author = {{Bally}, J.},
        title = "{Overview of the Orion Complex}",
    booktitle = {Handbook of Star Forming Regions, Volume I},
         year = 2008,
       editor = {{Reipurth}, B.},
    publisher = {Astronomical Society of the Pacific},
       volume = {4},
        pages = {459},
          doi = {10.48550/arXiv.0812.0046},
       adsurl = {https://ui.adsabs.harvard.edu/abs/2008hsf1.book..459B}
}

@ARTICLE{Brown_1994,
       author = {{Brown}, A.~G.~A. and {de Geus}, E.~J. and {de Zeeuw}, P.~T.},
        title = "{The Orion OB1 association. I. Stellar content.}",
      journal = {\aap},
         year = 1994,
        month = sep,
       volume = {289},
        pages = {101-120},
          doi = {10.48550/arXiv.astro-ph/9403051},
archivePrefix = {arXiv},
       eprint = {astro-ph/9403051},
 primaryClass = {astro-ph},
       adsurl = {https://ui.adsabs.harvard.edu/abs/1994A&A...289..101B}
}

@ARTICLE{Blaauw_1964,
       author = {{Blaauw}, Adriaan},
        title = "{The O Associations in the Solar Neighborhood}",
      journal = {\araa},
         year = 1964,
        month = jan,
       volume = {2},
        pages = {213},
          doi = {10.1146/annurev.aa.02.090164.001241},
       adsurl = {https://ui.adsabs.harvard.edu/abs/1964ARA&A...2..213B}
}

@ARTICLE{Elmegreen_1977,
       author = {{Elmegreen}, B.~G. and {Lada}, C.~J.},
        title = "{Sequential formation of subgroups in OB associations.}",
      journal = {\apj},
         year = 1977,
        month = jun,
       volume = {214},
        pages = {725-741},
          doi = {10.1086/155302},
       adsurl = {https://ui.adsabs.harvard.edu/abs/1977ApJ...214..725E}
}

@ARTICLE{Zari_2017,
       author = {{Zari}, E. and {Brown}, A.~G.~A. and {de Bruijne}, J. and {Manara}, C.~F. and {de Zeeuw}, P.~T.},
        title = "{Mapping young stellar populations toward Orion with Gaia DR1}",
      journal = {\aap},
         year = 2017,
        month = dec,
       volume = {608},
          eid = {A148},
        pages = {A148},
          doi = {10.1051/0004-6361/201731309},
archivePrefix = {arXiv},
       eprint = {1711.03815},
 primaryClass = {astro-ph.SR},
       adsurl = {https://ui.adsabs.harvard.edu/abs/2017A&A...608A.148Z}
}

@ARTICLE{Zari_2019,
       author = {{Zari}, E. and {Brown}, A.~G.~A. and {de Zeeuw}, P.~T.},
        title = "{Structure, kinematics, and ages of the young stellar populations in the Orion region}",
      journal = {\aap},
         year = 2019,
        month = aug,
       volume = {628},
          eid = {A123},
        pages = {A123},
          doi = {10.1051/0004-6361/201935781},
archivePrefix = {arXiv},
       eprint = {1906.07002},
 primaryClass = {astro-ph.SR},
       adsurl = {https://ui.adsabs.harvard.edu/abs/2019A&A...628A.123Z}
}

@ARTICLE{Jeffries_2006,
       author = {{Jeffries}, R.~D. and {Maxted}, P.~F.~L. and {Oliveira}, J.~M. and {Naylor}, Tim},
        title = "{Kinematic structure in the young {\ensuremath{\sigma}} Orionis association}",
      journal = {\mnras},
         year = 2006,
        month = sep,
       volume = {371},
       number = {1},
        pages = {L6-L10},
          doi = {10.1111/j.1745-3933.2006.00196.x},
archivePrefix = {arXiv},
       eprint = {astro-ph/0605616},
 primaryClass = {astro-ph},
       adsurl = {https://ui.adsabs.harvard.edu/abs/2006MNRAS.371L...6J}
}

@ARTICLE{Sacco_2008,
       author = {{Sacco}, G.~G. and {Franciosini}, E. and {Randich}, S. and {Pallavicini}, R.},
        title = "{FLAMES spectroscopy of low-mass stars in the young clusters {\ensuremath{\sigma}} Ori and {\ensuremath{\lambda}} Ori}",
      journal = {\aap},
         year = 2008,
        month = sep,
       volume = {488},
       number = {1},
        pages = {167-179},
          doi = {10.1051/0004-6361:20079049},
archivePrefix = {arXiv},
       eprint = {0805.2914},
 primaryClass = {astro-ph},
       adsurl = {https://ui.adsabs.harvard.edu/abs/2008A&A...488..167S}
}

@ARTICLE{Furesz_2008,
       author = {{F{\H{u}}r{\'e}sz}, G{\'a}bor and {Hartmann}, Lee W. and {Megeath}, S. Thomas and {Szentgyorgyi}, Andrew H. and {Hamden}, Erika T.},
        title = "{Kinematic Structure of the Orion Nebula Cluster and Its Surroundings}",
      journal = {\apj},
         year = 2008,
        month = apr,
       volume = {676},
       number = {2},
        pages = {1109-1122},
          doi = {10.1086/525844},
archivePrefix = {arXiv},
       eprint = {0711.0391},
 primaryClass = {astro-ph},
       adsurl = {https://ui.adsabs.harvard.edu/abs/2008ApJ...676.1109F}
}

@ARTICLE{Tobin_2009,
       author = {{Tobin}, John J. and {Hartmann}, Lee and {Furesz}, Gabor and {Mateo}, Mario and {Megeath}, S. Tom},
        title = "{Kinematics of the Orion Nebula Cluster: Velocity Substructure and Spectroscopic Binaries}",
      journal = {\apj},
         year = 2009,
        month = jun,
       volume = {697},
       number = {2},
        pages = {1103-1118},
          doi = {10.1088/0004-637X/697/2/1103},
archivePrefix = {arXiv},
       eprint = {0903.2775},
 primaryClass = {astro-ph.SR},
       adsurl = {https://ui.adsabs.harvard.edu/abs/2009ApJ...697.1103T}
}

@ARTICLE{Hernandez_2014,
       author = {{Hern{\'a}ndez}, Jes{\'u}s and {Calvet}, Nuria and {Perez}, Alice and {Brice{\~n}o}, Cesar and {Olguin}, Lorenzo and {Contreras}, Maria E. and {Hartmann}, Lee and {Allen}, Lori and {Espaillat}, Catherine and {Hernan}, Ram{\'\i}rez},
        title = "{A Spectroscopic Census in Young Stellar Regions: The {\ensuremath{\sigma}} Orionis Cluster}",
      journal = {\apj},
         year = 2014,
        month = oct,
       volume = {794},
       number = {1},
          eid = {36},
        pages = {36},
          doi = {10.1088/0004-637X/794/1/36},
archivePrefix = {arXiv},
       eprint = {1408.0225},
 primaryClass = {astro-ph.SR},
       adsurl = {https://ui.adsabs.harvard.edu/abs/2014ApJ...794...36H}
}

@ARTICLE{daRio_2016,
       author = {{Da Rio}, Nicola and {Tan}, Jonathan C. and {Covey}, Kevin R. and {Cottaar}, Michiel and {Foster}, Jonathan B. and {Cullen}, Nicholas C. and {Tobin}, John J. and {Kim}, Jinyoung S. and {Meyer}, Michael R. and {Nidever}, David L. and {Stassun}, Keivan G. and {Chojnowski}, S. Drew and {Flaherty}, Kevin M. and {Majewski}, Steve and {Skrutskie}, Michael F. and {Zasowski}, Gail and {Pan}, Kaike},
        title = "{IN-SYNC. IV. The Young Stellar Population in the Orion A Molecular Cloud}",
      journal = {\apj},
         year = 2016,
        month = feb,
       volume = {818},
       number = {1},
          eid = {59},
        pages = {59},
          doi = {10.3847/0004-637X/818/1/59},
archivePrefix = {arXiv},
       eprint = {1511.04147},
 primaryClass = {astro-ph.GA},
       adsurl = {https://ui.adsabs.harvard.edu/abs/2016ApJ...818...59D}
}

@ARTICLE{daRio_2017,
       author = {{Da Rio}, Nicola and {Tan}, Jonathan C. and {Covey}, Kevin R. and {Cottaar}, Michiel and {Foster}, Jonathan B. and {Cullen}, Nicholas C. and {Tobin}, John and {Kim}, Jinyoung S. and {Meyer}, Michael R. and {Nidever}, David L. and {Stassun}, Keivan G. and {Chojnowski}, S. Drew and {Flaherty}, Kevin M. and {Majewski}, Steven R. and {Skrutskie}, Michael F. and {Zasowski}, Gail and {Pan}, Kaike},
        title = "{IN-SYNC. V. Stellar Kinematics and Dynamics in the Orion A Molecular Cloud}",
      journal = {\apj},
         year = 2017,
        month = aug,
       volume = {845},
       number = {2},
          eid = {105},
        pages = {105},
          doi = {10.3847/1538-4357/aa7a5b},
archivePrefix = {arXiv},
       eprint = {1702.04113},
 primaryClass = {astro-ph.GA},
       adsurl = {https://ui.adsabs.harvard.edu/abs/2017ApJ...845..105D}
}

@ARTICLE{Kubiak_2017,
       author = {{Kubiak}, K. and {Alves}, J. and {Bouy}, H. and {Sarro}, L.~M. and {Ascenso}, J. and {Burkert}, A. and {Forbrich}, J. and {Gro{\ss}schedl}, J. and {Hacar}, A. and {Hasenberger}, B. and {Lombardi}, M. and {Meingast}, S. and {K{\"o}hler}, R. and {Teixeira}, P.~S.},
        title = "{Orion revisited. III. The Orion Belt population}",
      journal = {\aap},
         year = 2017,
        month = feb,
       volume = {598},
          eid = {A124},
        pages = {A124},
          doi = {10.1051/0004-6361/201628920},
archivePrefix = {arXiv},
       eprint = {1609.04948},
 primaryClass = {astro-ph.SR},
       adsurl = {https://ui.adsabs.harvard.edu/abs/2017A&A...598A.124K}
}

@ARTICLE{Alves_2012,
       author = {{Alves}, J. and {Bouy}, H.},
        title = "{Orion revisited. I. The massive cluster in front of the Orion nebula cluster}",
      journal = {\aap},
         year = 2012,
        month = nov,
       volume = {547},
          eid = {A97},
        pages = {A97},
          doi = {10.1051/0004-6361/201220119},
archivePrefix = {arXiv},
       eprint = {1209.3787},
 primaryClass = {astro-ph.GA},
       adsurl = {https://ui.adsabs.harvard.edu/abs/2012A&A...547A..97A}
}

@ARTICLE{Bouy_2014,
       author = {{Bouy}, H. and {Alves}, J. and {Bertin}, E. and {Sarro}, L.~M. and {Barrado}, D.},
        title = "{Orion revisited. II. The foreground population to Orion A}",
      journal = {\aap},
         year = 2014,
        month = apr,
       volume = {564},
          eid = {A29},
        pages = {A29},
          doi = {10.1051/0004-6361/201323191},
archivePrefix = {arXiv},
       eprint = {1402.1034},
 primaryClass = {astro-ph.SR},
       adsurl = {https://ui.adsabs.harvard.edu/abs/2014A&A...564A..29B}
}

@ARTICLE{Pillitteri_2013,
       author = {{Pillitteri}, I. and {Wolk}, S.~J. and {Megeath}, S.~T. and {Allen}, L. and {Bally}, J. and {Gagn{\'e}}, M. and {Gutermuth}, R.~A. and {Hartmann}, L. and {Micela}, G. and {Myers}, P. and {Oliveira}, J.~M. and {Sciortino}, S. and {Walter}, F. and {Rebull}, L. and {Stauffer}, J.},
        title = "{An X-Ray Survey of the Young Stellar Population of the Lynds 1641 and Iota Orionis Regions}",
      journal = {\apj},
         year = 2013,
        month = may,
       volume = {768},
       number = {2},
          eid = {99},
        pages = {99},
          doi = {10.1088/0004-637X/768/2/99},
archivePrefix = {arXiv},
       eprint = {1303.3996},
 primaryClass = {astro-ph.SR},
       adsurl = {https://ui.adsabs.harvard.edu/abs/2013ApJ...768...99P}
}

@ARTICLE{Fang_2017,
       author = {{Fang}, Min and {Kim}, Jinyoung Serena and {Pascucci}, Ilaria and {Apai}, D{\'a}niel and {Zhang}, Lan and {Sicilia-Aguilar}, Aurora and {Alonso-Mart{\'\i}nez}, Miguel and {Eiroa}, Carlos and {Wang}, Hongchi},
        title = "{NGC 1980 Is Not a Foreground Population of Orion: Spectroscopic Survey of Young Stars with Low Extinction in Orion A}",
      journal = {\aj},
         year = 2017,
        month = apr,
       volume = {153},
       number = {4},
          eid = {188},
        pages = {188},
          doi = {10.3847/1538-3881/aa647b},
archivePrefix = {arXiv},
       eprint = {1703.06948},
 primaryClass = {astro-ph.SR},
       adsurl = {https://ui.adsabs.harvard.edu/abs/2017AJ....153..188F}
}

@ARTICLE{Kounkel_2017a,
       author = {{Kounkel}, Marina and {Hartmann}, Lee and {Calvet}, Nuria and {Megeath}, Tom},
        title = "{Characterizing the Stellar Population of NGC 1980}",
      journal = {\aj},
         year = 2017,
        month = jul,
       volume = {154},
       number = {1},
          eid = {29},
        pages = {29},
          doi = {10.3847/1538-3881/aa74df},
archivePrefix = {arXiv},
       eprint = {1705.07922},
 primaryClass = {astro-ph.SR},
       adsurl = {https://ui.adsabs.harvard.edu/abs/2017AJ....154...29K}
}

@ARTICLE{Beccari_2017,
       author = {{Beccari}, G. and {Petr-Gotzens}, M.~G. and {Boffin}, H.~M.~J. and {Romaniello}, M. and {Fedele}, D. and {Carraro}, G. and {De Marchi}, G. and {de Wit}, W.-J. and {Drew}, J.~E. and {Kalari}, V.~M. and {Manara}, C.~F. and {Martin}, E.~L. and {Mieske}, S. and {Panagia}, N. and {Testi}, L. and {Vink}, J.~S. and {Walsh}, J.~R. and {Wright}, N.~J.},
        title = "{A tale of three cities. OmegaCAM discovers multiple sequences in the color-magnitude diagram of the Orion Nebula Cluster}",
      journal = {\aap},
         year = 2017,
        month = jul,
       volume = {604},
          eid = {A22},
        pages = {A22},
          doi = {10.1051/0004-6361/201730432},
archivePrefix = {arXiv},
       eprint = {1705.09496},
 primaryClass = {astro-ph.SR},
       adsurl = {https://ui.adsabs.harvard.edu/abs/2017A&A...604A..22B}
}

@ARTICLE{Jerabkova_2019,
       author = {{Jerabkova}, Tereza and {Beccari}, Giacomo and {Boffin}, Henri M.~J. and {Petr-Gotzens}, Monika G. and {Manara}, Carlo F. and {Prada Moroni}, Pier Giorgio and {Tognelli}, Emanuele and {Degl'Innocenti}, Scilla},
        title = "{When the tale comes true: multiple populations and wide binaries in the Orion Nebula Cluster}",
      journal = {\aap},
         year = 2019,
        month = jul,
       volume = {627},
          eid = {A57},
        pages = {A57},
          doi = {10.1051/0004-6361/201935016},
archivePrefix = {arXiv},
       eprint = {1905.06974},
 primaryClass = {astro-ph.SR},
       adsurl = {https://ui.adsabs.harvard.edu/abs/2019A&A...627A..57J}
}

@ARTICLE{Alzate_2023,
       author = {{Alzate}, Jairo A. and {Bruzual}, Gustavo and {Kounkel}, Marina and {Magris}, Gladis and {Hartmann}, Lee and {Calvet}, Nuria and {Cao}, Lyra},
        title = "{Constraints on star formation in Orion A from Gaia}",
      journal = {\mnras},
         year = 2023,
        month = aug,
       volume = {523},
       number = {4},
        pages = {4821-4840},
          doi = {10.1093/mnras/stad1482},
archivePrefix = {arXiv},
       eprint = {2305.11823},
 primaryClass = {astro-ph.SR},
       adsurl = {https://ui.adsabs.harvard.edu/abs/2023MNRAS.523.4821A}
}

@ARTICLE{Zacharias_2013,
       author = {{Zacharias}, N. and {Finch}, C.~T. and {Girard}, T.~M. and {Henden}, A. and {Bartlett}, J.~L. and {Monet}, D.~G. and {Zacharias}, M.~I.},
        title = "{The Fourth US Naval Observatory CCD Astrograph Catalog (UCAC4)}",
      journal = {\aj},
         year = 2013,
        month = feb,
       volume = {145},
       number = {2},
          eid = {44},
        pages = {44},
          doi = {10.1088/0004-6256/145/2/44},
archivePrefix = {arXiv},
       eprint = {1212.6182},
 primaryClass = {astro-ph.IM},
       adsurl = {https://ui.adsabs.harvard.edu/abs/2013AJ....145...44Z}
}

@ARTICLE{Dzib_2017,
       author = {{Dzib}, Sergio A. and {Loinard}, Laurent and {Rodr{\'\i}guez}, Luis F. and {G{\'o}mez}, Laura and {Forbrich}, Jan and {Menten}, Karl M. and {Kounkel}, Marina A. and {Mioduszewski}, Amy J. and {Hartmann}, Lee and {Tobin}, John J. and {Rivera}, Juana L.},
        title = "{Radio Measurements of the Stellar Proper Motions in the Core of the Orion Nebula Cluster}",
      journal = {\apj},
         year = 2017,
        month = jan,
       volume = {834},
       number = {2},
          eid = {139},
        pages = {139},
          doi = {10.3847/1538-4357/834/2/139},
archivePrefix = {arXiv},
       eprint = {1611.03767},
 primaryClass = {astro-ph.SR},
       adsurl = {https://ui.adsabs.harvard.edu/abs/2017ApJ...834..139D}
}

@ARTICLE{Megeath_2012,
       author = {{Megeath}, S.~T. and {Gutermuth}, R. and {Muzerolle}, J. and {Kryukova}, E. and {Flaherty}, K. and {Hora}, J.~L. and {Allen}, L.~E. and {Hartmann}, L. and {Myers}, P.~C. and {Pipher}, J.~L. and {Stauffer}, J. and {Young}, E.~T. and {Fazio}, G.~G.},
        title = "{The Spitzer Space Telescope Survey of the Orion A and B Molecular Clouds. I. A Census of Dusty Young Stellar Objects and a Study of Their Mid-infrared Variability}",
      journal = {\aj},
         year = 2012,
        month = dec,
       volume = {144},
       number = {6},
          eid = {192},
        pages = {192},
          doi = {10.1088/0004-6256/144/6/192},
archivePrefix = {arXiv},
       eprint = {1209.3826},
 primaryClass = {astro-ph.GA},
       adsurl = {https://ui.adsabs.harvard.edu/abs/2012AJ....144..192M}
}

@ARTICLE{Hacar_2016,
       author = {{Hacar}, A. and {Alves}, J. and {Forbrich}, J. and {Meingast}, S. and {Kubiak}, K. and {Gro{\ss}schedl}, J.},
        title = "{APOGEE strings: A fossil record of the gas kinematic structure}",
      journal = {\aap},
         year = 2016,
        month = may,
       volume = {589},
          eid = {A80},
        pages = {A80},
          doi = {10.1051/0004-6361/201527805},
archivePrefix = {arXiv},
       eprint = {1602.01854},
 primaryClass = {astro-ph.GA},
       adsurl = {https://ui.adsabs.harvard.edu/abs/2016A&A...589A..80H}
}

@ARTICLE{Hacar_2018,
       author = {{Hacar}, A. and {Tafalla}, M. and {Forbrich}, J. and {Alves}, J. and {Meingast}, S. and {Grossschedl}, J. and {Teixeira}, P.~S.},
        title = "{An ALMA study of the Orion Integral Filament. I. Evidence for narrow fibers in a massive cloud}",
      journal = {\aap},
         year = 2018,
        month = mar,
       volume = {610},
          eid = {A77},
        pages = {A77},
          doi = {10.1051/0004-6361/201731894},
archivePrefix = {arXiv},
       eprint = {1801.01500},
 primaryClass = {astro-ph.GA},
       adsurl = {https://ui.adsabs.harvard.edu/abs/2018A&A...610A..77H}
}

@ARTICLE{Hacar_2024,
       author = {{Hacar}, A. and {Socci}, A. and {Bonanomi}, F. and {Petry}, D. and {Tafalla}, M. and {Harsono}, D. and {Forbrich}, J. and {Alves}, J. and {Grossschedl}, J. and {Goicoechea}, J.~R. and {Pety}, J. and {Burkert}, A. and {Li}, G.~X.},
        title = "{Emergence of high-mass stars in complex fiber networks (EMERGE). I. Early ALMA Survey: Observations and massive data reduction}",
      journal = {\aap},
         year = 2024,
        month = jul,
       volume = {687},
          eid = {A140},
        pages = {A140},
          doi = {10.1051/0004-6361/202348565},
archivePrefix = {arXiv},
       eprint = {2403.08091},
 primaryClass = {astro-ph.GA},
       adsurl = {https://ui.adsabs.harvard.edu/abs/2024A&A...687A.140H}
}

@ARTICLE{Socci_2024,
       author = {{Socci}, A. and {Hacar}, A. and {Bonanomi}, F. and {Tafalla}, M. and {Suri}, S.},
        title = "{Emergence of high-mass stars in complex fiber networks (EMERGE): III. Fiber networks in Orion}",
      journal = {\aap},
         year = 2024,
        month = oct,
       volume = {690},
          eid = {A375},
        pages = {A375},
          doi = {10.1051/0004-6361/202449316},
archivePrefix = {arXiv},
       eprint = {2409.01321},
 primaryClass = {astro-ph.GA},
       adsurl = {https://ui.adsabs.harvard.edu/abs/2024A&A...690A.375S}
}

@ARTICLE{Kounkel_2018,
       author = {{Kounkel}, Marina and {Covey}, Kevin and {Su{\'a}rez}, Genaro and {Rom{\'a}n-Z{\'u}{\~n}iga}, Carlos and {Hernandez}, Jesus and {Stassun}, Keivan and {Jaehnig}, Karl O. and {Feigelson}, Eric D. and {Pe{\~n}a Ram{\'\i}rez}, Karla and {Roman-Lopes}, Alexandre and {Da Rio}, Nicola and {Stringfellow}, Guy S. and {Kim}, J. Serena and {Borissova}, Jura and {Fern{\'a}ndez-Trincado}, Jos{\'e} G. and {Burgasser}, Adam and {Garc{\'\i}a-Hern{\'a}ndez}, D.~A. and {Zamora}, Olga and {Pan}, Kaike and {Nitschelm}, Christian},
        title = "{The APOGEE-2 Survey of the Orion Star-forming Complex. II. Six-dimensional Structure}",
      journal = {\aj},
         year = 2018,
        month = sep,
       volume = {156},
       number = {3},
          eid = {84},
        pages = {84},
          doi = {10.3847/1538-3881/aad1f1},
archivePrefix = {arXiv},
       eprint = {1805.04649},
 primaryClass = {astro-ph.SR},
       adsurl = {https://ui.adsabs.harvard.edu/abs/2018AJ....156...84K}
}

@ARTICLE{Chen_2020,
       author = {{Chen}, Boquan and {D'Onghia}, Elena and {Alves}, Jo{\~a}o and {Adamo}, Angela},
        title = "{Discovery of new stellar groups in the Orion complex. Towards a robust unsupervised approach}",
      journal = {\aap},
         year = 2020,
        month = nov,
       volume = {643},
          eid = {A114},
        pages = {A114},
          doi = {10.1051/0004-6361/201935955},
archivePrefix = {arXiv},
       eprint = {1905.11429},
 primaryClass = {astro-ph.GA},
       adsurl = {https://ui.adsabs.harvard.edu/abs/2020A&A...643A.114C}
}

@ARTICLE{Kerr_2023,
       author = {{Kerr}, Ronan and {Kraus}, Adam L. and {Rizzuto}, Aaron C.},
        title = "{SPYGLASS. IV. New Stellar Survey of Recent Star Formation within 1 kpc}",
      journal = {\apj},
         year = 2023,
        month = sep,
       volume = {954},
       number = {2},
          eid = {134},
        pages = {134},
          doi = {10.3847/1538-4357/ace5b3},
archivePrefix = {arXiv},
       eprint = {2306.08150},
 primaryClass = {astro-ph.GA},
       adsurl = {https://ui.adsabs.harvard.edu/abs/2023ApJ...954..134K}
}

@ARTICLE{Sanchez_2024,
       author = {{S{\'a}nchez-Sanju{\'a}n}, Sergio and {Hern{\'a}ndez}, Jes{\'u}s and {P{\'e}rez-Villegas}, {\'A}ngeles and {Rom{\'a}n-Z{\'u}{\~n}iga}, Carlos and {Aguilar}, Luis and {Ballesteros-Paredes}, Javier and {Bonilla-Barroso}, Andrea},
        title = "{Kinematic study of the Orion Complex: analysing the young stellar clusters from big and small structures}",
      journal = {\mnras},
         year = 2024,
        month = nov,
       volume = {534},
       number = {3},
        pages = {2566-2584},
          doi = {10.1093/mnras/stae2157},
archivePrefix = {arXiv},
       eprint = {2409.09206},
 primaryClass = {astro-ph.GA},
       adsurl = {https://ui.adsabs.harvard.edu/abs/2024MNRAS.534.2566S}
}

@ARTICLE{Kounkel_2017b,
       author = {{Kounkel}, Marina and {Hartmann}, Lee and {Mateo}, Mario and {Bailey}, III, John I.},
        title = "{Kinematics of the Optically Visible YSOs toward the Orion B Molecular Cloud}",
      journal = {\apj},
         year = 2017,
        month = aug,
       volume = {844},
       number = {2},
          eid = {138},
        pages = {138},
          doi = {10.3847/1538-4357/aa7dea},
archivePrefix = {arXiv},
       eprint = {1707.01115},
 primaryClass = {astro-ph.SR},
       adsurl = {https://ui.adsabs.harvard.edu/abs/2017ApJ...844..138K}
}

@ARTICLE{Grossschedl_2018,
       author = {{Gro{\ss}schedl}, Josefa E. and {Alves}, Jo{\~a}o and {Meingast}, Stefan and {Ackerl}, Christine and {Ascenso}, Joana and {Bouy}, Herv{\'e} and {Burkert}, Andreas and {Forbrich}, Jan and {F{\"u}rnkranz}, Verena and {Goodman}, Alyssa and {Hacar}, {\'A}lvaro and {Herbst-Kiss}, Gabor and {Lada}, Charles J. and {Larreina}, Irati and {Leschinski}, Kieran and {Lombardi}, Marco and {Moitinho}, Andr{\'e} and {Mortimer}, Daniel and {Zari}, Eleonora},
        title = "{3D shape of Orion A from Gaia DR2}",
      journal = {\aap},
         year = 2018,
        month = nov,
       volume = {619},
          eid = {A106},
        pages = {A106},
          doi = {10.1051/0004-6361/201833901},
archivePrefix = {arXiv},
       eprint = {1808.05952},
 primaryClass = {astro-ph.GA},
       adsurl = {https://ui.adsabs.harvard.edu/abs/2018A&A...619A.106G}
}

@ARTICLE{Kounkel_2020,
       author = {{Kounkel}, Marina},
        title = "{Supernovae in Orion: The Missing Link in the Star-forming History of the Region}",
      journal = {\apj},
         year = 2020,
        month = oct,
       volume = {902},
       number = {2},
          eid = {122},
        pages = {122},
          doi = {10.3847/1538-4357/abb6e8},
archivePrefix = {arXiv},
       eprint = {2007.09160},
 primaryClass = {astro-ph.SR},
       adsurl = {https://ui.adsabs.harvard.edu/abs/2020ApJ...902..122K}
}

@ARTICLE{Swiggum_2021,
       author = {{Swiggum}, Cameren and {D'Onghia}, Elena and {Alves}, Jo{\~a}o and {Gro{\ss}schedl}, Josefa and {Foley}, Michael and {Zucker}, Catherine and {Meingast}, Stefan and {Chen}, Boquan and {Goodman}, Alyssa},
        title = "{Evidence for Radial Expansion at the Core of the Orion Complex with Gaia EDR3}",
      journal = {\apj},
         year = 2021,
        month = aug,
       volume = {917},
       number = {1},
          eid = {21},
        pages = {21},
          doi = {10.3847/1538-4357/ac0633},
archivePrefix = {arXiv},
       eprint = {2101.10380},
 primaryClass = {astro-ph.GA},
       adsurl = {https://ui.adsabs.harvard.edu/abs/2021ApJ...917...21S}
}

@ARTICLE{Grossschedl_2021,
       author = {{Gro{\ss}schedl}, Josefa E. and {Alves}, Jo{\~a}o and {Meingast}, Stefan and {Herbst-Kiss}, Gabor},
        title = "{3D dynamics of the Orion cloud complex. Discovery of coherent radial gas motions at the 100-pc scale}",
      journal = {\aap},
         year = 2021,
        month = mar,
       volume = {647},
          eid = {A91},
        pages = {A91},
          doi = {10.1051/0004-6361/202038913},
archivePrefix = {arXiv},
       eprint = {2007.07254},
 primaryClass = {astro-ph.SR},
       adsurl = {https://ui.adsabs.harvard.edu/abs/2021A&A...647A..91G}
}

@INPROCEEDINGS{Wright_2023,
       author = {{Wright}, N.~J. and {Kounkel}, M. and {Zari}, E. and {Goodwin}, S. and {Jeffries}, R.~D.},
        title = "{OB Associations}",
    booktitle = {Protostars and Planets VII},
         year = 2023,
       editor = {{Inutsuka}, S. and {Aikawa}, Y. and {Muto}, T. and {Tomida}, K. and {Tamura}, M.},
       series = {Astronomical Society of the Pacific Conference Series},
       volume = {534},
        month = jul,
        pages = {129},
       adsurl = {https://ui.adsabs.harvard.edu/abs/2023ASPC..534..129W}
}

@ARTICLE{Ratzenboeck_2023a,
       author = {{Ratzenb{\"o}ck}, Sebastian and {Gro{\ss}schedl}, Josefa E. and {M{\"o}ller}, Torsten and {Alves}, Jo{\~a}o and {Bomze}, Immanuel and {Meingast}, Stefan},
        title = "{Significance mode analysis (SigMA) for hierarchical structures. An application to the Sco-Cen OB association}",
      journal = {\aap},
         year = 2023,
        month = sep,
       volume = {677},
          eid = {A59},
        pages = {A59},
          doi = {10.1051/0004-6361/202243690},
archivePrefix = {arXiv},
       eprint = {2211.14225},
 primaryClass = {astro-ph.GA},
       adsurl = {https://ui.adsabs.harvard.edu/abs/2023A&A...677A..59R}
}

@ARTICLE{Ratzenboeck_2023b,
       author = {{Ratzenb{\"o}ck}, Sebastian and {Gro{\ss}schedl}, Josefa E. and {Alves}, Jo{\~a}o and {Miret-Roig}, N{\'u}ria and {Bomze}, Immanuel and {Forbes}, John and {Goodman}, Alyssa and {Hacar}, {\'A}lvaro and {Lin}, Doug and {Meingast}, Stefan and {M{\"o}ller}, Torsten and {Piecka}, Martin and {Posch}, Laura and {Rottensteiner}, Alena and {Swiggum}, Cameren and {Zucker}, Catherine},
        title = "{The star formation history of the Sco-Cen association. Coherent star formation patterns in space and time}",
      journal = {\aap},
         year = 2023,
        month = oct,
       volume = {678},
          eid = {A71},
        pages = {A71},
          doi = {10.1051/0004-6361/202346901},
archivePrefix = {arXiv},
       eprint = {2302.07853},
 primaryClass = {astro-ph.SR},
       adsurl = {https://ui.adsabs.harvard.edu/abs/2023A&A...678A..71R}
}

@ARTICLE{deBruijne_1999,
       author = {{de Bruijne}, Jos H.~J.},
        title = "{Structure and colour-magnitude diagrams of Scorpius OB2 based on kinematic modelling of Hipparcos data}",
      journal = {\mnras},
         year = 1999,
        month = dec,
       volume = {310},
       number = {3},
        pages = {585-617},
          doi = {10.1046/j.1365-8711.1999.02953.x},
       adsurl = {https://ui.adsabs.harvard.edu/abs/1999MNRAS.310..585D}
}

@ARTICLE{Tian_2020,
       author = {{Tian}, Hai-Jun},
        title = "{Discovery of a Young Stellar Snake with Two Dissolving Cores in the Solar Neighborhood}",
      journal = {\apj},
         year = 2020,
        month = dec,
       volume = {904},
       number = {2},
          eid = {196},
        pages = {196},
          doi = {10.3847/1538-4357/abbf4b},
archivePrefix = {arXiv},
       eprint = {2005.12265},
 primaryClass = {astro-ph.GA},
       adsurl = {https://ui.adsabs.harvard.edu/abs/2020ApJ...904..196T}
}

@ARTICLE{Wang_2022,
       author = {{Wang}, Fan and {Tian}, Haijun and {Qiu}, Dan and {Xu}, Qi and {Fang}, Min and {Tian}, Hao and {Di}, Li and {Bird}, Sarah A. and {Shi}, Jianrong and {Fu}, Xiaoting and {Liu}, Gaochao and {Cui}, Sheng and {Zhang}, Yong},
        title = "{The stellar 'Snake' - I. Whole structure and properties}",
      journal = {\mnras},
         year = 2022,
        month = jun,
       volume = {513},
       number = {1},
        pages = {503-515},
          doi = {10.1093/mnras/stac843},
archivePrefix = {arXiv},
       eprint = {2109.05999},
 primaryClass = {astro-ph.GA},
       adsurl = {https://ui.adsabs.harvard.edu/abs/2022MNRAS.513..503W}
}

@ARTICLE{Beccari_2020,
       author = {{Beccari}, Giacomo and {Boffin}, Henri M.~J. and {Jerabkova}, Tereza},
        title = "{Uncovering a 260 pc wide, 35-Myr-old filamentary relic of star formation}",
      journal = {\mnras},
         year = 2020,
        month = jan,
       volume = {491},
       number = {2},
        pages = {2205-2216},
          doi = {10.1093/mnras/stz3195},
archivePrefix = {arXiv},
       eprint = {1911.05709},
 primaryClass = {astro-ph.SR},
       adsurl = {https://ui.adsabs.harvard.edu/abs/2020MNRAS.491.2205B}
}

@ARTICLE{Rybizki_2022,
       author = {{Rybizki}, Jan and {Green}, Gregory M. and {Rix}, Hans-Walter and {El-Badry}, Kareem and {Demleitner}, Markus and {Zari}, Eleonora and {Udalski}, Andrzej and {Smart}, Richard L. and {Gould}, Andrew},
        title = "{A classifier for spurious astrometric solutions in Gaia eDR3}",
      journal = {\mnras},
         year = 2022,
        month = feb,
       volume = {510},
       number = {2},
        pages = {2597-2616},
          doi = {10.1093/mnras/stab3588},
archivePrefix = {arXiv},
       eprint = {2101.11641},
 primaryClass = {astro-ph.IM},
       adsurl = {https://ui.adsabs.harvard.edu/abs/2022MNRAS.510.2597R}
}

@ARTICLE{Zari_2021,
       author = {{Zari}, E. and {Rix}, H. -W. and {Frankel}, N. and {Xiang}, M. and {Poggio}, E. and {Drimmel}, R. and {Tkachenko}, A.},
        title = "{Mapping luminous hot stars in the Galaxy}",
      journal = {\aap},
         year = 2021,
        month = jun,
       volume = {650},
          eid = {A112},
        pages = {A112},
          doi = {10.1051/0004-6361/202039726},
archivePrefix = {arXiv},
       eprint = {2102.08684},
 primaryClass = {astro-ph.GA},
       adsurl = {https://ui.adsabs.harvard.edu/abs/2021A&A...650A.112Z}
}

@ARTICLE{2010Schoenrich,
       author = {{Sch{\"o}nrich}, Ralph and {Binney}, James and {Dehnen}, Walter},
        title = "{Local kinematics and the local standard of rest}",
      journal = {\mnras},
         year = 2010,
        month = apr,
       volume = {403},
       number = {4},
        pages = {1829-1833},
          doi = {10.1111/j.1365-2966.2010.16253.x},
archivePrefix = {arXiv},
       eprint = {0912.3693},
 primaryClass = {astro-ph.GA},
       adsurl = {https://ui.adsabs.harvard.edu/abs/2010MNRAS.403.1829S}
}

@ARTICLE{Bailer-Jones_2021,
       author = {{Bailer-Jones}, C.~A.~L. and {Rybizki}, J. and {Fouesneau}, M. and {Demleitner}, M. and {Andrae}, R.},
        title = "{Estimating Distances from Parallaxes. V. Geometric and Photogeometric Distances to 1.47 Billion Stars in Gaia Early Data Release 3}",
      journal = {\aj},
         year = 2021,
        month = mar,
       volume = {161},
       number = {3},
          eid = {147},
        pages = {147},
          doi = {10.3847/1538-3881/abd806},
archivePrefix = {arXiv},
       eprint = {2012.05220},
 primaryClass = {astro-ph.SR},
       adsurl = {https://ui.adsabs.harvard.edu/abs/2021AJ....161..147B}
}

@inproceedings{Wishart_1969,
    author={Wishart, D.},
    title={Mode analysis: a generalization of nearest neighbour which reduces chaining effects},
    year={1969},
    booktitle={Proceedings of the Colloquium in Numerical Taxonomy},
    pages={282--308},
    journal = {Numerical Taxonomy, Edited by A.~J.~Cole, Academic Press, New York},
    publisher={Edited by A.J. Cole, Academic Press, New York},
    location = {New York},
}

@ARTICLE{Koontz_1976,
    author={Koontz and Narendra and Fukunaga},
    journal={IEEE Transactions on Computers},
    title={A Graph-Theoretic Approach to Nonparametric Cluster Analysis},
    year={1976},
    volume={C-25},
    number={9},
    pages={936-944},
    doi={10.1109/TC.1976.1674719}
}

@incollection{Kirkpatrick_1985,
    title = {A Framework for Computational Morphology},
    editor = {Godfried T. TOUSSAINT},
    series = {Machine Intelligence and Pattern Recognition},
    publisher = {North-Holland},
    volume = {2},
    pages = {217-248},
    year = {1985},
    booktitle = {Computational Geometry},
    issn = {0923-0459},
    doi = {https://doi.org/10.1016/B978-0-444-87806-9.50013-X},
    url = {https://www.sciencedirect.com/science/article/pii/B978044487806950013X},
    author = {David G. Kirkpatrick and John D. Radke},
}

@article{Benjamini_1995,
    author = {Benjamini, Yoav and Hochberg, Yosef},
    title = {Controlling the False Discovery Rate: A Practical and Powerful Approach to Multiple Testing},
    journal = {Journal of the Royal Statistical Society: Series B (Methodological)},
    volume = {57},
    number = {1},
    pages = {289-300},
    year = {1995},
    month = {01},
    issn = {0035-9246},
    doi = {10.1111/j.2517-6161.1995.tb02031.x},
    url = {https://doi.org/10.1111/j.2517-6161.1995.tb02031.x},
    eprint = {https://academic.oup.com/jrsssb/article-pdf/57/1/289/49173396/jrsssb_57_1_289.pdf},
}

@article{Burman_2008,
    title = {Multivariate mode hunting: Data analytic tools with     measures of significance},
    journal = {Journal of Multivariate Analysis},
    volume = {100},
    number = {6},
    pages = {1198-1218},
    year = {2009},
    issn = {0047-259X},
    doi = {https://doi.org/10.1016/j.jmva.2008.10.015},
    author = {Prabir Burman and Wolfgang Polonik}
}

@article{Liu_2020,
  title={Cauchy combination test: a powerful test with analytic p-value calculation under arbitrary dependency structures},
  author={Liu, Yaowu and Xie, Jun},
  journal={Journal of the American Statistical Association},
  volume={115},
  number={529},
  pages={393--402},
  year={2020},
  publisher={Taylor \& Francis}
}

@ARTICLE{Hunt_2024,
       author = {{Hunt}, Emily L. and {Reffert}, Sabine},
        title = "{Improving the open cluster census. III. Using cluster masses, radii, and dynamics to create a cleaned open cluster catalogue}",
      journal = {\aap},
         year = 2024,
        month = jun,
       volume = {686},
          eid = {A42},
        pages = {A42},
          doi = {10.1051/0004-6361/202348662},
archivePrefix = {arXiv},
       eprint = {2403.05143},
 primaryClass = {astro-ph.GA},
       adsurl = {https://ui.adsabs.harvard.edu/abs/2024A&A...686A..42H}
}

@article{Sobol_1967,
  title={The distribution of points in a cube and the accurate evaluation of integrals.},
  author={Sobol, I. M.},
  journal={Vychisl. Mat. i Mat. Phys.},
  number={7},
  pages={784--802},
  year={1967},
}

@ARTICLE{Chen_2014,
       author = {{Chen}, Yang and {Girardi}, L{\'e}o and {Bressan}, Alessandro and {Marigo}, Paola and {Barbieri}, Mauro and {Kong}, Xu},
        title = "{Improving PARSEC models for very low mass stars}",
      journal = {\mnras},
         year = 2014,
        month = nov,
       volume = {444},
       number = {3},
        pages = {2525-2543},
          doi = {10.1093/mnras/stu1605},
archivePrefix = {arXiv},
       eprint = {1409.0322},
 primaryClass = {astro-ph.SR},
       adsurl = {https://ui.adsabs.harvard.edu/abs/2014MNRAS.444.2525C}
}

@ARTICLE{Marigo_2017,
       author = {{Marigo}, Paola and {Girardi}, L{\'e}o and {Bressan}, Alessandro and {Rosenfield}, Philip and {Aringer}, Bernhard and {Chen}, Yang and {Dussin}, Marco and {Nanni}, Ambra and {Pastorelli}, Giada and {Rodrigues}, Tha{\'\i}se S. and {Trabucchi}, Michele and {Bladh}, Sara and {Dalcanton}, Julianne and {Groenewegen}, Martin A.~T. and {Montalb{\'a}n}, Josefina and {Wood}, Peter R.},
        title = "{A New Generation of PARSEC-COLIBRI Stellar Isochrones Including the TP-AGB Phase}",
      journal = {\apj},
         year = 2017,
        month = jan,
       volume = {835},
       number = {1},
          eid = {77},
        pages = {77},
          doi = {10.3847/1538-4357/835/1/77},
archivePrefix = {arXiv},
       eprint = {1701.08510},
 primaryClass = {astro-ph.SR},
       adsurl = {https://ui.adsabs.harvard.edu/abs/2017ApJ...835...77M}
}

@ARTICLE{Bressan_2012,
       author = {{Bressan}, Alessandro and {Marigo}, Paola and {Girardi}, L{\'e}o. and {Salasnich}, Bernardo and {Dal Cero}, Claudia and {Rubele}, Stefano and {Nanni}, Ambra},
        title = "{PARSEC: stellar tracks and isochrones with the PAdova and TRieste Stellar Evolution Code}",
      journal = {\mnras},
         year = 2012,
        month = nov,
       volume = {427},
       number = {1},
        pages = {127-145},
          doi = {10.1111/j.1365-2966.2012.21948.x},
archivePrefix = {arXiv},
       eprint = {1208.4498},
 primaryClass = {astro-ph.SR},
       adsurl = {https://ui.adsabs.harvard.edu/abs/2012MNRAS.427..127B}
}

@ARTICLE{2024Rottensteiner,
       author = {{Rottensteiner}, Alena and {Meingast}, Stefan},
        title = "{An empirical isochrone archive for nearby open clusters}",
      journal = {\aap},
         year = 2024,
        month = oct,
       volume = {690},
          eid = {A16},
        pages = {A16},
          doi = {10.1051/0004-6361/202347701},
archivePrefix = {arXiv},
       eprint = {2406.06691},
 primaryClass = {astro-ph.GA},
       adsurl = {https://ui.adsabs.harvard.edu/abs/2024A&A...690A..16R}
}

@ARTICLE{2024Edenhofer,
       author = {{Edenhofer}, Gordian and {Zucker}, Catherine and {Frank}, Philipp and {Saydjari}, Andrew K. and {Speagle}, Joshua S. and {Finkbeiner}, Douglas and {En{\ss}lin}, Torsten A.},
        title = "{A parsec-scale Galactic 3D dust map out to 1.25 kpc from the Sun}",
      journal = {\aap},
         year = 2024,
        month = may,
       volume = {685},
          eid = {A82},
        pages = {A82},
          doi = {10.1051/0004-6361/202347628},
archivePrefix = {arXiv},
       eprint = {2308.01295},
 primaryClass = {astro-ph.GA},
       adsurl = {https://ui.adsabs.harvard.edu/abs/2024A&A...685A..82E}
}

@article{Rousseeuw_1984,
author = {Rousseeuw, Peter},
year = {1984},
month = {12},
pages = {871-880},
title = {Least Median of Squares Regression},
volume = {79},
journal = {Journal of the American statistical association},
doi = {10.2307/2288718}
}

@article{Butler_1993,
 ISSN = {00905364, 21688966},
 URL = {http://www.jstor.org/stable/2242201},
 author = {R. W. Butler and P. L. Davies and M. Jhun},
 journal = {The Annals of Statistics},
 number = {3},
 pages = {1385--1400},
 publisher = {Institute of Mathematical Statistics},
 title = {Asymptotics for the Minimum Covariance Determinant Estimator},
 urldate = {2026-07-17},
 volume = {21},
 year = {1993}
}

@article{Rousseeuw_1999,
 ISSN = {00401706},
 URL = {http://www.jstor.org/stable/1270566},
 author = {Peter J. Rousseeuw and Katrien van Driessen},
 journal = {Technometrics},
 number = {3},
 pages = {212--223},
 publisher = {[Taylor & Francis, Ltd., American Statistical Association, American Society for Quality]},
 title = {A Fast Algorithm for the Minimum Covariance Determinant Estimator},
 urldate = {2026-07-17},
 volume = {41},
 year = {1999}
}

@ARTICLE{Gagne2018,
       author = {{Gagn{\'e}}, Jonathan and {Roy-Loubier}, Olivier and {Faherty}, Jacqueline K. and {Doyon}, Ren{\'e} and {Malo}, Lison},
        title = "{BANYAN. XII. New Members of Nearby Young Associations from GAIA-Tycho Data}",
      journal = {\apj},
         year = {2018a},
        month = jun,
       volume = {860},
       number = {1},
          eid = {43},
        pages = {43},
          doi = {10.3847/1538-4357/aac2b8},
archivePrefix = {arXiv},
       eprint = {1804.03093},
 primaryClass = {astro-ph.SR},
       adsurl = {https://ui.adsabs.harvard.edu/abs/2018ApJ...860...43G}
}

@ARTICLE{Cantat-Gaudin2020,
       author = {{Cantat-Gaudin}, T. and {Anders}, F.},
        title = "{Clusters and mirages: cataloguing stellar aggregates in the Milky Way}",
      journal = {\aap},
         year = 2020,
        month = jan,
       volume = {633},
          eid = {A99},
        pages = {A99},
          doi = {10.1051/0004-6361/201936691},
archivePrefix = {arXiv},
       eprint = {1911.07075},
 primaryClass = {astro-ph.SR},
       adsurl = {https://ui.adsabs.harvard.edu/abs/2020A&A...633A..99C}
}

@ARTICLE{Hubert_1985,
       author = {{Hubert}, Lawrence and {Arabie}, Phipps},
        title = "{Comparing partitions}",
      journal = {Journal of Classification},
         year = 1985,
        month = dec,
       volume = {2},
       number = {1},
        pages = {193--218},
          doi = {10.1007/BF01908075},
         issn = {1432-1343},
          url = {https://doi.org/10.1007/BF01908075}
}

@ARTICLE{Baraffe_2015,
       author = {{Baraffe}, Isabelle and {Homeier}, Derek and {Allard}, France and {Chabrier}, Gilles},
        title = "{New evolutionary models for pre-main sequence and main sequence low-mass stars down to the hydrogen-burning limit}",
      journal = {\aap},
         year = 2015,
        month = may,
       volume = {577},
          eid = {A42},
        pages = {A42},
          doi = {10.1051/0004-6361/201425481},
archivePrefix = {arXiv},
       eprint = {1503.04107},
 primaryClass = {astro-ph.SR},
       adsurl = {https://ui.adsabs.harvard.edu/abs/2015A&A...577A..42B}
}

@ARTICLE{Aru_2024,
       author = {{Aru}, M.-L. and {Mauc{\'o}}, K. and {Manara}, C.~F. and {Haworth}, T.~J. and {Facchini}, S. and {McLeod}, A.~F. and {Miotello}, A. and {Petr-Gotzens}, M.~G. and {Robberto}, M. and {Rosotti}, G.~P. and {Vicente}, S. and {Winter}, A. and {Ansdell}, M.},
        title = "{Kaleidoscope of irradiated disks: MUSE observations of proplyds in the Orion Nebula Cluster. I. Sample presentation and ionization front sizes<xref rid=``FN2'' ref-type=``fn''/>}",
      journal = {\aap},
         year = 2024,
        month = jul,
       volume = {687},
          eid = {A93},
        pages = {A93},
          doi = {10.1051/0004-6361/202349004},
archivePrefix = {arXiv},
       eprint = {2403.12604},
 primaryClass = {astro-ph.SR},
       adsurl = {https://ui.adsabs.harvard.edu/abs/2024A&A...687A..93A}
}

@ARTICLE{Tobin_2020,
       author = {{Tobin}, John J. and {Sheehan}, Patrick D. and {Megeath}, S. Thomas and {D{\'\i}az-Rodr{\'\i}guez}, Ana Karla and {Offner}, Stella S.~R. and {Murillo}, Nadia M. and {van 't Hoff}, Merel L.~R. and {van Dishoeck}, Ewine F. and {Osorio}, Mayra and {Anglada}, Guillem and {Furlan}, Elise and {Stutz}, Amelia M. and {Reynolds}, Nickalas and {Karnath}, Nicole and {Fischer}, William J. and {Persson}, Magnus and {Looney}, Leslie W. and {Li}, Zhi-Yun and {Stephens}, Ian and {Chandler}, Claire J. and {Cox}, Erin and {Dunham}, Michael M. and {Tychoniec}, {\L}ukasz and {Kama}, Mihkel and {Kratter}, Kaitlin and {Kounkel}, Marina and {Mazur}, Brian and {Maud}, Luke and {Patel}, Lisa and {Perez}, Laura and {Sadavoy}, Sarah I. and {Segura-Cox}, Dominique and {Sharma}, Rajeeb and {Stephenson}, Brian and {Watson}, Dan M. and {Wyrowski}, Friedrich},
        title = "{The VLA/ALMA Nascent Disk and Multiplicity (VANDAM) Survey of Orion Protostars. II. A Statistical Characterization of Class 0 and Class I Protostellar Disks}",
      journal = {\apj},
         year = 2020,
        month = feb,
       volume = {890},
       number = {2},
          eid = {130},
        pages = {130},
          doi = {10.3847/1538-4357/ab6f64},
archivePrefix = {arXiv},
       eprint = {2001.04468},
 primaryClass = {astro-ph.GA},
       adsurl = {https://ui.adsabs.harvard.edu/abs/2020ApJ...890..130T}
}

@ARTICLE{Zerjal_2024,
       author = {{{\v{Z}}erjal}, M. and {Mart{\'\i}n}, E.~L. and {P{\'e}rez-Garrido}, A.},
        title = "{Fine structure in the Sigma Orionis cluster revealed by Gaia DR3}",
      journal = {\aap},
         year = 2024,
        month = jun,
       volume = {686},
          eid = {A161},
        pages = {A161},
          doi = {10.1051/0004-6361/202347817},
archivePrefix = {arXiv},
       eprint = {2404.16923},
 primaryClass = {astro-ph.GA},
       adsurl = {https://ui.adsabs.harvard.edu/abs/2024A&A...686A.161Z}
}

@article{Kun_2008,
    author = {Kun, M. and Balog, Z. and Mizuno, N. and Kawamura, A. and Gáspár, A. and Kenyon, S. J. and Fukui, Y.},
    title = {Lynds 1622: a nearby star-forming cloud projected on Orion B?},
    journal = {Monthly Notices of the Royal Astronomical Society},
    volume = {391},
    number = {1},
    pages = {84-94},
    year = {2008},
    month = {11},
    issn = {0035-8711},
    doi = {10.1111/j.1365-2966.2008.13898.x},
    url = {https://doi.org/10.1111/j.1365-2966.2008.13898.x},
    eprint = {https://academic.oup.com/mnras/article-pdf/391/1/84/3893902/mnras0391-0084.pdf},
}

@ARTICLE{Meingast_2018,
       author = {{Meingast}, Stefan and {Alves}, Jo{\~a}o and {Lombardi}, Marco},
        title = "{VISION - Vienna Survey in Orion. II. Infrared extinction in Orion A}",
      journal = {\aap},
         year = 2018,
        month = jun,
       volume = {614},
          eid = {A65},
        pages = {A65},
          doi = {10.1051/0004-6361/201731396},
archivePrefix = {arXiv},
       eprint = {1803.01004},
 primaryClass = {astro-ph.GA},
       adsurl = {https://ui.adsabs.harvard.edu/abs/2018A&A...614A..65M}
}

@ARTICLE{Rottensteiner_2026,
       author = {{Rottensteiner}, Alena and {Petr-Gotzens}, Monika G. and {Meingast}, Stefan and {Alves}, Jo{\~a}o and {Bertin}, Emmanuel and {Bouy}, Herv{\'e} and {Piecka}, Martin and {Ratzenb{\"o}ck}, Sebastian and {Socci}, Andrea},
        title = "{Kinematics of young stellar objects in NGC 2024 based on infrared proper motions}",
      journal = {\aap},
         year = 2026,
        month = feb,
       volume = {706},
          eid = {A210},
        pages = {A210},
          doi = {10.1051/0004-6361/202557533},
archivePrefix = {arXiv},
       eprint = {2512.02124},
 primaryClass = {astro-ph.GA},
       adsurl = {https://ui.adsabs.harvard.edu/abs/2026A&A...706A.210R}
}

@ARTICLE{Posch_2025,
       author = {{Posch}, Laura and {Alves}, Jo{\~a}o and {Miret-Roig}, N{\'u}ria and {Ratzenb{\"o}ck}, Sebastian and {Gro{\ss}schedl}, Josefa and {Meingast}, Stefan and {Swiggum}, Cameren and {Konietzka}, Ralf},
        title = "{The physical properties of cluster chains}",
      journal = {\aap},
         year = 2025,
        month = jan,
       volume = {693},
          eid = {A175},
        pages = {A175},
          doi = {10.1051/0004-6361/202451312},
archivePrefix = {arXiv},
       eprint = {2410.18080},
 primaryClass = {astro-ph.GA},
       adsurl = {https://ui.adsabs.harvard.edu/abs/2025A&A...693A.175P}
}

@ARTICLE{Posch_2023,
       author = {{Posch}, Laura and {Miret-Roig}, N{\'u}ria and {Alves}, Jo{\~a}o and {Ratzenb{\"o}ck}, Sebastian and {Gro{\ss}schedl}, Josefa and {Meingast}, Stefan and {Zucker}, Catherine and {Burkert}, Andreas},
        title = "{The Corona Australis star formation complex is accelerating away from the Galactic plane}",
      journal = {\aap},
         year = 2023,
        month = nov,
       volume = {679},
          eid = {L10},
        pages = {L10},
          doi = {10.1051/0004-6361/202347186},
archivePrefix = {arXiv},
       eprint = {2310.14373},
 primaryClass = {astro-ph.GA},
       adsurl = {https://ui.adsabs.harvard.edu/abs/2023A&A...679L..10P}
}

@ARTICLE{Kerr_2021,
       author = {{Kerr}, Ronan M.~P. and {Rizzuto}, Aaron C. and {Kraus}, Adam L. and {Offner}, Stella S.~R.},
        title = "{Stars with Photometrically Young Gaia Luminosities Around the Solar System (SPYGLASS). I. Mapping Young Stellar Structures and Their Star Formation Histories}",
      journal = {\apj},
         year = 2021,
        month = aug,
       volume = {917},
       number = {1},
          eid = {23},
        pages = {23},
          doi = {10.3847/1538-4357/ac0251},
archivePrefix = {arXiv},
       eprint = {2105.09338},
 primaryClass = {astro-ph.GA},
       adsurl = {https://ui.adsabs.harvard.edu/abs/2021ApJ...917...23K}
}

@article{Rawat2024,
author  = {Rawat, Vineet and Samal, M. R. and Ojha, D. K. and Kumar, Brajesh and Sharma, Saurabh and Jose, J. and Sagar, Ram and Yadav, R. K.},
title   = {Peering into the Heart of the Giant Molecular Cloud {G148.24+00.41}: A Deep Near-infrared View of the Newly Hatched Cluster {FSR 655}},
journal = {The Astronomical Journal},
   year = {2024},
 volume = {168},
 number = {3},
  pages = {136},
    doi = {10.3847/1538-3881/ad630d},
    url = {https://doi.org/10.3847/1538-3881/ad630d}
}

@article{Getman2014,
    author = {Getman, Konstantin V. and Feigelson, Eric D. and Kuhn, Michael A. and Broos, Patrick S. and Townsley, Leisa K. and Naylor, Tim and Povich,              Matthew S. and Luhman, Kevin L. and Garmire, Gordon P.},
     title = {Age Gradients in the Stellar Populations of Massive Star-forming Regions Based on a New Stellar Chronometer},
   journal = {The Astrophysical Journal},
      year = {2014},
    volume = {787},
    number = {2},
     pages = {108},
       doi = {10.1088/0004-637X/787/2/108},
       url = {https://doi.org/10.1088/0004-637X/787/2/108}
}

@article{Hannon2022,
    author = {Hannon, Stephen and Lee, Janice C. and Whitmore, Bradley C. and Mobasher, Bahram and Thilker, David and Chandar, Rupali and Adamo, Angela and
              Wofford, Aida and Orozco-Duarte, Rogelio and Calzetti, Daniela and Della Bruna, Lorenza and Kreckel, Kathryn and Groves, Brent and Barnes, Ashley T. and Boquien, M{\'e}d{\'e}ric and Belfiore, Francesco and Linden, Sean},
     title = {{$\mathrm{H}\alpha$} Morphologies of Star Clusters in 16 {LEGUS} Galaxies: Constraints on {H\,{\sc ii}} Region Evolution Time-scales},
   journal = {Monthly Notices of the Royal Astronomical Society},
      year = {2022},
    volume = {512},
    number = {1},
     pages = {1294--1316},
       doi = {10.1093/mnras/stac550},
       url = {https://doi.org/10.1093/mnras/stac550}
}

@article{Messa2021,
    author = {Messa, Matteo and Calzetti, Daniela and Adamo, Angela and Grasha, Kathryn and Johnson, Kelsey E. and Sabbi, Elena and Smith, Linda J. and Bajaj, Varun and Finn, Molly K. and Lin, Zesen},
     title = {Looking for Obscured Young Star Clusters in {NGC 1313}},
   journal = {The Astrophysical Journal},
      year = {2021},
    volume = {909},
     pages = {121},
       doi = {10.3847/1538-4357/abe0b5},
       url = {https://doi.org/10.3847/1538-4357/abe0b5}
}

@article{Grossi2010,
    author = {Grossi, M. and Corbelli, E. and Giovanardi, C. and Magrini, L.},
     title = {Young Stellar Clusters and Associations in {M33}},
   journal = {Astronomy \& Astrophysics},
      year = {2010},
    volume = {521},
     pages = {A41},
       doi = {10.1051/0004-6361/200913513},
       url = {https://doi.org/10.1051/0004-6361/200913513}
}

@ARTICLE{deZeeuw_1999,
       author = {{de Zeeuw}, P.~T. and {Hoogerwerf}, R. and {de Bruijne}, J.~H.~J. and {Brown}, A.~G.~A. and {Blaauw}, A.},
        title = "{A HIPPARCOS Census of the Nearby OB Associations}",
      journal = {\aj},
         year = 1999,
        month = jan,
       volume = {117},
       number = {1},
        pages = {354-399},
          doi = {10.1086/300682},
archivePrefix = {arXiv},
       eprint = {astro-ph/9809227},
 primaryClass = {astro-ph},
       adsurl = {https://ui.adsabs.harvard.edu/abs/1999AJ....117..354D}
}

@INPROCEEDINGS{Hillenbrand_2001,
       author = {{Hillenbrand}, L.~A. and {Carpenter}, J.~M. and {Feigelson}, E.~D.},
        title = "{The Orion Star-Forming Region}",
    booktitle = {From Darkness to Light: Origin and Evolution of Young Stellar Clusters},
         year = 2001,
       editor = {{Montmerle}, Thierry and {Andr{\'e}}, Philippe},
       series = {Astronomical Society of the Pacific Conference Series},
       volume = {243},
        month = jan,
        pages = {439},
          doi = {10.48550/arXiv.astro-ph/0010627},
archivePrefix = {arXiv},
       eprint = {astro-ph/0010627},
 primaryClass = {astro-ph},
       adsurl = {https://ui.adsabs.harvard.edu/abs/2001ASPC..243..439H}
}

@ARTICLE{Schoettler_2020,
       author = {{Schoettler}, Christina and {de Bruijne}, Jos and {Vaher}, Eero and {Parker}, Richard J.},
        title = "{Runaway and walkaway stars from the ONC with Gaia DR2}",
      journal = {\mnras},
         year = 2020,
        month = jul,
       volume = {495},
       number = {3},
        pages = {3104-3123},
          doi = {10.1093/mnras/staa1228},
archivePrefix = {arXiv},
       eprint = {2004.13730},
 primaryClass = {astro-ph.SR},
       adsurl = {https://ui.adsabs.harvard.edu/abs/2020MNRAS.495.3104S}
}

@ARTICLE{HDBSCAN_2017,
       author = {{McInnes}, Leland and {Healy}, John and {Astels}, Steve},
        title = "{hdbscan: Hierarchical density based clustering}",
      journal = {The Journal of Open Source Software},
         year = 2017,
        month = mar,
       volume = {2},
       number = {11},
          eid = {205},
        pages = {205},
          doi = {10.21105/joss.00205},
       adsurl = {https://ui.adsabs.harvard.edu/abs/2017JOSS....2..205M}
}

@ARTICLE{Meingast_2023a,
       author = {{Meingast}, S. and {Alves}, J. and {Bouy}, H. and {VISIONS Collaboration}},
        title = "{The VISTA Star Formation Atlas (VISIONS)}",
      journal = {The Messenger},
         year = 2023,
        month = sep,
       volume = {191},
        pages = {18-22},
          doi = {10.18727/0722-6691/5336},
       adsurl = {https://ui.adsabs.harvard.edu/abs/2023Msngr.191...18M}
}

@ARTICLE{HIPPARCOS_1997,
       author = {{Perryman}, M.~A.~C. and {Lindegren}, L. and {Kovalevsky}, J. and {Hoeg}, E. and {Bastian}, U. and {Bernacca}, P.~L. and {Cr{\'e}z{\'e}}, M. and {Donati}, F. and {Grenon}, M. and {Grewing}, M. and {van Leeuwen}, F. and {van der Marel}, H. and {Mignard}, F. and {Murray}, C.~A. and {Le Poole}, R.~S. and {Schrijver}, H. and {Turon}, C. and {Arenou}, F. and {Froeschl{\'e}}, M. and {Petersen}, C.~S.},
        title = "{The HIPPARCOS Catalogue}",
      journal = {\aap},
         year = 1997,
        month = jul,
       volume = {323},
        pages = {L49-L52},
       adsurl = {https://ui.adsabs.harvard.edu/abs/1997A&A...323L..49P}
}

@ARTICLE{Gaia_DR1_2016,
       author = {{Gaia Collaboration} and {Brown}, A.~G.~A. and {Vallenari}, A. and {Prusti}, T. and {de Bruijne}, J.~H.~J. and {Mignard}, F. and {Drimmel}, R. and {Babusiaux}, C. and {Bailer-Jones}, C.~A.~L. and {Bastian}, U. and {Biermann}, M. and {Evans}, D.~W. and {Eyer}, L. and {Jansen}, F. and {Jordi}, C. and {Katz}, D. and {Klioner}, S.~A. and {Lammers}, U. and {Lindegren}, L. and {Luri}, X. and {O'Mullane}, W. and {Panem}, C. and {Pourbaix}, D. and {Randich}, S. and {Sartoretti}, P. and {Siddiqui}, H.~I. and {Soubiran}, C. and {Valette}, V. and {van Leeuwen}, F. and {Walton}, N.~A. and {Aerts}, C. and {Arenou}, F. and {Cropper}, M. and {H{\o}g}, E. and {Lattanzi}, M.~G. and {Grebel}, E.~K. and {Holland}, A.~D. and {Huc}, C. and {Passot}, X. and {Perryman}, M. and {Bramante}, L. and {Cacciari}, C. and {Casta{\~n}eda}, J. and {Chaoul}, L. and {Cheek}, N. and {De Angeli}, F. and {Fabricius}, C. and {Guerra}, R. and {Hern{\'a}ndez}, J. and {Jean-Antoine-Piccolo}, A. and {Masana}, E. and {Messineo}, R. and {Mowlavi}, N. and {Nienartowicz}, K. and {Ord{\'o}{\~n}ez-Blanco}, D. and {Panuzzo}, P. and {Portell}, J. and {Richards}, P.~J. and {Riello}, M. and {Seabroke}, G.~M. and {Tanga}, P. and {Th{\'e}venin}, F. and {Torra}, J. and {Els}, S.~G. and {Gracia-Abril}, G. and {Comoretto}, G. and {Garcia-Reinaldos}, M. and {Lock}, T. and {Mercier}, E. and {Altmann}, M. and {Andrae}, R. and {Astraatmadja}, T.~L. and {Bellas-Velidis}, I. and {Benson}, K. and {Berthier}, J. and {Blomme}, R. and {Busso}, G. and {Carry}, B. and {Cellino}, A. and {Clementini}, G. and {Cowell}, S. and {Creevey}, O. and {Cuypers}, J. and {Davidson}, M. and {De Ridder}, J. and {de Torres}, A. and {Delchambre}, L. and {Dell'Oro}, A. and {Ducourant}, C. and {Fr{\'e}mat}, Y. and {Garc{\'\i}a-Torres}, M. and {Gosset}, E. and {Halbwachs}, J.-L. and {Hambly}, N.~C. and {Harrison}, D.~L. and {Hauser}, M. and {Hestroffer}, D. and {Hodgkin}, S.~T. and {Huckle}, H.~E. and {Hutton}, A. and {Jasniewicz}, G. and {Jordan}, S. and {Kontizas}, M. and {Korn}, A.~J. and {Lanzafame}, A.~C. and {Manteiga}, M. and {Moitinho}, A. and {Muinonen}, K. and {Osinde}, J. and {Pancino}, E. and {Pauwels}, T. and {Petit}, J.-M. and {Recio-Blanco}, A. and {Robin}, A.~C. and {Sarro}, L.~M. and {Siopis}, C. and {Smith}, M. and {Smith}, K.~W. and {Sozzetti}, A. and {Thuillot}, W. and {van Reeven}, W. and {Viala}, Y. and {Abbas}, U. and {Abreu Aramburu}, A. and {Accart}, S. and {Aguado}, J.~J. and {Allan}, P.~M. and {Allasia}, W. and {Altavilla}, G. and {{\'A}lvarez}, M.~A. and {Alves}, J. and {Anderson}, R.~I. and {Andrei}, A.~H. and {Anglada Varela}, E. and {Antiche}, E. and {Antoja}, T. and {Ant{\'o}n}, S. and {Arcay}, B. and {Bach}, N. and {Baker}, S.~G. and {Balaguer-N{\'u}{\~n}ez}, L. and {Barache}, C. and {Barata}, C. and {Barbier}, A. and {Barblan}, F. and {Barrado y Navascu{\'e}s}, D. and {Barros}, M. and {Barstow}, M.~A. and {Becciani}, U. and {Bellazzini}, M. and {Bello Garc{\'\i}a}, A. and {Belokurov}, V. and {Bendjoya}, P. and {Berihuete}, A. and {Bianchi}, L. and {Bienaym{\'e}}, O. and {Billebaud}, F. and {Blagorodnova}, N. and {Blanco-Cuaresma}, S. and {Boch}, T. and {Bombrun}, A. and {Borrachero}, R. and {Bouquillon}, S. and {Bourda}, G. and {Bouy}, H. and {Bragaglia}, A. and {Breddels}, M.~A. and {Brouillet}, N. and {Br{\"u}semeister}, T. and {Bucciarelli}, B. and {Burgess}, P. and {Burgon}, R. and {Burlacu}, A. and {Busonero}, D. and {Buzzi}, R. and {Caffau}, E. and {Cambras}, J. and {Campbell}, H. and {Cancelliere}, R. and {Cantat-Gaudin}, T. and {Carlucci}, T. and {Carrasco}, J.~M. and {Castellani}, M. and {Charlot}, P. and {Charnas}, J. and {Chiavassa}, A. and {Clotet}, M. and {Cocozza}, G. and {Collins}, R.~S. and {Costigan}, G. and {Crifo}, F. and {Cross}, N.~J.~G. and {Crosta}, M. and {Crowley}, C. and {Dafonte}, C. and {Damerdji}, Y. and {Dapergolas}, A. and {David}, P. and {David}, M. and {De Cat}, P.},
        title = "{Gaia Data Release 1. Summary of the astrometric, photometric, and survey properties}",
      journal = {\aap},
         year = 2016,
        month = nov,
       volume = {595},
          eid = {A2},
        pages = {A2},
          doi = {10.1051/0004-6361/201629512},
archivePrefix = {arXiv},
       eprint = {1609.04172},
 primaryClass = {astro-ph.IM},
       adsurl = {https://ui.adsabs.harvard.edu/abs/2016A&A...595A...2G}
}

@ARTICLE{Gaia_DR2_2018,
       author = {{Gaia Collaboration} and {Brown}, A.~G.~A. and {Vallenari}, A. and {Prusti}, T. and {de Bruijne}, J.~H.~J. and {Babusiaux}, C. and {Bailer-Jones}, C.~A.~L. and {Biermann}, M. and {Evans}, D.~W. and {Eyer}, L. and {Jansen}, F. and {Jordi}, C. and {Klioner}, S.~A. and {Lammers}, U. and {Lindegren}, L. and {Luri}, X. and {Mignard}, F. and {Panem}, C. and {Pourbaix}, D. and {Randich}, S. and {Sartoretti}, P. and {Siddiqui}, H.~I. and {Soubiran}, C. and {van Leeuwen}, F. and {Walton}, N.~A. and {Arenou}, F. and {Bastian}, U. and {Cropper}, M. and {Drimmel}, R. and {Katz}, D. and {Lattanzi}, M.~G. and {Bakker}, J. and {Cacciari}, C. and {Casta{\~n}eda}, J. and {Chaoul}, L. and {Cheek}, N. and {De Angeli}, F. and {Fabricius}, C. and {Guerra}, R. and {Holl}, B. and {Masana}, E. and {Messineo}, R. and {Mowlavi}, N. and {Nienartowicz}, K. and {Panuzzo}, P. and {Portell}, J. and {Riello}, M. and {Seabroke}, G.~M. and {Tanga}, P. and {Th{\'e}venin}, F. and {Gracia-Abril}, G. and {Comoretto}, G. and {Garcia-Reinaldos}, M. and {Teyssier}, D. and {Altmann}, M. and {Andrae}, R. and {Audard}, M. and {Bellas-Velidis}, I. and {Benson}, K. and {Berthier}, J. and {Blomme}, R. and {Burgess}, P. and {Busso}, G. and {Carry}, B. and {Cellino}, A. and {Clementini}, G. and {Clotet}, M. and {Creevey}, O. and {Davidson}, M. and {De Ridder}, J. and {Delchambre}, L. and {Dell'Oro}, A. and {Ducourant}, C. and {Fern{\'a}ndez-Hern{\'a}ndez}, J. and {Fouesneau}, M. and {Fr{\'e}mat}, Y. and {Galluccio}, L. and {Garc{\'\i}a-Torres}, M. and {Gonz{\'a}lez-N{\'u}{\~n}ez}, J. and {Gonz{\'a}lez-Vidal}, J.~J. and {Gosset}, E. and {Guy}, L.~P. and {Halbwachs}, J.-L. and {Hambly}, N.~C. and {Harrison}, D.~L. and {Hern{\'a}ndez}, J. and {Hestroffer}, D. and {Hodgkin}, S.~T. and {Hutton}, A. and {Jasniewicz}, G. and {Jean-Antoine-Piccolo}, A. and {Jordan}, S. and {Korn}, A.~J. and {Krone-Martins}, A. and {Lanzafame}, A.~C. and {Lebzelter}, T. and {L{\"o}ffler}, W. and {Manteiga}, M. and {Marrese}, P.~M. and {Mart{\'\i}n-Fleitas}, J.~M. and {Moitinho}, A. and {Mora}, A. and {Muinonen}, K. and {Osinde}, J. and {Pancino}, E. and {Pauwels}, T. and {Petit}, J.-M. and {Recio-Blanco}, A. and {Richards}, P.~J. and {Rimoldini}, L. and {Robin}, A.~C. and {Sarro}, L.~M. and {Siopis}, C. and {Smith}, M. and {Sozzetti}, A. and {S{\"u}veges}, M. and {Torra}, J. and {van Reeven}, W. and {Abbas}, U. and {Abreu Aramburu}, A. and {Accart}, S. and {Aerts}, C. and {Altavilla}, G. and {{\'A}lvarez}, M.~A. and {Alvarez}, R. and {Alves}, J. and {Anderson}, R.~I. and {Andrei}, A.~H. and {Anglada Varela}, E. and {Antiche}, E. and {Antoja}, T. and {Arcay}, B. and {Astraatmadja}, T.~L. and {Bach}, N. and {Baker}, S.~G. and {Balaguer-N{\'u}{\~n}ez}, L. and {Balm}, P. and {Barache}, C. and {Barata}, C. and {Barbato}, D. and {Barblan}, F. and {Barklem}, P.~S. and {Barrado}, D. and {Barros}, M. and {Barstow}, M.~A. and {Bartholom{\'e} Mu{\~n}oz}, S. and {Bassilana}, J.-L. and {Becciani}, U. and {Bellazzini}, M. and {Berihuete}, A. and {Bertone}, S. and {Bianchi}, L. and {Bienaym{\'e}}, O. and {Blanco-Cuaresma}, S. and {Boch}, T. and {Boeche}, C. and {Bombrun}, A. and {Borrachero}, R. and {Bossini}, D. and {Bouquillon}, S. and {Bourda}, G. and {Bragaglia}, A. and {Bramante}, L. and {Breddels}, M.~A. and {Bressan}, A. and {Brouillet}, N. and {Br{\"u}semeister}, T. and {Brugaletta}, E. and {Bucciarelli}, B. and {Burlacu}, A. and {Busonero}, D. and {Butkevich}, A.~G. and {Buzzi}, R. and {Caffau}, E. and {Cancelliere}, R. and {Cannizzaro}, G. and {Cantat-Gaudin}, T. and {Carballo}, R. and {Carlucci}, T. and {Carrasco}, J.~M. and {Casamiquela}, L. and {Castellani}, M. and {Castro-Ginard}, A. and {Charlot}, P. and {Chemin}, L. and {Chiavassa}, A. and {Cocozza}, G. and {Costigan}, G. and {Cowell}, S. and {Crifo}, F. and {Crosta}, M. and {Crowley}, C. and {Cuypers}, J. and {Dafonte}, C. and {Damerdji}, Y. and {Dapergolas}, A. and {David}, P. and {David}, M. and {de Laverny}, P. and {De Luise}, F.},
        title = "{Gaia Data Release 2. Summary of the contents and survey properties}",
      journal = {\aap},
         year = 2018,
        month = aug,
       volume = {616},
          eid = {A1},
        pages = {A1},
          doi = {10.1051/0004-6361/201833051},
archivePrefix = {arXiv},
       eprint = {1804.09365},
 primaryClass = {astro-ph.GA},
       adsurl = {https://ui.adsabs.harvard.edu/abs/2018A&A...616A...1G}
}

@ARTICLE{Gaia_DR3_2023,
       author = {{Gaia Collaboration} and {Vallenari}, A. and {Brown}, A.~G.~A. and {Prusti}, T. and {de Bruijne}, J.~H.~J. and {Arenou}, F. and {Babusiaux}, C. and {Biermann}, M. and {Creevey}, O.~L. and {Ducourant}, C. and {Evans}, D.~W. and {Eyer}, L. and {Guerra}, R. and {Hutton}, A. and {Jordi}, C. and {Klioner}, S.~A. and {Lammers}, U.~L. and {Lindegren}, L. and {Luri}, X. and {Mignard}, F. and {Panem}, C. and {Pourbaix}, D. and {Randich}, S. and {Sartoretti}, P. and {Soubiran}, C. and {Tanga}, P. and {Walton}, N.~A. and {Bailer-Jones}, C.~A.~L. and {Bastian}, U. and {Drimmel}, R. and {Jansen}, F. and {Katz}, D. and {Lattanzi}, M.~G. and {van Leeuwen}, F. and {Bakker}, J. and {Cacciari}, C. and {Casta{\~n}eda}, J. and {De Angeli}, F. and {Fabricius}, C. and {Fouesneau}, M. and {Fr{\'e}mat}, Y. and {Galluccio}, L. and {Guerrier}, A. and {Heiter}, U. and {Masana}, E. and {Messineo}, R. and {Mowlavi}, N. and {Nicolas}, C. and {Nienartowicz}, K. and {Pailler}, F. and {Panuzzo}, P. and {Riclet}, F. and {Roux}, W. and {Seabroke}, G.~M. and {Sordo}, R. and {Th{\'e}venin}, F. and {Gracia-Abril}, G. and {Portell}, J. and {Teyssier}, D. and {Altmann}, M. and {Andrae}, R. and {Audard}, M. and {Bellas-Velidis}, I. and {Benson}, K. and {Berthier}, J. and {Blomme}, R. and {Burgess}, P.~W. and {Busonero}, D. and {Busso}, G. and {C{\'a}novas}, H. and {Carry}, B. and {Cellino}, A. and {Cheek}, N. and {Clementini}, G. and {Damerdji}, Y. and {Davidson}, M. and {de Teodoro}, P. and {Nu{\~n}ez Campos}, M. and {Delchambre}, L. and {Dell'Oro}, A. and {Esquej}, P. and {Fern{\'a}ndez-Hern{\'a}ndez}, J. and {Fraile}, E. and {Garabato}, D. and {Garc{\'\i}a-Lario}, P. and {Gosset}, E. and {Haigron}, R. and {Halbwachs}, J.-L. and {Hambly}, N.~C. and {Harrison}, D.~L. and {Hern{\'a}ndez}, J. and {Hestroffer}, D. and {Hodgkin}, S.~T. and {Holl}, B. and {Jan{\ss}en}, K. and {Jevardat de Fombelle}, G. and {Jordan}, S. and {Krone-Martins}, A. and {Lanzafame}, A.~C. and {L{\"o}ffler}, W. and {Marchal}, O. and {Marrese}, P.~M. and {Moitinho}, A. and {Muinonen}, K. and {Osborne}, P. and {Pancino}, E. and {Pauwels}, T. and {Recio-Blanco}, A. and {Reyl{\'e}}, C. and {Riello}, M. and {Rimoldini}, L. and {Roegiers}, T. and {Rybizki}, J. and {Sarro}, L.~M. and {Siopis}, C. and {Smith}, M. and {Sozzetti}, A. and {Utrilla}, E. and {van Leeuwen}, M. and {Abbas}, U. and {{\'A}brah{\'a}m}, P. and {Abreu Aramburu}, A. and {Aerts}, C. and {Aguado}, J.~J. and {Ajaj}, M. and {Aldea-Montero}, F. and {Altavilla}, G. and {{\'A}lvarez}, M.~A. and {Alves}, J. and {Anders}, F. and {Anderson}, R.~I. and {Anglada Varela}, E. and {Antoja}, T. and {Baines}, D. and {Baker}, S.~G. and {Balaguer-N{\'u}{\~n}ez}, L. and {Balbinot}, E. and {Balog}, Z. and {Barache}, C. and {Barbato}, D. and {Barros}, M. and {Barstow}, M.~A. and {Bartolom{\'e}}, S. and {Bassilana}, J.-L. and {Bauchet}, N. and {Becciani}, U. and {Bellazzini}, M. and {Berihuete}, A. and {Bernet}, M. and {Bertone}, S. and {Bianchi}, L. and {Binnenfeld}, A. and {Blanco-Cuaresma}, S. and {Blazere}, A. and {Boch}, T. and {Bombrun}, A. and {Bossini}, D. and {Bouquillon}, S. and {Bragaglia}, A. and {Bramante}, L. and {Breedt}, E. and {Bressan}, A. and {Brouillet}, N. and {Brugaletta}, E. and {Bucciarelli}, B. and {Burlacu}, A. and {Butkevich}, A.~G. and {Buzzi}, R. and {Caffau}, E. and {Cancelliere}, R. and {Cantat-Gaudin}, T. and {Carballo}, R. and {Carlucci}, T. and {Carnerero}, M.~I. and {Carrasco}, J.~M. and {Casamiquela}, L. and {Castellani}, M. and {Castro-Ginard}, A. and {Chaoul}, L. and {Charlot}, P. and {Chemin}, L. and {Chiaramida}, V. and {Chiavassa}, A. and {Chornay}, N. and {Comoretto}, G. and {Contursi}, G. and {Cooper}, W.~J. and {Cornez}, T. and {Cowell}, S. and {Crifo}, F. and {Cropper}, M. and {Crosta}, M. and {Crowley}, C. and {Dafonte}, C. and {Dapergolas}, A. and {David}, M. and {David}, P. and {de Laverny}, P. and {De Luise}, F. and {De March}, R.},
        title = "{Gaia Data Release 3. Summary of the content and survey properties}",
      journal = {\aap},
         year = 2023,
        month = jun,
       volume = {674},
          eid = {A1},
        pages = {A1},
          doi = {10.1051/0004-6361/202243940},
archivePrefix = {arXiv},
       eprint = {2208.00211},
 primaryClass = {astro-ph.GA},
       adsurl = {https://ui.adsabs.harvard.edu/abs/2023A&A...674A...1G}
}

@ARTICLE{APOGEE-2_2020,
       author = {{Ahumada}, Romina and {Allende Prieto}, Carlos and {Almeida}, Andr{\'e}s and {Anders}, Friedrich and {Anderson}, Scott F. and {Andrews}, Brett H. and {Anguiano}, Borja and {Arcodia}, Riccardo and {Armengaud}, Eric and {Aubert}, Marie and {Avila}, Santiago and {Avila-Reese}, Vladimir and {Badenes}, Carles and {Balland}, Christophe and {Barger}, Kat and {Barrera-Ballesteros}, Jorge K. and {Basu}, Sarbani and {Bautista}, Julian and {Beaton}, Rachael L. and {Beers}, Timothy C. and {Benavides}, B. Izamar T. and {Bender}, Chad F. and {Bernardi}, Mariangela and {Bershady}, Matthew and {Beutler}, Florian and {Bidin}, Christian Moni and {Bird}, Jonathan and {Bizyaev}, Dmitry and {Blanc}, Guillermo A. and {Blanton}, Michael R. and {Boquien}, M{\'e}d{\'e}ric and {Borissova}, Jura and {Bovy}, Jo and {Brandt}, W.~N. and {Brinkmann}, Jonathan and {Brownstein}, Joel R. and {Bundy}, Kevin and {Bureau}, Martin and {Burgasser}, Adam and {Burtin}, Etienne and {Cano-D{\'\i}az}, Mariana and {Capasso}, Raffaella and {Cappellari}, Michele and {Carrera}, Ricardo and {Chabanier}, Sol{\`e}ne and {Chaplin}, William and {Chapman}, Michael and {Cherinka}, Brian and {Chiappini}, Cristina and {Doohyun Choi}, Peter and {Chojnowski}, S. Drew and {Chung}, Haeun and {Clerc}, Nicolas and {Coffey}, Damien and {Comerford}, Julia M. and {Comparat}, Johan and {da Costa}, Luiz and {Cousinou}, Marie-Claude and {Covey}, Kevin and {Crane}, Jeffrey D. and {Cunha}, Katia and {Ilha}, Gabriele da Silva and {Dai}, Yu Sophia and {Damsted}, Sanna B. and {Darling}, Jeremy and {Davidson}, Jr., James W. and {Davies}, Roger and {Dawson}, Kyle and {De}, Nikhil and {de la Macorra}, Axel and {De Lee}, Nathan and {Queiroz}, Anna B{\'a}rbara de Andrade and {Deconto Machado}, Alice and {de la Torre}, Sylvain and {Dell'Agli}, Flavia and {du Mas des Bourboux}, H{\'e}lion and {Diamond-Stanic}, Aleksandar M. and {Dillon}, Sean and {Donor}, John and {Drory}, Niv and {Duckworth}, Chris and {Dwelly}, Tom and {Ebelke}, Garrett and {Eftekharzadeh}, Sarah and {Davis Eigenbrot}, Arthur and {Elsworth}, Yvonne P. and {Eracleous}, Mike and {Erfanianfar}, Ghazaleh and {Escoffier}, Stephanie and {Fan}, Xiaohui and {Farr}, Emily and {Fern{\'a}ndez-Trincado}, Jos{\'e} G. and {Feuillet}, Diane and {Finoguenov}, Alexis and {Fofie}, Patricia and {Fraser-McKelvie}, Amelia and {Frinchaboy}, Peter M. and {Fromenteau}, Sebastien and {Fu}, Hai and {Galbany}, Llu{\'\i}s and {Garcia}, Rafael A. and {Garc{\'\i}a-Hern{\'a}ndez}, D.~A. and {Garma Oehmichen}, Luis Alberto and {Ge}, Junqiang and {Geimba Maia}, Marcio Antonio and {Geisler}, Doug and {Gelfand}, Joseph and {Goddy}, Julian and {Gonzalez-Perez}, Violeta and {Grabowski}, Kathleen and {Green}, Paul and {Grier}, Catherine J. and {Guo}, Hong and {Guy}, Julien and {Harding}, Paul and {Hasselquist}, Sten and {Hawken}, Adam James and {Hayes}, Christian R. and {Hearty}, Fred and {Hekker}, S. and {Hogg}, David W. and {Holtzman}, Jon A. and {Horta}, Danny and {Hou}, Jiamin and {Hsieh}, Bau-Ching and {Huber}, Daniel and {Hunt}, Jason A.~S. and {Ider Chitham}, J. and {Imig}, Julie and {Jaber}, Mariana and {Jimenez Angel}, Camilo Eduardo and {Johnson}, Jennifer A. and {Jones}, Amy M. and {J{\"o}nsson}, Henrik and {Jullo}, Eric and {Kim}, Yerim and {Kinemuchi}, Karen and {Kirkpatrick}, IV, Charles C. and {Kite}, George W. and {Klaene}, Mark and {Kneib}, Jean-Paul and {Kollmeier}, Juna A. and {Kong}, Hui and {Kounkel}, Marina and {Krishnarao}, Dhanesh and {Lacerna}, Ivan and {Lan}, Ting-Wen and {Lane}, Richard R. and {Law}, David R. and {Le Goff}, Jean-Marc and {Leung}, Henry W. and {Lewis}, Hannah and {Li}, Cheng and {Lian}, Jianhui and {Lin}, Lihwai and {Long}, Dan and {Longa-Pe{\~n}a}, Pen{\'e}lope and {Lundgren}, Britt and {Lyke}, Brad W. and {Mackereth}, J. Ted and {MacLeod}, Chelsea L. and {Majewski}, Steven R. and {Manchado}, Arturo and {Maraston}, Claudia and {Martini}, Paul and {Masseron}, Thomas and {Masters}, Karen L. and {Mathur}, Savita and {McDermid}, Richard M. and {Merloni}, Andrea and {Merrifield}, Michael and {M{\'e}sz{\'a}ros}, Szabolcs and {Miglio}, Andrea and {Minniti}, Dante and {Minsley}, Rebecca and {Miyaji}, Takamitsu and {Mohammad}, Faizan Gohar and {Mosser}, Benoit and {Mueller}, Eva-Maria and {Muna}, Demitri and {Mu{\~n}oz-Guti{\'e}rrez}, Andrea and {Myers}, Adam D. and {Nadathur}, Seshadri and {Nair}, Preethi and {Nandra}, Kirpal and {Correa do Nascimento}, Janaina and {Nevin}, Rebecca Jean and {Newman}, Jeffrey A. and {Nidever}, David L. and {Nitschelm}, Christian and {Noterdaeme}, Pasquier and {O'Connell}, Julia E. and {Olmstead}, Matthew D. and {Oravetz}, Daniel and {Oravetz}, Audrey and {Osorio}, Yeisson and {Pace}, Zachary J. and {Padilla}, Nelson and {Palanque-Delabrouille}, Nathalie and {Palicio}, Pedro A.},
        title = "{The 16th Data Release of the Sloan Digital Sky Surveys: First Release from the APOGEE-2 Southern Survey and Full Release of eBOSS Spectra}",
      journal = {\apjs},
         year = 2020,
        month = jul,
       volume = {249},
       number = {1},
          eid = {3},
        pages = {3},
          doi = {10.3847/1538-4365/ab929e},
archivePrefix = {arXiv},
       eprint = {1912.02905},
 primaryClass = {astro-ph.GA},
       adsurl = {https://ui.adsabs.harvard.edu/abs/2020ApJS..249....3A}
}

@Article{numpy,
 title         = {Array programming with {NumPy}},
 author        = {Charles R. Harris and K. Jarrod Millman and St{\'{e}}fan J.
                 van der Walt and Ralf Gommers and Pauli Virtanen and David
                 Cournapeau and Eric Wieser and Julian Taylor and Sebastian
                 Berg and Nathaniel J. Smith and Robert Kern and Matti Picus
                 and Stephan Hoyer and Marten H. van Kerkwijk and Matthew
                 Brett and Allan Haldane and Jaime Fern{\'{a}}ndez del
                 R{\'{i}}o and Mark Wiebe and Pearu Peterson and Pierre
                 G{\'{e}}rard-Marchant and Kevin Sheppard and Tyler Reddy and
                 Warren Weckesser and Hameer Abbasi and Christoph Gohlke and
                 Travis E. Oliphant},
 year          = {2020},
 month         = sep,
 journal       = {Nature},
 volume        = {585},
 number        = {7825},
 pages         = {357--362},
 doi           = {10.1038/s41586-020-2649-2},
 publisher     = {Springer Science and Business Media {LLC}},
 url           = {https://doi.org/10.1038/s41586-020-2649-2}
}

@ARTICLE{scipy,
  author  = {Virtanen, Pauli and Gommers, Ralf and Oliphant, Travis E. and
            Haberland, Matt and Reddy, Tyler and Cournapeau, David and
            Burovski, Evgeni and Peterson, Pearu and Weckesser, Warren and
            Bright, Jonathan and {van der Walt}, St{\'e}fan J. and
            Brett, Matthew and Wilson, Joshua and Millman, K. Jarrod and
            Mayorov, Nikolay and Nelson, Andrew R. J. and Jones, Eric and
            Kern, Robert and Larson, Eric and Carey, C J and
            Polat, {\.I}lhan and Feng, Yu and Moore, Eric W. and
            {VanderPlas}, Jake and Laxalde, Denis and Perktold, Josef and
            Cimrman, Robert and Henriksen, Ian and Quintero, E. A. and
            Harris, Charles R. and Archibald, Anne M. and
            Ribeiro, Ant{\^o}nio H. and Pedregosa, Fabian and
            {van Mulbregt}, Paul and {SciPy 1.0 Contributors}},
  title   = {{{SciPy} 1.0: Fundamental Algorithms for Scientific
            Computing in Python}},
  journal = {Nature Methods},
  year    = {2020},
  volume  = {17},
  pages   = {261--272},
  adsurl  = {https://rdcu.be/b08Wh},
  doi     = {10.1038/s41592-019-0686-2},
}

@article{astropy:2013,
        Adsurl = {https://adsabs.harvard.edu/abs/2013A%26A...558A..33A},
        Archiveprefix = {arXiv},
        Author = {{Astropy Collaboration} and {Robitaille}, T.~P. and {Tollerud}, E.~J. and {Greenfield}, P. and {Droettboom}, M. and {Bray}, E. and {Aldcroft}, T. and {Davis}, M. and {Ginsburg}, A. and {Price-Whelan}, A.~M. and {Kerzendorf}, W.~E. and {Conley}, A. and {Crighton}, N. and {Barbary}, K. and {Muna}, D. and {Ferguson}, H. and {Grollier}, F. and {Parikh}, M.~M. and {Nair}, P.~H. and {Unther}, H.~M. and {Deil}, C. and {Woillez}, J. and {Conseil}, S. and {Kramer}, R. and {Turner}, J.~E.~H. and {Singer}, L. and {Fox}, R. and {Weaver}, B.~A. and {Zabalza}, V. and {Edwards}, Z.~I. and {Azalee Bostroem}, K. and {Burke}, D.~J. and {Casey}, A.~R. and {Crawford}, S.~M. and {Dencheva}, N. and {Ely}, J. and {Jenness}, T. and {Labrie}, K. and {Lim}, P.~L. and {Pierfederici}, F. and {Pontzen}, A. and {Ptak}, A. and {Refsdal}, B. and {Servillat}, M. and {Streicher}, O.},
        Doi = {10.1051/0004-6361/201322068},
        Eid = {A33},
        Eprint = {1307.6212},
        Journal = {\aap},
        Month = oct,
        Pages = {A33},
        Primaryclass = {astro-ph.IM},
        Title = {{Astropy: A community Python package for astronomy}},
        Volume = 558,
        Year = 2013}

@ARTICLE{astropy:2018,
               author = {{Astropy Collaboration} and {Price-Whelan}, A.~M. and
                 {Sip{\H{o}}cz}, B.~M. and {G{\"u}nther}, H.~M. and {Lim}, P.~L. and
                 {Crawford}, S.~M. and {Conseil}, S. and {Shupe}, D.~L. and
                 {Craig}, M.~W. and {Dencheva}, N. and {Ginsburg}, A. and {Vand
                erPlas}, J.~T. and {Bradley}, L.~D. and {P{\'e}rez-Su{\'a}rez}, D. and
                 {de Val-Borro}, M. and {Aldcroft}, T.~L. and {Cruz}, K.~L. and
                 {Robitaille}, T.~P. and {Tollerud}, E.~J. and {Ardelean}, C. and
                 {Babej}, T. and {Bach}, Y.~P. and {Bachetti}, M. and {Bakanov}, A.~V. and
                 {Bamford}, S.~P. and {Barentsen}, G. and {Barmby}, P. and
                 {Baumbach}, A. and {Berry}, K.~L. and {Biscani}, F. and {Boquien}, M. and
                 {Bostroem}, K.~A. and {Bouma}, L.~G. and {Brammer}, G.~B. and
                 {Bray}, E.~M. and {Breytenbach}, H. and {Buddelmeijer}, H. and
                 {Burke}, D.~J. and {Calderone}, G. and {Cano Rodr{\'\i}guez}, J.~L. and
                 {Cara}, M. and {Cardoso}, J.~V.~M. and {Cheedella}, S. and {Copin}, Y. and
                 {Corrales}, L. and {Crichton}, D. and {D'Avella}, D. and {Deil}, C. and
                 {Depagne}, {\'E}. and {Dietrich}, J.~P. and {Donath}, A. and
                 {Droettboom}, M. and {Earl}, N. and {Erben}, T. and {Fabbro}, S. and
                 {Ferreira}, L.~A. and {Finethy}, T. and {Fox}, R.~T. and
                 {Garrison}, L.~H. and {Gibbons}, S.~L.~J. and {Goldstein}, D.~A. and
                 {Gommers}, R. and {Greco}, J.~P. and {Greenfield}, P. and
                 {Groener}, A.~M. and {Grollier}, F. and {Hagen}, A. and {Hirst}, P. and
                 {Homeier}, D. and {Horton}, A.~J. and {Hosseinzadeh}, G. and {Hu}, L. and
                 {Hunkeler}, J.~S. and {Ivezi{\'c}}, {\v{Z}}. and {Jain}, A. and
                 {Jenness}, T. and {Kanarek}, G. and {Kendrew}, S. and {Kern}, N.~S. and
                 {Kerzendorf}, W.~E. and {Khvalko}, A. and {King}, J. and {Kirkby}, D. and
                 {Kulkarni}, A.~M. and {Kumar}, A. and {Lee}, A. and {Lenz}, D. and
                 {Littlefair}, S.~P. and {Ma}, Z. and {Macleod}, D.~M. and
                 {Mastropietro}, M. and {McCully}, C. and {Montagnac}, S. and
                 {Morris}, B.~M. and {Mueller}, M. and {Mumford}, S.~J. and {Muna}, D. and
                 {Murphy}, N.~A. and {Nelson}, S. and {Nguyen}, G.~H. and
                 {Ninan}, J.~P. and {N{\"o}the}, M. and {Ogaz}, S. and {Oh}, S. and
                 {Parejko}, J.~K. and {Parley}, N. and {Pascual}, S. and {Patil}, R. and
                 {Patil}, A.~A. and {Plunkett}, A.~L. and {Prochaska}, J.~X. and
                 {Rastogi}, T. and {Reddy Janga}, V. and {Sabater}, J. and
                 {Sakurikar}, P. and {Seifert}, M. and {Sherbert}, L.~E. and
                 {Sherwood-Taylor}, H. and {Shih}, A.~Y. and {Sick}, J. and
                 {Silbiger}, M.~T. and {Singanamalla}, S. and {Singer}, L.~P. and
                 {Sladen}, P.~H. and {Sooley}, K.~A. and {Sornarajah}, S. and
                 {Streicher}, O. and {Teuben}, P. and {Thomas}, S.~W. and
                 {Tremblay}, G.~R. and {Turner}, J.~E.~H. and {Terr{\'o}n}, V. and
                 {van Kerkwijk}, M.~H. and {de la Vega}, A. and {Watkins}, L.~L. and
                 {Weaver}, B.~A. and {Whitmore}, J.~B. and {Woillez}, J. and
                 {Zabalza}, V. and {Astropy Contributors}},
                title = "{The Astropy Project: Building an Open-science Project and Status of the v2.0 Core Package}",
              journal = {\aj},
                 year = 2018,
                month = sep,
               volume = {156},
               number = {3},
                  eid = {123},
                pages = {123},
                  doi = {10.3847/1538-3881/aabc4f},
        archivePrefix = {arXiv},
               eprint = {1801.02634},
         primaryClass = {astro-ph.IM},
               adsurl = {https://ui.adsabs.harvard.edu/abs/2018AJ....156..123A}
        }

@ARTICLE{astropy:2022,
               author = {{Astropy Collaboration} and {Price-Whelan}, Adrian M. and {Lim}, Pey Lian and {Earl}, Nicholas and {Starkman}, Nathaniel and {Bradley}, Larry and {Shupe}, David L. and {Patil}, Aarya A. and {Corrales}, Lia and {Brasseur}, C.~E. and {N{"o}the}, Maximilian and {Donath}, Axel and {Tollerud}, Erik and {Morris}, Brett M. and {Ginsburg}, Adam and {Vaher}, Eero and {Weaver}, Benjamin A. and {Tocknell}, James and {Jamieson}, William and {van Kerkwijk}, Marten H. and {Robitaille}, Thomas P. and {Merry}, Bruce and {Bachetti}, Matteo and {G{"u}nther}, H. Moritz and {Aldcroft}, Thomas L. and {Alvarado-Montes}, Jaime A. and {Archibald}, Anne M. and {B{'o}di}, Attila and {Bapat}, Shreyas and {Barentsen}, Geert and {Baz{'a}n}, Juanjo and {Biswas}, Manish and {Boquien}, M{'e}d{'e}ric and {Burke}, D.~J. and {Cara}, Daria and {Cara}, Mihai and {Conroy}, Kyle E. and {Conseil}, Simon and {Craig}, Matthew W. and {Cross}, Robert M. and {Cruz}, Kelle L. and {D'Eugenio}, Francesco and {Dencheva}, Nadia and {Devillepoix}, Hadrien A.~R. and {Dietrich}, J{"o}rg P. and {Eigenbrot}, Arthur Davis and {Erben}, Thomas and {Ferreira}, Leonardo and {Foreman-Mackey}, Daniel and {Fox}, Ryan and {Freij}, Nabil and {Garg}, Suyog and {Geda}, Robel and {Glattly}, Lauren and {Gondhalekar}, Yash and {Gordon}, Karl D. and {Grant}, David and {Greenfield}, Perry and {Groener}, Austen M. and {Guest}, Steve and {Gurovich}, Sebastian and {Handberg}, Rasmus and {Hart}, Akeem and {Hatfield-Dodds}, Zac and {Homeier}, Derek and {Hosseinzadeh}, Griffin and {Jenness}, Tim and {Jones}, Craig K. and {Joseph}, Prajwel and {Kalmbach}, J. Bryce and {Karamehmetoglu}, Emir and {Ka{l}uszy{'n}ski}, Miko{l}aj and {Kelley}, Michael S.~P. and {Kern}, Nicholas and {Kerzendorf}, Wolfgang E. and {Koch}, Eric W. and {Kulumani}, Shankar and {Lee}, Antony and {Ly}, Chun and {Ma}, Zhiyuan and {MacBride}, Conor and {Maljaars}, Jakob M. and {Muna}, Demitri and {Murphy}, N.~A. and {Norman}, Henrik and {O'Steen}, Richard and {Oman}, Kyle A. and {Pacifici}, Camilla and {Pascual}, Sergio and {Pascual-Granado}, J. and {Patil}, Rohit R. and {Perren}, Gabriel I. and {Pickering}, Timothy E. and {Rastogi}, Tanuj and {Roulston}, Benjamin R. and {Ryan}, Daniel F. and {Rykoff}, Eli S. and {Sabater}, Jose and {Sakurikar}, Parikshit and {Salgado}, Jes{'u}s and {Sanghi}, Aniket and {Saunders}, Nicholas and {Savchenko}, Volodymyr and {Schwardt}, Ludwig and {Seifert-Eckert}, Michael and {Shih}, Albert Y. and {Jain}, Anany Shrey and {Shukla}, Gyanendra and {Sick}, Jonathan and {Simpson}, Chris and {Singanamalla}, Sudheesh and {Singer}, Leo P. and {Singhal}, Jaladh and {Sinha}, Manodeep and {Sip{H{o}}cz}, Brigitta M. and {Spitler}, Lee R. and {Stansby}, David and {Streicher}, Ole and {{{S}}umak}, Jani and {Swinbank}, John D. and {Taranu}, Dan S. and {Tewary}, Nikita and {Tremblay}, Grant R. and {Val-Borro}, Miguel de and {Van Kooten}, Samuel J. and {Vasovi{'c}}, Zlatan and {Verma}, Shresth and {de Miranda Cardoso}, Jos{'e} Vin{'i}cius and {Williams}, Peter K.~G. and {Wilson}, Tom J. and {Winkel}, Benjamin and {Wood-Vasey}, W.~M. and {Xue}, Rui and {Yoachim}, Peter and {Zhang}, Chen and {Zonca}, Andrea and {Astropy Project Contributors}},
                title = "{The Astropy Project: Sustaining and Growing a Community-oriented Open-source Project and the Latest Major Release (v5.0) of the Core Package}",
              journal = {\apj},
                 year = 2022,
                month = aug,
               volume = {935},
               number = {2},
                  eid = {167},
                pages = {167},
                  doi = {10.3847/1538-4357/ac7c74},
        archivePrefix = {arXiv},
               eprint = {2206.14220},
         primaryClass = {astro-ph.IM},
               adsurl = {https://ui.adsabs.harvard.edu/abs/2022ApJ...935..167A}
        }

@Article{matplotlib,
  Author    = {Hunter, J. D.},
  Title     = {Matplotlib: A 2D graphics environment},
  Journal   = {Computing in Science \& Engineering},
  Volume    = {9},
  Number    = {3},
  Pages     = {90--95},
  publisher = {IEEE COMPUTER SOC},
  doi       = {10.1109/MCSE.2007.55},
  year      = 2007
}

@misc{plotly, 
author = {Plotly Technologies Inc.}, 
title = {Collaborative data science}, 
publisher = {Plotly Technologies Inc.}, 
address = {Montreal, QC}, 
year = {2015}, 
url = {https://plot.ly} 
}

@InProceedings{pandas1,
  author    = { {W}es {M}c{K}inney },
  title     = { {D}ata {S}tructures for {S}tatistical {C}omputing in {P}ython },
  booktitle = { {P}roceedings of the 9th {P}ython in {S}cience {C}onference },
  pages     = { 56 - 61 },
  year      = { 2010 },
  editor    = { {S}t\'efan van der {W}alt and {J}arrod {M}illman },
  doi       = { 10.25080/Majora-92bf1922-00a }
}

@misc{Pandas2,
  author       = {Pandas Development Team},
  title        = {pandas-dev/pandas: Pandas},
  month        = sep,
  year         = 2024,
  publisher    = {Zenodo},
  version      = {v2.2.3},
  doi          = {10.5281/zenodo.13819579},
  url          = {https://doi.org/10.5281/zenodo.13819579},
}

@article{scikit-learn,
  title={Scikit-learn: Machine Learning in {P}ython},
  author={Pedregosa, F. and Varoquaux, G. and Gramfort, A. and Michel, V.
          and Thirion, B. and Grisel, O. and Blondel, M. and Prettenhofer, P.
          and Weiss, R. and Dubourg, V. and Vanderplas, J. and Passos, A. and
          Cournapeau, D. and Brucher, M. and Perrot, M. and Duchesnay, E.},
  journal={Journal of Machine Learning Research},
  volume={12},
  pages={2825--2830},
  year={2011}
}

@misc{skopt,
  author       = {Head, Tim and
                  Kumar, Manoj and
                  Nahrstaedt, Holger and
                  Louppe, Gilles and
                  Shcherbatyi, Iaroslav},
  title        = {scikit-optimize/scikit-optimize},
  month        = oct,
  year         = 2021,
  publisher    = {Zenodo},
  version      = {v0.9.0},
  doi          = {10.5281/zenodo.5565057},
  url          = {https://doi.org/10.5281/zenodo.5565057},
}

@InProceedings{networkx,
  author =       {Aric A. Hagberg and Daniel A. Schult and Pieter J. Swart},
  title =        {Exploring Network Structure, Dynamics, and Function using NetworkX},
  booktitle =   {Proceedings of the 7th Python in Science Conference},
  pages =     {11 - 15},
  address = {Pasadena, CA USA},
  year =      {2008},
  editor =    {Ga\"el Varoquaux and Travis Vaught and Jarrod Millman},
}

@ARTICLE{astroquery:2019,
   author = {{Ginsburg}, A. and {Sip{\H o}cz}, B.~M. and {Brasseur}, C.~E. and
	{Cowperthwaite}, P.~S. and {Craig}, M.~W. and {Deil}, C. and
	{Guillochon}, J. and {Guzman}, G. and {Liedtke}, S. and {Lian Lim}, P. and
	{Lockhart}, K.~E. and {Mommert}, M. and {Morris}, B.~M. and
	{Norman}, H. and {Parikh}, M. and {Persson}, M.~V. and {Robitaille}, T.~P. and
	{Segovia}, J.-C. and {Singer}, L.~P. and {Tollerud}, E.~J. and
	{de Val-Borro}, M. and {Valtchanov}, I. and {Woillez}, J. and
	{The Astroquery collaboration} and {a subset of the astropy collaboration}
	},
    title = "{astroquery: An Astronomical Web-querying Package in Python}",
  journal = {\aj},
archivePrefix = "arXiv",
   eprint = {1901.04520},
 primaryClass = "astro-ph.IM",
     year = 2019,
    month = mar,
   volume = 157,
      eid = {98},
    pages = {98},
      doi = {10.3847/1538-3881/aafc33},
   adsurl = {https://adsabs.harvard.edu/abs/2019AJ....157...98G}
}

@INPROCEEDINGS{pyvo_TAP_2019,
       author = {{Becker}, Stefan and {Demleitner}, Markus},
        title = "{TAP Support in PyVO}",
    booktitle = {Astronomical Data Analysis Software and Systems XXVI},
         year = 2019,
       editor = {{Molinaro}, Marco and {Shortridge}, Keith and {Pasian}, Fabio},
       series = {Astronomical Society of the Pacific Conference Series},
       volume = {521},
        month = oct,
        pages = {483},
       adsurl = {https://ui.adsabs.harvard.edu/abs/2019ASPC..521..483B}
}

@inproceedings{numba,
author = {Lam, Siu Kwan and Pitrou, Antoine and Seibert, Stanley},
title = {Numba: a LLVM-based Python JIT compiler},
year = {2015},
isbn = {9781450340052},
publisher = {Association for Computing Machinery},
address = {New York, NY, USA},
url = {https://doi.org/10.1145/2833157.2833162},
doi = {10.1145/2833157.2833162},
booktitle = {Proceedings of the Second Workshop on the LLVM Compiler Infrastructure in HPC},
articleno = {7},
numpages = {6},
location = {Austin, Texas},
series = {LLVM '15}
}

@article{arviz,
doi = {10.21105/joss.09889},
url = {https://doi.org/10.21105/joss.09889},
year = {2026},
publisher = {The Open Journal},
volume = {11},
number = {119},
pages = {9889},
author = {Martin, Osvaldo A. and Abril-Pla, Oriol and Deklerk, Jordan and Axen, Seth D. and Carroll, Colin and Hartikainen, Ari and Vehtari, Aki},
title = {ArviZ: a modular and flexible library for exploratory analysis of Bayesian models},
journal = {Journal of Open Source Software}}
